\documentclass{emulateapj}
\usepackage{graphicx}
\usepackage{multirow}

\shorttitle{Evolution of SINGS spiral galaxies}
\shortauthors{Mu\~{n}oz-Mateos et al.}

\begin{document}

\title{Radial distribution of stars, gas and dust in SINGS galaxies. III. Modeling the evolution of the stellar component in galaxy disks.}

\author{J. C. Mu\~noz-Mateos\altaffilmark{1,2},
S. Boissier\altaffilmark{3},
A. Gil de Paz\altaffilmark{2},
J. Zamorano\altaffilmark{2},
R. C. Kennicutt, Jr.\altaffilmark{4,5},
J. Moustakas\altaffilmark{6},
N. Prantzos\altaffilmark{7},
J. Gallego\altaffilmark{2}
}
\altaffiltext{1}{National Radio Astronomy Observatory, 520 Edgemont
  Road, Charlottesville, VA 22903-2475, USA; jmunoz@nrao.edu}
\altaffiltext{2}{Departamento de Astrof\'{\i}sica y CC$.$ de la
  Atm\'osfera, Universidad Complutense de Madrid, Avda$.$ de la
  Complutense, s/n, E-28040 Madrid, Spain; gildepaz@gmail.com;
  jzamorano@fis.ucm.es; j.gallego@fis.ucm.es}
\altaffiltext{3}{Laboratoire d'Astrophysique de Marseille, OAMP,
  Universit\'e Aix-Marseille \& CNRS UMR 6110, 38 rue Fr\'ed\'eric
  Joliot-Curie, 13388 Marseille cedex 13, France;
  samuel.boissier@oamp.fr}
\altaffiltext{4}{Institute of Astronomy, University of Cambridge,
  Madingley Road, Cambridge CB3 0HA, UK; robk@ast.cam.ac.uk}
\altaffiltext{5}{Steward Observatory, University of Arizona, Tucson,
  AZ 85721, USA;}
\altaffiltext{6}{Center for Astrophysics and Space Sciences,
  University of California, San Diego, 9500 Gilman Drive, La Jolla, CA
  92093, USA; jmoustakas@ucsd.edu }
\altaffiltext{7}{CNRS, UMR7095, UMPC and Institut d'Astrophysique de
  Paris, F-75014, Paris, France; prantzos@iap.fr}

\begin{abstract}
  We analyze the evolution of 42 spiral galaxies in the {\it Spitzer}
  Infrared Nearby Galaxies Survey (SINGS). We make use of ultraviolet
  (UV), optical and near-infrared radial profiles, corrected for
  internal extinction using the total-infrared to UV ratio, to probe
  the emission of stellar populations of different ages as a function
  of galactocentric distance. We fit these radial profiles with models
  that describe the chemical and spectro-photometric evolution of
  spiral disks within a self-consistent framework. These backward
  evolutionary models succesfully reproduce the multi-wavelength
  profiles of our galaxies, except the UV profiles of some early-type
  disks for which the models seem to retain too much gas. From the
  model fitting we infer the maximum circular velocity of the rotation
  curve $V_{\mathrm{C}}$ and the dimensionless spin parameter
  $\lambda$. The values of $V_{\mathrm{C}}$ are in good agreement with
  the velocities measured in HI rotation curves. Even though our
  sample is not volume-limited, the resulting distribution of
  $\lambda$ is close to the lognormal function obtained in
  cosmological $N$-body simulations, peaking at $\lambda \sim 0.03$
  regardless of the total halo mass. We do not find any evident trend
  between $\lambda$ and Hubble type, besides an increase in the
  scatter for the latest types. According to the model, galaxies
  evolve along a roughly constant mass-size relation, increasing their
  scale-lengths as they become more massive. The radial scale-length
  of most disks in our sample seems to have increased at a rate of
  0.05-0.06\,kpc\,Gyr$^{-1}$, although the same cannot be said of a
  volume-limited sample. In relative terms, the scale-length has grown
  by 20-25\% since $z=1$ and, unlike the former figure, we argue that
  this relative growth rate can be indeed representative of a complete
  galaxy sample.

\end{abstract}

\keywords{galaxies: abundances --- galaxies: evolution --- galaxies: photometry --- galaxies: spiral}

\section{Introduction}
Unveiling the details that govern the formation of disk galaxies is
paramount for our understanding of the evolution of the universe as a
whole. In the currently accepted paradigm of galaxy formation,
rotating protogalactic clouds collapse within the gravitational wells
of dark matter haloes. Gas cools via radiative processes and, if it
keeps enough angular momentum, a rotationally supported gaseous disk
will eventually form (Fall \& Efstathiou 1980; White \& Frenk 1991; Mo
et al$.$ 1998). Dark matter haloes themselves grow from primordial
density fluctuations, and they are supposed to merge and evolve
according to the $\Lambda$ Cold Dark Matter model ($\Lambda$CDM;
Springel et al$.$ 2005; Spergel et al$.$ 2007).

Because gas takes longer to settle onto the disk in the outer parts,
given its larger angular momentum and the longer gravitational
collapse time, star formation should proceed on longer timescales in
the outskirts of disks than in the inner regions. Therefore, a natural
consequence of such a scenario is that disk galaxies should be
assembled from inside out (Samland \& Gerhard 2003). In particular,
the radial scale-length of exponential disks is expected to increase
with time (Brook et al. 2006; Brooks et al$.$ 2010). In principle, the
mass and size evolution of galaxies can be probed with observations at
different redshifts (see, e.g$.$ Trujillo et al$.$ 2004, 2006; Barden
et al$.$ 2005 and references therein), provided that one can properly
deal with cosmological and selection effects.

In a way complementary to this look-back approach, galactic
archaeology in the local universe has also proven useful to infer the
past evolution of galaxies. Each particular scenario of galactic
evolution should have left characteristic imprints in the radial
variation of the properties of stars, gas and dust in present-day
galaxies. If galaxies do indeed grow from inside out, stars should be
younger on average in the outer parts, leading to radial color
gradients such as those we actually observe (de Jong 1996; Bell \& de
Jong 2000; MacArthur et al$.$ 2004; Taylor et al$.$ 2005;
Mu\~noz-Mateos et al$.$ 2007). Measuring age gradients in disks from
color profiles is not straightforward, since the radial decrease in
the internal extinction and metallicity also conspire to yield bluer
colors at larger radius.

More recently, color-magnitude diagrams of resolved stellar fields in
nearby galaxies have also favored an inside-out scenario of galactic
growth (Gogarten et al$.$ 2010; Barker et al$.$ 2010). There is also
ample observational evidence that chemical abundances decrease with
galactocentric distance, both in the Milky Way (MW; Shaver et al$.$
1983; Smartt \& Rolleston 1997) and in external galaxies (Zaritsky et
al$.$ 1994; van Zee et al$.$ 1998; Pilyugin et al$.$ 2004; Moustakas
\& Kennicutt 2006; Moustakas et al$.$ 2010). Multi-zone chemical
evolution models based on the inside-out scenario usually invoke a
radially-increasing timescale of gas infall to reproduce these
metallicity gradients (Matteucci \& Francois 1989; Molla, Ferrini \&
Diaz 1996; Prantzos \& Boissier 2000; Chiappini et al$.$ 2001; Carigi
et al$.$ 2005).

In this paper, the third in a series devoted to the spatial
distribution of stars, gas and dust in nearby galaxies, we will test
the predictions of the multi-zone model of Boissier \& Prantzos (1999,
2000; BP99 and BP00 hereafter). This model describes the chemical and
spectro-photometric evolution of spiral disks in a self-consistent
framework, taking into account the radially-varying gas infall rate, a
physically motivated star formation law and a full treatment of
chemical evolution. The model is first calibrated to reproduce
observables of the MW (BP99) and then extended to other disk-like
galaxies through scaling laws resulting from the $\Lambda$CDM model
(BP00). Using only two free parameters, the maximum rotational
velocity of the rotation curve $V_{\mathrm{C}}$ and the dimensionless
spin parameter $\lambda$, the model is able to predict radial profiles
of several quantities, including multi-wavelength photometry,
metallicity and gas density among others.

In Mu\~noz-Mateos et al$.$ (2009a, Paper~I hereafter) we derived
multi-wavelength profiles from the far-ultraviolet (FUV) to the
far-infrared (FIR) for the galaxies in the {\it Spitzer} Infrared
Nearby Galaxies Survey (SINGS; Kennicutt et al$.$ 2003), which
comprises 75 objects representative of the typical galaxy population
in the Local Universe. In Mu\~noz-Mateos et al$.$ (2009b, Paper~II
hereafter) we centered our attention on the radial variation of
several physical properties of dust, such as the internal extinction,
the dust mass surface density, the abundance of polycyclic aromatic
hydrocarbons (PAHs) and the dust-to-gas ratio. In the present paper we
have focused on a subsample of 42 spiral galaxies within the SINGS
sample. We have combined the UV, optical and near-IR profiles measured
in Paper~I with the extinction profiles obtained in Paper~II, in order
to recover the intrinsic emission of stars of different ages across
the galactic disks. These stellar, extinction-free profiles have been
fitted with the models of BP00, thus testing the ability of the models
to simultaneously reproduce the multi-band profiles, while at the same
time inferring the circular velocities and spin parameters of each
object.

The purpose of the present paper is thus twofold: first, we will
verify whether the models are able to reproduce the present-day
profiles of nearby disks. Second, we will indirectly obtain the values
of $V_{\mathrm{C}}$ and $\lambda$ for each galaxy. The spin parameter
is particularly important in cosmological studies. While
$V_{\mathrm{C}}$ can be easily determined from rotation curves,
$\lambda$ is not a directly measurable quantity. Previous studies
(Syer et al$.$ 1999; Hernandez et al$.$ 2007; Cervantes-Sodi et al$.$
2008) have shown that $\lambda$ can be empirically estimated from a
combination of observed galactic properties such as the disk
scale-length, provided that some $\Lambda$CDM-based assumptions are
made. When applying this methodology to large optical datasets of
nearby galaxies, these authors found an excellent agreement between
the empiricial distribution of $\lambda$ values and the one obtained
in $N$-body simulations of hierarchical clustering. However, optical
measurements of disk scale-lengths might be biased by radial
variations of the mass-to-light ratio or the internal extinction. The
BP00 models incorporate the radial variation of gas infall (inside-out
formation) and of the star formation law, thereby accounting for
wavelength variations in the disk scale-length in a natural
way. Moreover, the surface brightness profiles we use to constrain the
models are corrected for internal extinction using robust methods (see
Paper~II). Besides, apart from UV and optical profiles, we incorporate
in our analysis near-IR ones, which are less sensitive to
mass-to-light variations and dust attenuation. This extra wavelength
coverage obviously comes at the expense of using a much more reduced
sample than in the aforementioned studies. Therefore, our analysis
cannot reach the same levels of statistical completeness, but it
should serve nevertheless as a robust foundation for future works.

Features like bulges, bars and radial mass flows are not considered in
the BP00 models. Accounting for all these effects require $N$-body
simulations, which given their complexity are usually limited to a
handful of objects. Therefore, despite their somewhat simplified
underlying assumptions, models such as the BP00 ones allow simulating
large grids of galaxies that cover a wide range of properties, and are
therefore better suited for our purposes.

This paper is organized as follows. In Section~\ref{model_description}
we provide a brief summary of the inner working of the BP00 models. In
Section~\ref{sam_data} we describe the main properties of our
subsample of SINGS spirals, and summarize the procedure used in
Paper~I to derive the surface brightness profiles. In
Section~\ref{sec_sandwich} we explain how the multi-wavelength
profiles were corrected for the effects of internal
extinction. Section~\ref{fitting} deals with the details of the
fitting procedure, and in Section~\ref{results} we present the main
results of our analysis. The main conclusions of this work are
summarized in Section~\ref{conclusions}. Finally, the two-dimensional
distribution of $\chi^2$ values resulting from the model fitting are
compiled in Appendix~\ref{app_chi2}

\section{Description of the models}\label{model_description}
In this section we broadly outline the main ingredients and underlying
assumptions of the chemo-spectrophotometric models used to fit the
multi-wavelength profiles of the SINGS galaxies. The reader is
referred to BP99 and BP00 for a more in-depth description of the
physical details of the models. Briefly, an initial model was first
developed and calibrated to reproduce several observed properties of
the MW (BP99). This model was then generalized to other spiral disks
of different sizes and masses by means of several scaling laws deduced
from the $\Lambda$CDM scenario of disk formation (BP00).

\subsection{The Milky Way model}\label{MW_model}
The Milky Way disk is modeled as several concentric rings, which are
progressively built up by accretion of primordial gas from the
halo. In the model, these annuli evolve independently from one
another, in the sense that no radial mass flows are allowed. Such
flows can actually take place in real galaxies as a result, for
instance, of the presence of bars (Sellwood \& Wilkinson 1993),
redistribution of angular momentum due to viscosity (Yoshii \&
Sommer-Larsen 1989; Ferguson \& Clarke 2001) and radial stellar
migration (Ro\u{s}kar et al$.$ 2008; Mart\'inez-Serrano et al$.$ 2009;
S\'anchez-Bl\'azquez et al$.$ 2009). In particular, stellar migration
has been proposed as a likely mechanism to explain the observed
U-shaped color profiles in galaxies (Azzollini et al$.$ 2008a; Bakos et
al$.$ 2008). Moreover, gas outflows are not included in the models,
while observations suggest that they play a role in the chemical
evolution of low-mass galaxies (Garnett 2002). Nevertheless, this is
likely not a concern in our analysis, since most of our disks are
large and massive enough. Finally, the model does not include the
bulge nor it does differentiate between the thin and thick
disk. Implementing all these phenomena in an analytic way is not
straightforward, and it would introduce many additional free
parameters whose values might be difficult to constrain. Despite the
simplifying assumption of independently-evolving rings, the model is
still successful at reproducing the radial structure of the Milky Way.

The star formation rate (SFR) surface density at each radius $r$ and
time $t$, $\Sigma_{\mathrm{SFR}}(t,r)$, depends on the local gas
density $\Sigma_{\mathrm{g}}(t,r)$ following a Schmidt law modulated
by a dynamical term:
\begin{equation}
\Sigma_{\mathrm{SFR}}(t,r)=\alpha \Sigma_{\mathrm{g}}(t,r)^{n}V(r)r^{-1}\label{eq_SFR_sam_models}
\end{equation}
Here $V(r)$ is the rotational velocity at radius $r$. The term
$V(r)r^{-1}$ is intended to mimic the conversion of gas into stars by
the periodic passage of spiral density waves (Wyse \& Silk 1989); it
can be also seen as the inverse of a dynamical timescale. Originally,
$\alpha$ was fixed in order to reproduce the local gas fraction in the
solar neighbourhood at $T=13.5$\,Gyr, and $n=1.5$ was chosen to
reproduce radial trends. Later on, in Boissier et al$.$ (2003) the
star formation law was empirically determined using a sample of nearby
galaxies. Here we adopt the values of $\alpha=0.00263$\footnote{The
  units of $\alpha$ are such that $\Sigma_{\mathrm{SFR}}(r)$ is
  measured in M$_{\odot}$\,pc$^{-2}$\,Gyr$^{-1}$,
  $\Sigma_{\mathrm{gas}}(r)$ in M$_{\odot}$\,pc$^{-2}$, $r$ in kpc and
  $V(r)$ in km\,s$^{-1}$. See also Fig.~\ref{sf_efficiency}} and
$n=1.48$ found in that work. Note that these values are very close to
the ones originally adopted in BP99 and BP00 ($\alpha=0.00364$ and
$n=1.5$).

The gas infall rate $f(r,t)$ decreases exponentially with time:
\begin{equation}
f(t,r)=A(r)e^{-t/\tau(r)}
\end{equation}
The timescale of gas accretion $\tau(r)$ is assumed to increase with
radius, from 1\,Gyr at $r=1$\,kpc to 15\,Gyr at $r=17$\,kpc. This
allows reproducing the inside-out formation of disks, since gas
settles onto the disk on longer timescales in the outer regions due to
having larger angular momentum. At $r=8$\,kpc $\tau$ is set equal to
7\,Gyr to reproduce the metallicity distribution of G-dwarf stars in
the solar neighbourhood. The normalizing factor $A(r)$ can be deduced
by integrating the infall rate until $T=13.5$\,Gyr, and then matching
the result to the current stellar mass profile of the disk:
\begin{equation}
\int_0^T f(t,r) dt=\Sigma_{\mathrm{solar\ neigh.}} e^{-(r-8\,\mathrm{kpc})/R_{\mathrm{dG}}}=\Sigma_{0\mathrm{G}} e^{-r/R_{\mathrm{dG}}}
\end{equation}
The subscript G refers to the parameters of our galaxy. According to
observations of the Milky Way, the radial scale-length is fixed to
$R_{\mathrm{dG}}=2.6$\,kpc, and the central mass density (extrapolated
from the one in the solar neighborhood) is set to
$\Sigma_{0\mathrm{G}}=1150$\,M$_{\odot}$\,pc$^{-2}$ (see BP99 for
references).

The distribution of stars for a given SFR follows a user-specified
initial mass function (IMF). Even though we are explicitly assuming
the existence of a universal IMF, this might not be necessarily the
case (see Bastian et al$.$ 2010 for a review on the
subject). Therefore, in this work we will compare the results obtained
with the IMFs of Kroupa et al$.$ (1993; K93 hereafter) and Kroupa
(2001; K01 hereafter). The K93 IMF was used in the original models,
but here we are also interested in analyzing the results yielded by a
more recent version of the IMF. The optical and near-IR fluxes, as
well as the gas quantities, change by less than 20\% between these two
IMFs, which justifies not to totally recalibrate the model. It is the
UV fluxes and metallicities that vary significantly, given the
different content in high-mass stars of both IMFs (see
section~\ref{IMF}). Of course, there exist several other
parameterizations of the IMF which are widely used in extragalactic
studies, but the comparison between the K93 and K01 IMFs presented
here should suffice for the purpose of showing the impact of varying
the relative amount of high- and low-mass stars.

Stars of different masses enrich the ISM with varying amounts of
different elements; in this regard, the model does not assume the {\it
  instantaneous recycling approximation} (Tinsley 1980), according to
which stars more massive than 1\,M$_{\odot}$ die instantly, whereas
less massive ones live forever. On the contrary, the model takes into
account the finite lifetimes of stars of different masses when
computing the chemical evolution within each ring. Moreover, the
properties of each new generation of stars (lifetimes, stellar yields,
evolutionary tracks and spectra) depend on the local metallicity at
the corresponding radius and time of formation (see BP99). The
spectrum of a given ring at time $t$ can be then computed as the sum
(both in time and mass) of the individual spectra of previously formed
stars which are still alive at time $t$.

With the assumptions outlined above, BP99 showed that their Milky Way
model is able to reproduce not only observables in the solar
neighbourhood, but also radially-dependent ones, such as profiles of
gas surface density, gas-phase oxygen abundance, SFR and supernova
rates, as well as luminosity profiles at different bands.

\subsection{Extension to other disk-like galaxies}\label{model_extension}
The previous model for the Milky Way was generalized to other disks in
BP00, by making use of the scaling laws derived by Mo et al$.$ (1998) within the
$\Lambda$CDM scenario. In this theoretical framework, galaxy formation
is usually split into two different processes: the growth of
non-baryonic dark matter haloes and the assembly of baryonic
structures within them. Gravitational instabilities amplify the
primordial density fluctuations, yielding dark matter clumps that
merge and interact with each other, acquiring angular torques during
the process. Meanwhile, baryonic gas cools and condenses within these
haloes, leading to self-gravitating structures that are able to form
stars, thus eventually giving rise to present-day galaxies.

The models of BP00 build on the mathematical formalism of Mo et al$.$ (1998), which
establishes that under certain assumptions the scaling properties of
disks depend only on two parameters: the maximum circular velocity of
the rotation curve $V_{\mathrm{C}}$ and the dimensionless spin
parameter $\lambda$:
\begin{eqnarray}
V_{\mathrm{C}}&=&[10 G H(z) M]^{1/3}\label{eq_vrot_def}\\
\lambda&=&J|E|^{1/2}G^{-1}M^{-5/2}\label{eq_lambda_def}
\end{eqnarray}

In the equations above, $M$, $J$ and $E$ are the total mass, angular
momentum and energy of the halo, $G$ is the gravitational constant and
$H(z)$ is the Hubble parameter at the redshift $z$ of halo
formation. In order to express the properties of disks in terms of
$V_{\mathrm{C}}$ and $\lambda$ alone, the following assumptions need
to be made:
\begin{enumerate}
\item The masses of disks $M_{\mathrm{d}}$ are just a few percent of
  those of their corresponding haloes. The precise value of this ratio
  is unclear, but it must conform to the baryonic fraction of the
  universe and the efficiency of disk formation. Following Mo et al$.$ (1998), the
  BP00 models assume $M_{\mathrm{d}}=0.05M$ for all disks.

\item The specific angular momentum of the disk and halo are equal
  (i.e., $J_{\mathrm{d}}/M_{\mathrm{d}}=J/M$). While this commonly
  used assumption is not strictly supported by numerical simulations,
  it is apparently required to produce disk sizes that match
  observations.

\item Variations in the formation time of the disks are ignored. It is
  now believed that the thin component of disks is assembled at $z
  \sim 1$ (Brook et al$.$ 2006), and its evolution dominates the
  inside-out growth of spirals until $z=0$ (Chiappini et al$.$
  1997). However, disks might contain stellar populations that formed
  much earlier. The concept of `formation time' is thus somehow
  ill-defined, and the BP00 models simply assume that all disks
  started forming stars at the same time, having today a fixed age of
  13.5\,Gyr.
\end{enumerate}

\begin{figure}
\begin{center}
\resizebox{1\hsize}{!}{\includegraphics{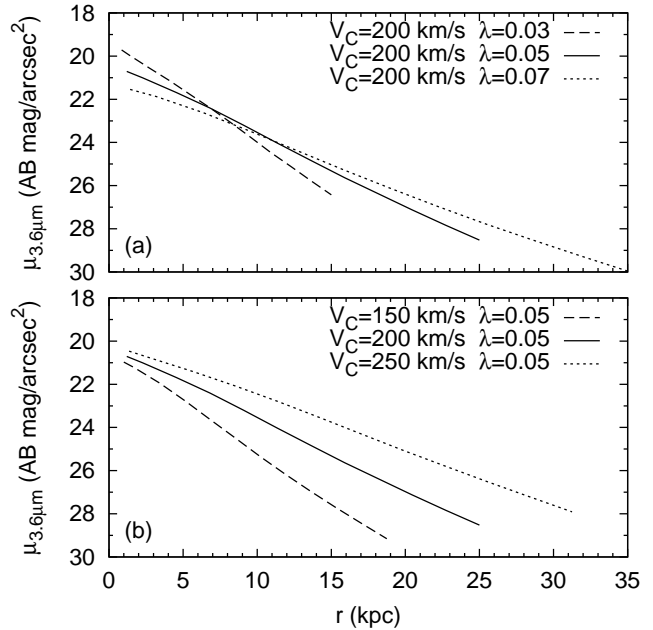}}
\caption{Sample surface brightness profiles generated by the model at
  3.6\,$\micron$, showing the effect of varying the spin parameter (a)
  and the circular velocity (b).\label{sample_model_profiles}}
\end{center}
\end{figure}

Under these assumptions, BP00 showed that the scale-length
$R_{\mathrm{d}}$ and central mass density $\Sigma_0$ of a given disk
can be derived from those of the Milky Way by means of their relative
spins and circular velocities:
\begin{eqnarray}
\frac{R_{\mathrm{d}}}{R_{\mathrm{dG}}} &=& \frac{\lambda}{\lambda_{\mathrm{G}}}\frac{V_{\mathrm{C}}}{V_{\mathrm{CG}}}\\
\frac{\Sigma_0}{\Sigma_{0\mathrm{G}}} &=& \left(\frac{\lambda}{\lambda_{\mathrm{G}}}\right)^{-2}\frac{V_{\mathrm{C}}}{V_{\mathrm{CG}}}
\end{eqnarray}
For the case of the Milky Way, the BP00 models assume that
$V_{\mathrm{CG}}=220$\,km\,s$^{-1}$ and
${\lambda_{\mathrm{G}}}=0.03$. Although both $V_{\mathrm{C}}$ and
$\lambda$ affect the final scale-length of a disk, they do it in
different ways, as can be seen in
Fig.~\ref{sample_model_profiles}. Larger values of $V_{\mathrm{C}}$
yield more extended and massive disks, while modifying $\lambda$
alters the scale-length alone.

Note that when calibrating our model on the MW, we are implicitly
assuming some sort of homology between the evolution of the galaxies
we study and the MW. This homology is observationally motivated by the
fact that several properties of galaxies such as metallicity gradients
(Garnett et al$.$ 1997; Henry \& Worthey 1999) or rotation curves
(Salucci \& Persic 1997) become 'universal' once normalized by the
optical size. Nevertheless, homology constitutes a symplifying
assumption that may be partly responsible of some of the discrepancies
that we will show later.

The scaling laws described above affect the way in which the SFR and
the gas infall time-scale depend on galactocentric distance. The final
rotation curve of a given simulated galaxy is computed as the sum of
the contributions of the halo and the disk. The resulting function
$V(r)$ is then used through Eq.~\ref{eq_SFR_sam_models} to determine
the radial variation of the SFR. The time-scale for the gas infall is
parameterized as a function of both the local mass surface surface
density and the total galaxy's mass, in the sense that a deeper
gravitational well leads to a more rapid infall of gas onto the disk
(see Eq.~21 and Fig.~3 in BP00).

\section{The data}\label{sam_data}
From the original 75 objects of the SINGS sample we first exclude all
ellipticals, lenticulars and dwarf irregulars, leaving only those
galaxies with morphological types $1 \leq T \leq 9$. From the
remaining list of galaxies we also exclude the following objects:

\begin{enumerate}
\item NGC~2798, a Sa galaxy with a severely distorted morphology due
  to its interaction with the neighbor galaxy NGC~2799.

\item NGC~3190 and NGC~4594 (The Sombrero Galaxy). These Sa galaxies
  are seen almost edge-on, with dense dust lanes heavily obscuring
  part of their disks. Besides, their prominent bulges modify the
  ellipticity of the isophotes used to measure their surface
  brightness profiles, which may not be representative of their disk
  components alone.

\item NGC~4631, an edge-on Sd galaxy for which our elliptical
  isophotes probably mix light emitted at very different
  galactocentric distances.

\item NGC~5474, a Scd galaxy with a disturbed morphology, probably
  due to a tidal interaction with M~101. In the optical and near-IR,
  its main disk is significantly shifted southwards with respect to
  the bulge.
\end{enumerate}

Even though NGC~5194 (M51a) is cleary interacting with NGC~5195
(M51b), we do not exclude it from our analysis, since it still retains
it axial symmetry. After applying these criteria we are left with 42
disk-like galaxies, whose main properties are summarized in
Table~\ref{table1}.

We refer the reader to Paper~I for a detailed description of the
imaging dataset employed here. UV images were taken with the GALEX
space telescope (Martin et al$.$ 2005) in the FUV and NUV bands, and
belong to the sample compiled in the GALEX Atlas of Nearby Galaxies
(Gil de Paz et al$.$ 2007). We employed two sets of optical images. On
one hand, we relied on $ugriz$ images from the Sloan Digital Sky
Survey DR6 (SDSS; York et al$.$ 2000; Adelman-McCarthy et al$.$ 2008)
when available. For objects without SDSS imaging, we used the original
SINGS optical images (Dale et al$.$ 2007), taken with the Kitt Peak
National Observatory (KPNO) 2.1\,m telescope and the Cerro Tololo
Inter-American Observatory (CTIO) 1.5\,m telescope. Some of these
$BVRI$ optical images were affected by zero-point offsets that were
corrected in Paper~I. Note that this recalibration procedure adds an
extra component to the photometric uncertainty of the $BVRI$ fluxes
that is not present in the SDSS ones.

Near-IR images in the $J$, $H$ and $K_S$ bands were compiled from the
2MASS Large Galaxy Atlas (LGA; Jarrett et al$.$ 2003). Finally, we
also used images at 3.6\,$\micron$ and 4.5\,$\micron$, taken with the
Infrared Array Camera (IRAC; Fazio et al$.$ 2004) onboard {\it
  Spitzer} (Werner et al$.$ 2004). Since we are interested in tracing
the stellar emission, we did not consider the other two IRAC bands at
5.8\,$\micron$ and 8.0\,$\micron$, which contain significant emission
from hot dust and PAHs. However, we did use those bands, together with
the FIR bands at 24, 70 and 160\,$\micron$ from the Multi-band Imaging
Photometer (MIPS, Rieke et al$.$ 2004) to compute the radial variation
of internal extinction from the TIR/FUV and TIR/NUV ratios (see
Paper~II and Section~\ref{sec_sandwich}).

Technical details on how the radial profiles were obtained are also
given in Paper~I. Briefly, we used the IRAF\footnote{IRAF is
  distributed by the National Optical Astronomy Observatories, which
  are operated by the Association of Universities for Research in
  Astronomy, Inc., under cooperative agreement with the National
  Science Foundation.} task {\sc ellipse} to measure the mean surface
brightness along concentric elliptical isophotes, using the same sets
of ellipses at all bands for each galaxy. The ellipticity and position
angle were kept fixed and equal to those of the $\mu_{B}$ = 25 mag
arcsec$^{-2}$ isophote from the RC3 catalog (de Vaucouleurs et al$.$
1991), or from the NASA Extragalactic Database (NED) when these
parameters were not included in the RC3. They are quoted in
Table~\ref{table1} together with the central coordinates of the
ellipses, which were also maintained fixed. We used radial increments
of 6\arcsec\ along the semimajor axis (similar to the FWHM of the
GALEX PSF) up to a final radius at least 1.5 times the diameter at
$\mu_{B}$ = 25 mag arcsec$^{-2}$ (D25). This upper limit was increased
when significant emission was seen beyond that radius (especially in
the UV bands). The uncertainty of the mean surface brightness at each
radius comprises the Poisson noise in the source flux and the error in
the sky level, the latter including both local and large-scale
background variations (see Paper~I).

\section{Internal extinction correction}\label{sec_sandwich}
Prior to fitting the multi-wavelength profiles, we must first correct
them for the radial variation of the internal
attenuation\footnote{Note that when interacting with dust grains,
  photons can be absorbed, scattered out of the line of sight and
  scattered back into it. The term ``extinction'' refers to the first
  two processes, while ``attenuation'' encompasses all of them. When
  talking about external galaxies the term ``attenuation'' is
  preferred, since it takes into account the complex radiative
  transfer processes resulting form the relative geometry of stars and
  dust. Nevertheless, with this caveat in mind we will use both terms
  interchangeably throughout this paper.}. In Paper~II we computed
internal attenuation profiles in the FUV and NUV bands independently
from the TIR/FUV and TIR/NUV ratios, respectively.  We followed the
prescriptions of Cortese et al$.$ (2008), which take into account the
varying extra dust heating due to evolved stellar populations. After
$A_\mathrm{FUV}$ and $A_\mathrm{NUV}$ have been obtained, the
attenuation at other wavelengths can be derived after assuming a given
extinction law and a geometry for the distribution of stars and
dust. Here we follow the prescriptions of Boselli et al$.$ (2003) and
adopt a sandwich model, where a thin layer of dust is embedded in a
thicker layer of stars:
\begin{eqnarray}
A_i(\lambda)&=&-2.5 \log \Bigg(\left[\frac{1-\zeta(\lambda)}{2}\right]\left(1+e^{-\tau(\lambda)\sec(i)}\right)\nonumber\\
&&+\left[\frac{\zeta(\lambda)}{\tau(\lambda)\sec(i)}\right]\left(1-e^{-\tau(\lambda)\sec(i)}\right)\Bigg)\label{eq_sandwich}
\end{eqnarray}
Here $\tau(\lambda)$ is the face-on optical depth and $i$ is the
inclination angle. Note that we need not to care about them separately
as $\tau(\lambda) \sec(i)$ is a joint quantity. The variable
$\zeta(\lambda)$ denotes the ratio between the thickness of the dust
and stars layers. Young stars, which dominate the emission in the UV
range, are likely immersed in a thin dust layer. More evolved ones,
which emit most of their light predominantly in the optical and
near-IR bands, migrate with time out of the galactic plane, and are
thus assumed to lie within a thicker layer, partly above and below the
thin dust layer. Therefore, Boselli et al$.$ (2003) parameterize the
dust-to-stars scale-height ratio as a decreasing function of $\lambda$:
\begin{equation}
\zeta(\lambda)=1.0867-5.501 \times 10^{-5}\lambda\label{eq_scaleheight}
\end{equation}
where $\lambda$ is measured in \AA. Equation~\ref{eq_scaleheight} was
obtained from the $\lambda$-dependent scale-height ratios given in
Boselli \& Gavazzi (1994), by averaging the optically thin and
optically thick cases. In the UV ($\lambda \simeq 2000$\AA), this
ratio is $\zeta = 1$, so Eq.~\ref{eq_sandwich} reduces to a slab model
and can be numerically inverted:
\begin{eqnarray}
\tau(UV)\sec(i)&=&0.0259+1.2002\times A_i(UV) + 1.5543\times A_i(UV)^2\nonumber\\
&&-0.7409\times A_i(UV)^3+0.2246\times A_i(UV)^4\label{eq_slab_invert}
\end{eqnarray}

Once the optical depth in the UV is known, the corresponding value at
any other wavelength is given by a particular extinction law
$k(\lambda)$:
\begin{equation}
\tau(\lambda)=\tau(UV) \times k(\lambda)/k(UV)\label{eq_ext_law}
\end{equation}
which plugged into Eq.~\ref{eq_sandwich} gives us the attenuation at
the desired band.

The conversion between the TIR/FUV and TIR/NUV ratios into
$A_\mathrm{FUV}$ and $A_\mathrm{NUV}$, respectively, is rather
insensitive to the adopted extinction law (Cortese et al$.$
2008). Nevertheless, to compute the extinction at other wavelengths we
must not only choose a particular extinction law, but also decide
whether to determine $A(\lambda)$ by extrapolating from
$A_\mathrm{FUV}$ or $A_\mathrm{NUV}$ in the equations above. Here we
use the MW extinction law of Li \& Draine (2001), assuming
$R_{V}=3.1$. Other extinction curves are possible, but most of them
agree pretty well from the near-IR to the NUV bands (Gordon et al$.$
2003). It is beyond the 2175\,\AA\ bump that large differences
arise. Therefore, in order to minimize the impact of our particular
choice of extinction law, we use the NUV band rather than the FUV one
in Eqs.~\ref{eq_slab_invert} and \ref{eq_ext_law}.

\section{Fitting procedure}\label{fitting}
In order to find the model that best fits the observed multiwavelength
profiles for each galaxy, a $\chi^2$ minimization procedure was
followed. We generated a grid of models with velocities ranging
between 130 and 250\,km\,s$^{-1}$ in steps of 10\,km\,s$^{-1}$, plus
extra values of 40, 80, 290 and 360\,km\,s$^{-1}$.  As for the spin
parameter, we sampled the interval $0.02 \leq \lambda \leq 0.09$ in
steps of 0.01; besides, we added $\lambda=0.10,0.15$ and $0.20$ to our
grid in order to account for possible low surface brightness galaxies
(LSBs).

Taking this set of 187 pre-computed models as our starting point, we
used a 2D interpolation algorithm to generate a finer grid of models
with steps of 0.001 in $\lambda$ and 1\,km\,s$^{-1}$ in
$V_{\mathrm{C}}$. We verified that any given property of a model
galaxy at a certain radius varies smoothly enough with $\lambda$ and
$V_{\mathrm{C}}$, so that the corresponding value for a model with an
intermediate spin and velocity can be indeed approximated by means of
a 2D interpolation.

The total $\chi^2$ of each model was computed by summing over
data-points at all bands and galactocentric distances. By visually
inspecting the multi-wavelength profiles, we excluded from the fit
those radial ranges in which the overall emission is dominated by the
bulge. In those galaxies with sharp outer truncations or
anti-truncations $-$which the BP00 models cannot reproduce by
construction$-$ the outermost regions were excluded as well. This
affects NGC~1512, NGC~3621, NGC~4625 and NGC~4736, so for these
objects the results of the fitting only concern the bright inner
disk\footnote{More galaxies in the sample exhibit multi-sloped
  profiles, but their breaks are considerably smeared out in the
  optical and especially in the near-IR. Indeed, excluding the outer
  disks does not perceptively change the output of the fit in those
  cases. For NGC~3031, NGC~5194 and NGC~6946, though, we did exclude
  the very outermost parts of the profiles, due to contamination of
  noisy background structures.} . The radial range used for the fit in
each galaxy is quoted in Table~\ref{table1}.

The resulting distribution of $\chi^2$ values are shown in
Appendix~\ref{app_chi2}. If they are to be used to derive confidence
intervals for the fitted parameters, rather than just to find the
best-fitting values, then a proper determination of the uncertainties
of each data-point and of the models is mandatory. In principle, the
$\chi^2$ method assumes that any deviation of the observed values with
respect to the model predictions is entirely due to measurement
errors. If these errors are properly accounted for when computing
$\chi^2$, then the confidence intervals for the fitted parameters are
defined by all models with $\chi^2 <
\chi^2_{\mathrm{min}}+\Delta\chi^2$, where $\Delta\chi^2$ depends on
the confidence level and the number of parameters that are being
estimated simultaneously (see e.g$.$ Avni 1976; Press et al$.$ 1992).

However, we cannot strictly follow this approach in our case, because
the models do not reproduce the small-scale structures of real
disks. In an attempt to overcome this problem, we first run our
fitting code assuming that the total uncertainty for each data-point
is the quadratic sum of the photometric and zero-point errors, plus an
extra uncertainty of 10\%. This additional term serves as an initial
guess for the intrinsic error of the model, and also avoids giving
excessive weight to any particular band and/or data-point. The typical
reduced $\chi^2$ at this stage is of the order of $\sim 5$. We then
compute the relative $rms$ of the best-fitting model with respect to
the galaxy's profiles, both as a function of radius and wavelength. In
this way we can estimate how well we can expect the model to fit that
particular galaxy at each band. These `error profiles' are then fed to
the code in a second run, in place of the initial uncertainties. The
new reduced $\chi^2$ values are now close to 1, by
construction. However, the purpose of this two-stage fitting process
is not to artificially bring the reduced $\chi^2$ closer to unity, but
to properly take into account deviations due to small-scale features
that the models, by construction, are not able to reproduce.

Even after following this process, we found that the technique of
adding a constant $\Delta\chi^2$ offset to the total (i.e. not
reduced) $\chi^2$ still yielded unrealistically small confidence
intervals for $V_{\mathrm{C}}$ and $\lambda$. A visual inspection
confirmed that indeed many models outside these confidence regions
were still in very good agreement with the observed profiles. Thus, we
finally opted for defining the boundaries of the confidence intervals
with those models whose total $\chi^2$ is twice the minimum
one. Therefore, although the resulting errors in $\lambda$ and
$V_{\mathrm{C}}$ cover the range of models that visually agree with
the galaxy's profiles, they are indicative and should not be
interpreted in a strict statistical way (see also Boselli et al$.$
2006 in this regard).

\begin{figure}
\begin{center}
\resizebox{1\hsize}{!}{\includegraphics{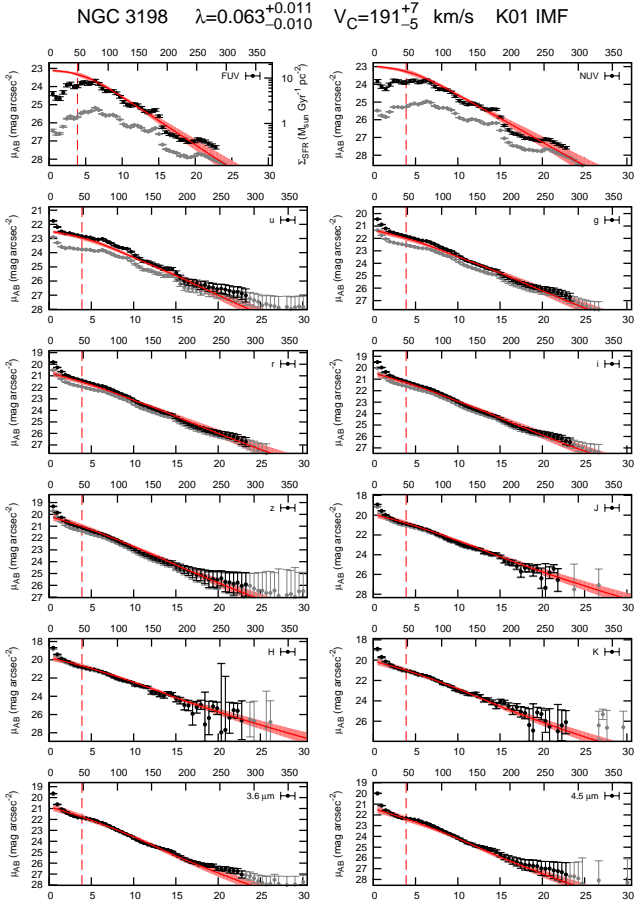}}
\caption{Best-fitting model for the Sc galaxy NGC~3198, using the K01
  IMF. The gray points show the observed profiles, corrected only for
  Milky way extinction, while the black ones also include a correction
  for the radial variation of internal extinction. Both profiles have
  been deprojected to their face-on values using the galaxy's
  morphological axis ratio. In each panel, the radius along the
  semimajor axis is expressed both in arcseconds (top $x$ axis) and in
  kpc (bottom $x$ axis). The fit is applied to those points in the
  extinction-corrected profiles beyond the dashed red line, in order
  to exclude the bulge. The red curve corresponds to the best-fitting
  model, and the shaded area contains all models with $\chi^2 \leq
  2\chi^2_{\mathrm{min}}$.\label{ngc3198_fit}}
\end{center}
\end{figure}

As an example, in Fig.~\ref{ngc3198_fit} we show the resulting fit for
the Sc spiral NGC~3198 using the K01 IMF (see the on-line edition of
the journal for similar plots for the remaining galaxies). The gray
data-points show the observed profiles, corrected only for MW
extinction, while the black ones are also corrected for internal
extinction. Both profiles have been deprojected to their face-on
values by means of the galaxy's morphological axis ratio\footnote{Note
  that, strictly speaking, this deprojection is only valid for the
  profiles corrected for internal extinction, which are the ones used
  in the fit. In the observed profiles, the difference between the
  inclined and face-on values would not just owe to a simple
  geometrical projection effect, since the interaction between
  starlight and dust along a different line of sight would also play a
  role.}. The fit is applied to the profiles corrected for internal
extinction, and only to those points beyond the red dashed line, which
separates the bulge- and disk-dominated regions of the profiles. In
the few cases where we had to exclude the outer regions (due to strong
up-bendings, for instance), the outer limit is marked with a blue
dashed line. The best-fitting model is shown with a red line, and the
band with a lighter shade of red contains all models with $\chi^2 \leq
2\chi^2_{\mathrm{min}}$.

\begin{figure}
\begin{center}
\resizebox{1\hsize}{!}{\includegraphics{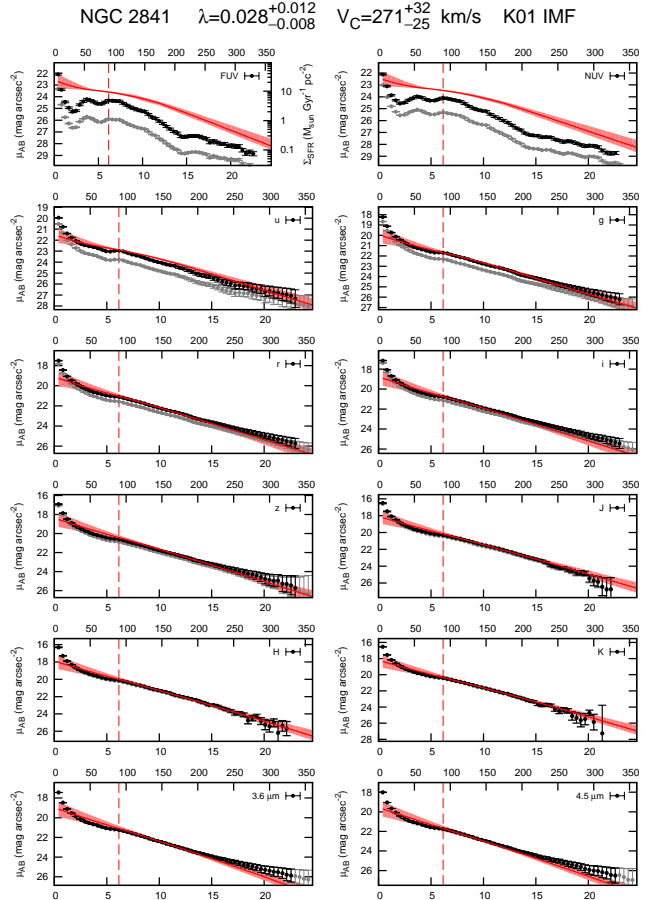}}
\caption{Same as Fig.~\ref{ngc3198_fit}, but for the Sb spiral
  NGC~2841. The fit is excellent at all wavelengths except in the FUV
  and NUV bands.\label{ngc2841_fit}}
\end{center}
\end{figure}

However, the fit is not always equally good at all wavelengths. In
Fig.~\ref{ngc2841_fit} we show the best fitting model for the Sb
galaxy NGC~2841. Even though the quality of the fit is excellent all
the way from 4.5\,$\micron$ to the $u$ band, the model overpredicts
the luminosity of the galaxy in the GALEX bands. This tends to happen
mostly in early-type spirals, as will be discussed in
section~\ref{IMF}.

\section{Results}\label{results}
\subsection{Global properties}
The results of the fitting procedure are quoted in
Table~\ref{table2}. Prior to going further in our analysis, we will
briefly describe the statistical distribution of the model parameters
$\lambda$ and $V_{\mathrm{C}}$. As mentioned before, for each galaxy
we have run our fitting code using both the K93 and K01 IMFs. Given
their different content in high-mass stars at a fixed total mass, the
resulting profiles differ in the UV bands, but agree in the optical
and near-IR ones. The effects of choosing one IMF or another will be
discussed in detail in section~\ref{IMF}, but for now it will suffice
to say that neither $\lambda$ nor $V_{\mathrm{C}}$ are significantly
affected by our particular choice of IMF. Therefore, hereafter we use
the K93 IMF as our default choice, unless otherwise mentioned.

\subsubsection{Statistical distribution of the model parameters}\label{stat_properties}

\begin{figure*}
\begin{center}
\resizebox{0.45\hsize}{!}{\includegraphics{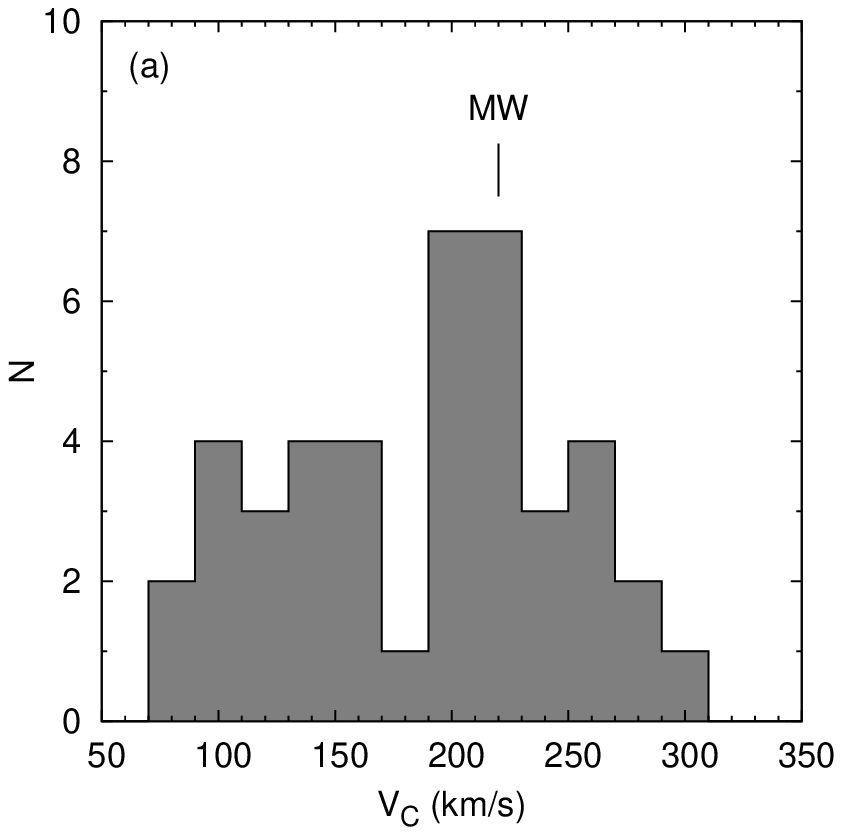}}
\resizebox{0.45\hsize}{!}{\includegraphics{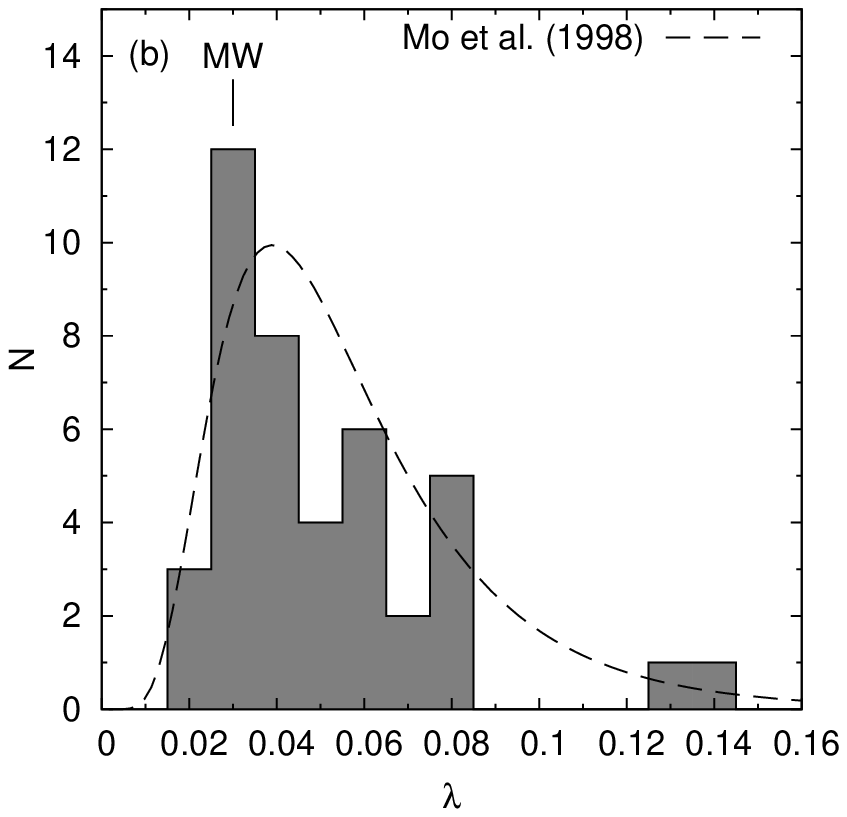}}
\caption{Distribution of the circular velocity (a) and spin (b)
  in our sample. The values adopted for the Milky-Way ($\lambda=0.03$
  and $V_{\mathrm{C}}=220$\,km\,s$^{-1}$) have been marked in both
  panels. The dashed curve corresponds to the probability distribution
  of $\lambda$ proposed by Mo et al$.$ (1998), scaled to match our
  histogram.\label{Vrot_spin_hist}}
\end{center}
\end{figure*} 

In Fig.~\ref{Vrot_spin_hist} we show the resulting histograms of both
fitting parameters. It can be seen that most galaxies exhibit values
of $\lambda$ and $V_{\mathrm{C}}$ similar to those of the Milky
Way. In particular, the distribution of rotational velocities peaks at
200-220\,km\,s$^{-1}$. It should be noted, however, that neither the
SINGS sample nor the smaller subsample of disks considered here are
complete. The well-known Schechter (1976) function can be used to fit
not only the mass and luminosity functions of nearby galaxies (Bell et
al$.$ 2003), but also their circular velocity distribution (Gonzalez
et al$.$ 2000; Boissier et al$.$ 2000). In this sense, low-mass
slow-rotating disks are known to outnumber more massive and
faster-rotating ones. Therefore, the velocity distribution shown in
Fig.~\ref{Vrot_spin_hist} obviously underestimates the number of
low-velocity galaxies that would be found in a volume-limited sample.

Regarding the spin parameter, most disks in our sample have $\lambda
\sim 0.03$, the same spin we have adopted for the MW. Had we chosen a
different MW spin, the resulting distribution of $\lambda$ would have
been shifted accordingly. Such a peaked histogram is in agreement with a key
prediction of $\Lambda$CDM simulations of galaxy formation: the fact
that most haloes exhibit the same angular momentum per unit of mass at
any epoch, regardless of their total mass and their particular history
of mass assembly. In this sense, the following analytic expression is
known to fit the distribution of $\lambda$ obtained in $N$-body
simulations simulations (see e.g$.$ Barnes \& Efstathiou 1987; Warren
et al$.$ 1992; Gardner 2001; Bullock et al$.$ 2001; Vitvitska et al$.$
2002):
\begin{equation}
p(\lambda)d\lambda=\frac{1}{\sqrt{2\pi}\sigma_\lambda}\exp\left[-\frac{\ln^2(\lambda/\overline{\lambda})}{2\sigma^2_\lambda}\right]\frac{d\lambda}{\lambda}\label{spin_function}
\end{equation}

In particular, Mo et al$.$ (1998) adopt $\overline{\lambda}=0.05$ and
$\sigma_\lambda=0.5$. Since it is a log-normal function, these values
should not be understood as the mean and width of the distribution. In
fact, this function peaks around $\lambda \sim 0.04$, and has a width
of $\sim 0.05$. Interestingly, the distribution of spin values of our
galaxies agrees well with Eq.~\ref{spin_function}. This implies that
even though our sample is not representative of a complete one, to
some extent it behaves as if it was with regard to $\lambda$ and, by
extension, to any other quantity that depends primarily on $\lambda$
rather than on $V_{\mathrm{C}}$.

\subsubsection{Comparison with observed values}
Before further proceeding with any detailed analysis, we must first
verify that the values of $\lambda$ and $V_{\mathrm{C}}$ that our
fitting code yields for each galaxy agree with the observed ones. This
comparison is not a straightforward task in the case of the spin
parameter: being a model-dependent quantity, it cannot be directly
measured in real galaxies in the same fashion as the rotational
velocity. For instance, the scaling laws adopted by Hern\'andez \&
Cervantes-Sodi (2006) would yield somewhat smaller velocities and
spins than the BP00 models for galaxies more massive than the MW, and
viceversa (assuming the same reference values of $\lambda$ and
$V_\mathrm{C}$ for the MW in both models). Nevertheless, as commented
above, Fig.~\ref{Vrot_spin_hist} shows that the distribution of
$\lambda$ in our sample resembles the one usually found in numerical
N-body simulations.

\begin{figure*}
\begin{center}
\resizebox{0.49\hsize}{!}{\includegraphics*[67,50][322,300]{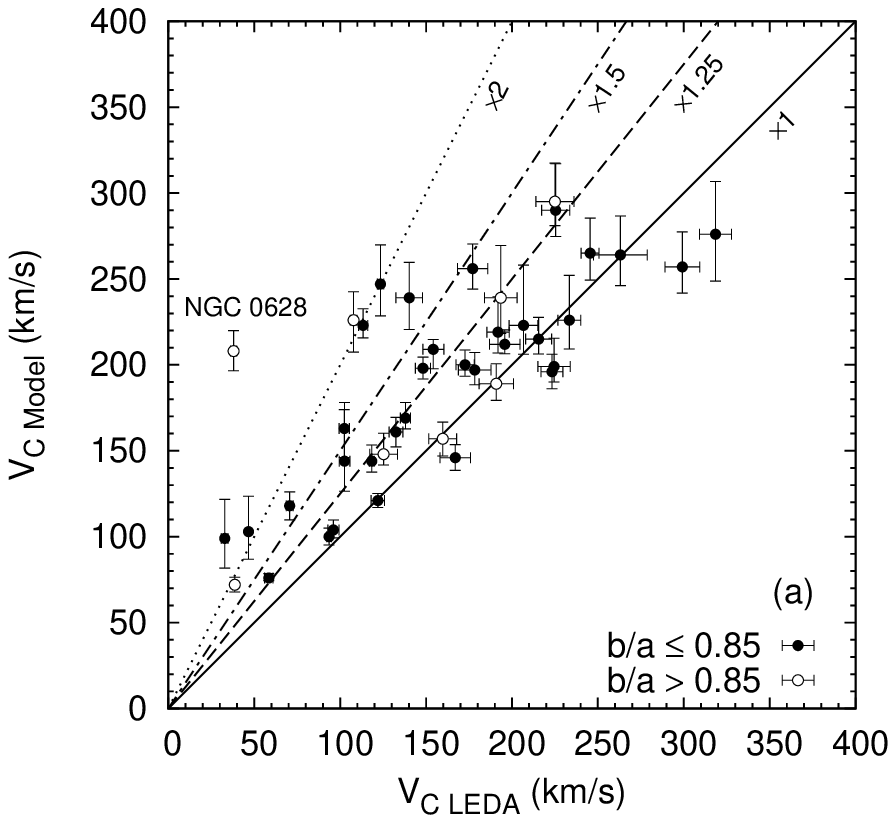}}
\resizebox{0.49\hsize}{!}{\includegraphics*[67,50][322,300]{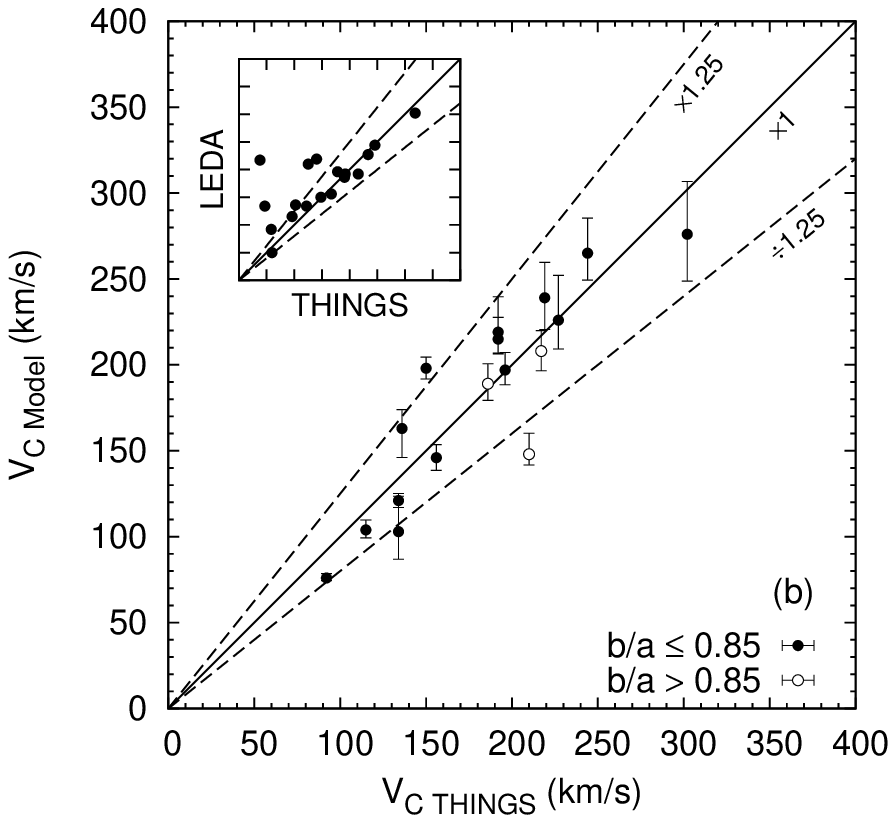}}\\
\caption{(a): Comparison between the circular velocities derived from
  the model fitting and the observed ones compiled in the LEDA
  database. Filled symbols correspond to galaxies with a minor to
  major axis ratio $b/a \leq 0.85$, while open ones show almost
  face-on galaxies with $b/a > 0.85$. (b): Same as panel (a), but
  using as a reference the circular velocities derived from the THINGS
  rotation curves. These were fitted by Leroy et al$.$ (2008) with an
  analytic expression, one of whose parameters is the velocity in the
  flat regime of the rotation curve, which is the value we plot
  here. The small inset shows a comparison between the THINGS
  velocities and those quoted in LEDA for the same
  objects.\label{Vrot_check}}
\end{center}
\end{figure*}

In order to check the accuracy of our circular velocities, in
Fig.~\ref{Vrot_check}a we compare the theoretical values given by the
model with the observed rotational velocities retrieved from the
Lyon-Meudon Extragalactic
Database\footnote{\url{http://leda.univ-lyon1.fr}} (LEDA; Paturel et
al$.$ 2003). The latter are determined from the width of the 21\,cm
hydrogen line at different levels and/or from rotation curves, usually
H$\alpha$ ones. The final values provided by LEDA are homogenized and
corrected for inclination. In general, our theoretical values for the
circular velocity are in agreement with the observed ones, which in
some sense is expected, given that the model incorporates a
Tully-Fisher (TF) relation through the adopted $\Lambda$CDM scaling
laws (see discussion below).

However, for many galaxies our velocities tend to be about 25\% larger
than those quoted in LEDA. The most discrepant outlier in this plot is
the Sc spiral NGC~0628, for which our fitting code yields
$V_{\mathrm{C}} = 208$\,km\,s$^{-1}$, while in LEDA we find a much
lower value of 38\,km\,s$^{-1}$. This latter velocity is surprisingly
small given that NGC~0628 has an absolute $K_S$-band magnitude of
$-21.64$\,mags, for which one should expect a rotational velocity of
$\sim 175$\,km\,s$^{-1}$ according to the TF relation (see below). The
fact that this galaxy is almost face-on might introduce large
uncertainties in the inclination correction, thus possibly making the
LEDA velocity very uncertain for this galaxy. The open circles in
Fig.~\ref{Vrot_check}a show, however, that the LEDA values for many
other face-on galaxies in the sample agree well with the ones obtained
from the model.

There exist in the literature more accurate kinematical data for some
of our galaxies. Daigle et al$.$ (2006) and Dicaire et al$.$ (2008)
obtained H$\alpha$ rotation curves for the SINGS galaxies using
Fabry-P\'erot interferometry. Also, the SINGS sample overlaps with The
HI Nearby Galaxies Survey (THINGS; Walter et al$.$ 2008), for which HI
rotation curves were derived by de Blok et al$.$ (2008). Here we make
use of the results of Leroy et al$.$ (2008), who parameterized the
rotation curves of the THINGS galaxies with the following analytical
expression:
\begin{equation}
v_{\mathrm{rot}}(r)=v_{\mathrm{flat}}\left[1-\exp\left(-r/l_{\mathrm{flat}}\right)\right]\label{eq_vrot}
\end{equation}
where $v_{\mathrm{rot}}$ is the circular rotational velocity at a
radius $r$, $v_{\mathrm{flat}}$ is the asymptotic velocity where the
rotation curve is flat, and $l_{\mathrm{flat}}$ is the radial scale
length over which $v_{\mathrm{flat}}$ is reached. Leroy et al$.$ (2008) derived
$v_{\mathrm{flat}}$ and $l_{\mathrm{flat}}$ from the high resolution
rotation curves presented in de Blok et al$.$ (2008), as well as from
the first moment maps for those low-inclination galaxies not included
in de Blok et al$.$ (2008).

In Fig.~\ref{Vrot_check}b we compare the circular velocities derived
from the model fitting with the $v_{\mathrm{flat}}$ values computed by
Leroy et al$.$ (2008), for the 17 galaxies we have in common. Contrary
to what happens with the LEDA velocities in Fig.~\ref{Vrot_check}a, no
systematic shift appears. Our velocities are within 20-25\% from the
THINGS ones. What is more, this scatter is uncorrelated with the one
between the predicted and observed photometric profiles. The
velocities quoted in LEDA for some galaxies in the sample of Leroy et
al$.$ (2008) tend to be lower than the values derived by those
authors, and part of the discrepancy seems to be associated with
differences in the corrections for inclination. Given the exquisite
quality of the THINGS data and the homogeneity in the derivation of
the rotation curves, we conclude that most of the systematic offset
seen in Fig.~\ref{Vrot_check}a is likely an issue of the LEDA values.

Another way to check the validity of our model rotational velocities
consists of trying to reproduce the Tully-Fisher relation (Tully
\& Fisher 1977). This tight empirical relation links the intrinsic
luminosity of a galaxy with the amplitude of its rotation curve. The
former quantity traces the stellar mass, while the latter probes the
total gravitational mass. Therefore, any successful model of disk
evolution must be able to reproduce this observed correlation. From a
observational point of view, numerous studies have shown how the slope
and zero-point of the TF relation vary with wavelength in the optical
and near-IR, mainly due to changes in the mass-to-light ratio,
together with extinction if it is not properly accounted for (Bell \&
de Jong 2001; Courteau et al$.$ 2007; Pizagno et al$.$ 2007; Blanton
\& Moustakas 2009).

The scaling laws adopted by BP00 imprint a built-in TF relation in the
models through Eq.~\ref{eq_vrot_def}. However, once star formation is
implemented in a self-consistent way, the resulting slope and
zero-point of the TF relation might vary with wavelength due to
changes in the mass-to-light ratio (see also Ferreras \& Silk
2001). When comparing the TF relation resulting from the models with
several empirical ones in the $I$ band from different authors, BP00
found a good agreement, although their theoretical TF relation yielded
somewhat larger velocities for a given absolute $I$-band magnitude.

\begin{figure}
\begin{center}
\resizebox{0.95\hsize}{!}{\includegraphics*[67,50][322,300]{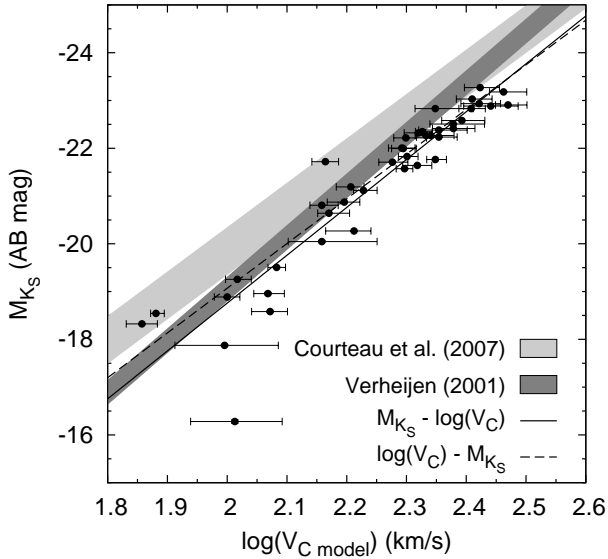}}\\
\caption{Absolute $K_S$-band magnitude of our galaxies as a function
  of the circular velocity of the corresponding best-fitting
  model. The absolute magnitudes have been derived from the asymptotic
  values, and are corrected for internal extinction. The shaded bands
  corresponds to the empirical TF relations of Verheijen (2001) and
  Courteau et al$.$ (2007). The width of the bands shows the observed
  1$\sigma$ scatter of those fits. The solid and dashed lines show
  direct and reverse linear fits to our data, respectively.\label{tf}}
\end{center}
\end{figure} 

In Fig.~\ref{tf} we plot the $K_S$-band absolute magnitude of our
galaxies as a function of the circular velocity resulting from the
model fitting. The absolute magnitudes were computed from the
asymptotic values presented in Paper~I, and were corrected for
internal extinction from the global TIR/UV ratio as described
above. The resulting median extinction in the $K_S$ band is just $\sim
0.03$\,mag. We compare our results with the empirical $K_S$-band TF
fits of Verheijen (2001) and Courteau et al$.$ (2007). The model TF
relation lies slightly below the observed ones, with values of
$V_\mathrm{C}$ roughly 0.05-0.1\,dex larger for a given absolute
magnitude, especially for the most massive galaxies. This translates
into a relative offset of 10-20\%, which is much lower than the
systematic scatter seen in Fig.~\ref{Vrot_check}a. After applying a
direct and a reverse weighted linear fit to our data-points, the
following relations are obtained:
\begin{eqnarray}
M_{K_S} &=& 1.3\pm1.4-(10.01\pm0.60)\times\log V_\mathrm{C}\\
 (rms&=&0.61\mathrm{\,mag})\nonumber\\
\log V_\mathrm{C} &=& -0.04\pm0.10-(0.1069\pm0.0049)\times M_{K_S}\\
(rms&=&0.068\mathrm{\,dex})\nonumber
\end{eqnarray}
where $M_{K_S}$ is expressed in AB magnitudes and $V_\mathrm{C}$ in
km\,s$^{-1}$. Here we rely on the $K_S$ band rather than on optical
ones in order to minimize the effects of internal extinction, but
comparisons between models and observations at other wavelengths can
be found in BP00.

\subsubsection{Trends along the Hubble sequence}
\begin{figure}
\begin{center}
\resizebox{1\hsize}{!}{\includegraphics{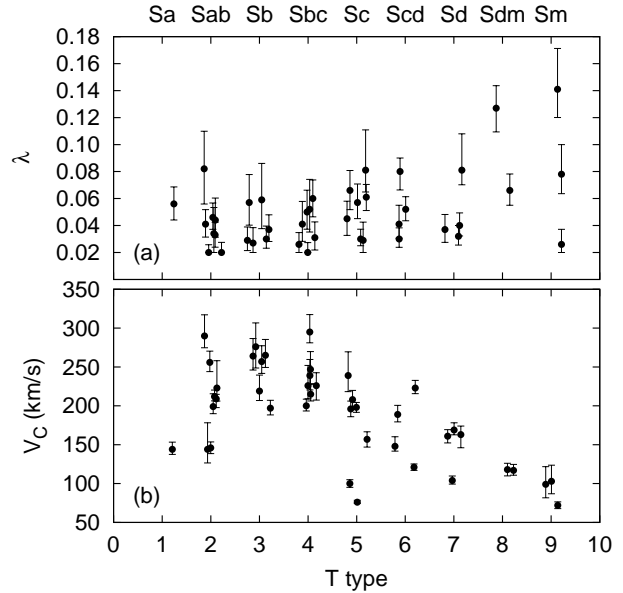}}
\caption{Dependence of the spin parameter (a) and the circular
  velocity (b) on the morphological type $T$. Given that the $T$ types
  are integer values, small random horizontal offsets have been
  applied for the sake of clarity.\label{params_vs_morpho}}
\end{center}
\end{figure}

It is illustrative to discuss whether the derived values of $\lambda$
and $V_\mathrm{C}$ depend on the morphological type. In
Fig.~\ref{params_vs_morpho} we plot the model parameters as a function
of the Hubble type. The rotational velocity is clearly correlated with
the morphology of the galaxies, with early-type disks rotating faster
$-$and hence being more massive$-$ than late-type ones (Bosma 1978;
Roberts 1978; Rubin et al$.$ 1985). However, there is no apparent
trend between the spin parameter and the Hubble type, except maybe an
increased scatter in Sdm-Sm galaxies.

Although the fact that the Hubble type does not seem to depend on the
spin parameter might look surprising at first glance, it supports the
findings of numerical simulations. The number density of galaxies per
velocity interval can be parameterized with a generalized form of the
Schechter function, resulting from the combination of the luminosity
function and the TF relation at a given band (Gonzalez et al$.$
2000). This distribution is obviously monotonic, with massive disks
being rather scarce compared to low-mass ones. Given that
$V_\mathrm{C}$ depends on the Hubble type, the probability
distribution of $\lambda$ would be also monotonic if this parameter
was also correlated with the morphological type. However, we have seen
that numerical N-body simulations conclude that most haloes usually
exhibit the same `universal' spin value $\lambda \sim 0.04$ quite
irrespective of their mass. Therefore, the lack of correlation between
$\lambda$ and the Hubble type agrees with this result.

\subsection{Potential sources of systematic errors and discrepancies in the UV}
In this section we will explore different issues that could
potentially introduce biases in our results. We will pay special
attention to those mechanisms that could be responsible of the excess
in the UV luminosity predicted by the model in early-type disks (see
Fig.~\ref{ngc2841_fit}).
\newpage
\subsubsection{Inclination}
After the profiles have been corrected for internal extinction, they
are also deprojected prior to fitting them with the models. Correcting
for inclination modifies the overall surface brightness level, so the
resulting values of $\lambda$ and $V_\mathrm{C}$ could be potentially
affected by the precise value of the inclination angle. For
consistency with the way in which the surface photometry was done, the
deprojection was done by means of the observed axial ratio, $q=b/a$,
so that $\cos i = q$. However, if disks are modeled as oblate
spheroids, then the inclination angle can be corrected for the effects
of the intrinsic thickness of the disk:
\begin{equation}
\cos^2 i' = (q^2-q_0^2) (1-q_0^2)^{-1}\label{eq_inclination}
\end{equation}
Here $q_0$ is the intrinsic axial ratio of the disk when observed
edge-on. It is customary to assume a constant value of $q_0 \simeq
0.2$ (see e.g$.$ Courteau et al$.$ 2007), although some authors argue
that this value could decrease towards $q_0 \simeq 0.1$ in late-type
spirals (Haynes \& Giovanelli 1984; Dale et al$.$ 1997). Taking into
account the thickness of the disk slightly increases the inclination
angle. As a result, the surface brightness $\mu'$ deprojected via
Eq.~\ref{eq_inclination} is fainter than the value $\mu$ deprojected
assuming $\cos i = q$:
\begin{equation}
\Delta \mu = \mu' - \mu = -2.5 \log \frac{\cos i'}{\cos i} = -2.5 \log \sqrt \frac{1-(q_0/q)^2}{1-q_0^2}
\end{equation}

It can be seen that $\Delta \mu$ lies well below 0.01\,mag for
$i<50^\circ$. At the largest inclination angles in our sample
($i\simeq 80^\circ$ for a couple of objects) the offset in the surface
brightness reaches 0.2\,mag. By inspecting the output profiles of the
BP00 models, we determined that $\Delta \mu \simeq 0.2$\,mag
corresponds to a decrease of just 10\,km\,s$^{-1}$ in $V_\mathrm{C}$
with respect to the case in which the disk thickness is ignored. The
spin parameter may increase slightly to compensate for the decrease in
the radial scale-length due to the lower circular velocity (see
Fig.~\ref{sample_model_profiles}).

The most inclined spirals in our sample are NGC~0024 (Sc; $b/a=0.22$)
and NGC~7331 (Sb; $b/a=0.35$). Our fitting code yields
$V_\mathrm{C}=93^{+6}_{-5}$\,km\,s$^{-1}$ and
$\lambda=0.071^{+0.019}_{-0.015}$ for NGC~0024, if we assume that
$q_0=0.13$ (Dale et al$.$ 1997), whereas we get
$V_\mathrm{C}=98^{+5}_{-5}$\,km\,s$^{-1}$ and
$\lambda=0.067^{+0.015}_{-0.014}$ when $q_0=0$. As for NGC~7331, we
obtain $V_\mathrm{C}=250^{+19}_{-13}$\,km\,s$^{-1}$ and
$\lambda=0.064^{+0.026}_{-0.024}$ when assuming $q_0=0.20$ (Dale et
al$.$ 1997), whereas ignoring the thickness leads to
$V_\mathrm{C}=263^{+22}_{-16}$\,km\,s$^{-1}$ and
$\lambda=0.059^{+0.029}_{-0.023}$. Therefore, employing a non-zero
value of $q_0$ in disks close to edge-on leads to differences of
merely $\sim 5$\% in both $\lambda$ and $V_\mathrm{C}$, which lie
within the estimated uncertainties. The vast majority of the galaxies
in our sample are not so inclined as the two extreme examples
discussed here, so including or not the disk thickness has an entirely
negligible impact in the model parameters.

\subsubsection{Dust attenuation}
Inaccuracies in the internal extinction correction might be
responsible of the large offset between the observed and predicted UV
profiles that we find in some disks. As explained in
Section~\ref{sec_sandwich}, we first compute $A_\mathrm{FUV}$ and
$A_\mathrm{NUV}$ from the TIR/FUV and TIR/NUV ratios,
respectively. The extinction in the optical and near-IR bands is then
derived from $A_\mathrm{NUV}$ assuming a sandwhich geometry and a
MW-like extinction curve.

By means of radiative transfer models, several studies (Buat \& Xu
1996; Meurer et al$.$ 1999; Gordon et al$.$ 2000; Witt \& Gordon 2000;
Buat et al$.$ 2005) have shown that the TIR/UV ratio constitutes a
robust proxy for the internal extinction in galaxies. In particular,
it appears to be quite insensitive to the relative distribution of
stars and dust and to the extinction curve (see, e.g., Fig.~12b in
Witt \& Gordon 2000). In the particular case of the recipes of Cortese
et al$.$ (2008) $-$the ones employed here to compute $A_\mathrm{FUV}$
and $A_\mathrm{NUV}$$-$, variations in the dust geometry and the
extinction law modify the derived attenuations by less than 0.2\,mag
(see their Fig.~9).

When computing the extinction $A_\lambda$ at other wavelengths by
extrapolating from $A_\mathrm{NUV}$, our particular choice of dust
geometry end extinction curve will indeed play a role. However, in the
case of the near-IR bands the impact will be minimal: at those
wavelengths the internal extinction will be still close to zero even if the
sandwhich model is not appropriate or if the galaxy's extinction curve
does not resemble that of the MW. In brief, for a given galaxy where
the model that best fits the near-IR profiles significantly
overestimates the UV ones, switching to another dust-to-stars geometry
or extinction law would not fix the problem. The extinction in the UV
would not vary noticeably due to the robustness of the TIR/UV ratio
against these changes, and the extinction in the near-IR would remain
close to zero. Variations would show up at intermediate wavelengths,
in the optical range, but the model seems to match the optical
profiles even in those cases where it fails in the UV (Fig.~\ref{ngc2841_fit}).

Another aspect of the internal extinction correction that should be
taken into account is the relative role of young and old stars
in heating the dust. In galaxies with low SFR per unit of mass
(specific SFR, or sSFR), the contribution of old stars to the dust
heating will be larger than in more actively star-forming galaxies,
thus biasing the energy balance argument on which the TIR/UV method
relies. Therefore, a given calibration between TIR/UV and
$A_\mathrm{UV}$ that is valid for late-type spirals will overestimate
the actual extinction in early-type ones.

In Paper~II we used two different recipes to derive attenuation
profiles for the SINGS galaxies: the calibration of Buat et al$.$
(2005), which does not account for the varying extra dust heating due
to old stars, and the age-dependent calibration of Cortese et al$.$
(2008), the one finally adopted here. The difference between both
prescriptions is negligible in disk-dominated galaxies. However, in
S0/a-Sab galaxies and the bulges of later-type spirals, the method of
Buat et al$.$ (2005) yields extinction values $\sim0.5$\,mag larger
than those obtained with the recipes of Cortese et al$.$ (2008), since
the extra contribution of old stars to the dust heating is not
``removed'' in the former. Therefore, using the calibration of Buat et
al$.$ (2005) would bring the observed profiles closer to the model
predictions. However, this better agreement would be achieved at the
expense of using an extinction recipe that is known not to be valid in
early-type spirals. Nevertheless, the difference of $\sim0.5$\,mag is
not large enough to account for the UV discrepancies found in many of
our disks.

\subsubsection{Initial Mass Function}\label{IMF}
The IMF, usually denoted as $\Phi(M)$, indicates the number of stars
in the mass interval $M$ to $M+dM$ formed in a given burst. The
classical IMF of Salpeter (1955) consists of a single power-law across
the whole stellar mass range, $\Phi(M) \propto m^{-\alpha}$, where
$\alpha=2.35$. In this work we have used two grid of models with the
K93 and K01 IMFs, which are multi-sloped:
\begin{eqnarray}
0.1 \leq M/M_{\odot} < 0.5 \;\;\;\;\;\;\;\; \alpha_{\mathrm{K93}}=1.3 \;;\; \alpha_{\mathrm{K01}}=1.3\nonumber\\
0.5 \leq M/M_{\odot} < 1.0 \;\;\;\;\;\;\;\; \alpha_{\mathrm{K93}}=2.2 \;;\; \alpha_{\mathrm{K01}}=2.3\nonumber\\
1.0 \leq M/M_{\odot} \;\;\;\;\;\;\;\; \alpha_{\mathrm{K93}}=2.7 \;;\; \alpha_{\mathrm{K01}}=2.3
\end{eqnarray}

\begin{figure}
\begin{center}
\resizebox{1\hsize}{!}{\includegraphics{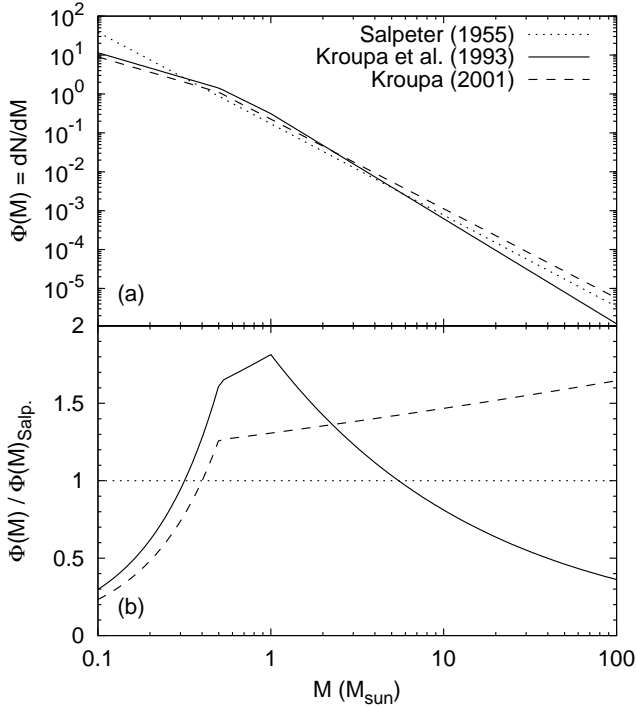}}
\caption{(a): The IMFs of Salpeter (1955), Kroupa et al$.$ (1993) and
  Kroupa (2001), normalized to a total stellar mass of 1\,M$_{\odot}$.
  (b): Ratio of the K93 and K01 IMFs to the Salpeter
  one.\label{imf_comp}}
\end{center}
\end{figure}

These IMFs are shown in Fig.~\ref{imf_comp}, where they have been
normalized such that:
\begin{equation}
\int_{M_{\mathrm{min}}}^{M_{\mathrm{max}}} \Phi(M) M dM = 1
\end{equation}
where the integration is carried out between
$M_{\mathrm{min}}=0.1$\,M$_{\odot}$ and
$M_{\mathrm{max}}=100$\,M$_{\odot}$\footnote{Note that while the
  Salpeter IMF is normally assumed to hold over the range
  0.1-100\,$M_{\odot}$, the original Salpeter (1955) study only
  included stars in the range 0.4-10\,$M_{\odot}$}. The K01 IMF has
more short-lived massive stars for a given total stellar mass, so it
yields larger UV fluxes for the same near-IR and optical
luminosities. Therefore, choosing one IMF or another obviously affects
the ability of the model to simultaneously fit all the
multi-wavelength profiles.

\begin{figure}
\begin{center}
\resizebox{1\hsize}{!}{\includegraphics{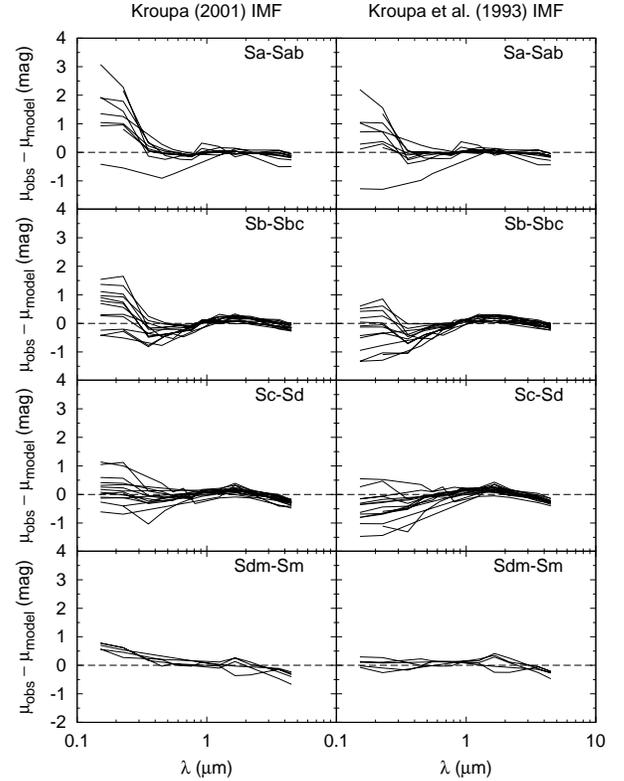}}
\caption{Average difference between the observed and model surface
  brightness profiles for each galaxy at different wavelengths. The
  average offset was computed within the radial range used during the
  fitting procedure. Note that the ``observed'' profiles are already
  corrected for internal extinction.\label{magdiff_lambda}}
\end{center}
\end{figure}

In order to quantify the discrepancy between the model predictions and
the actual profiles, for each galaxy and wavelength we have computed
the average difference between the observed and model surface
brightness within the radial range considered in the fitting
procedure. ``Observed'' is used here in contrast to ``model'', but the
``observed'' profiles we consider are of course those corrected from
internal extinction. The results are shown in
Fig.~\ref{magdiff_lambda} for each IMF and morphological type.

It is clear that neither IMF provides a satisfactory fit at all
wavelengths and for all morphological types simultaneously. The
largest discrepancies are found in the UV range, as expected. The K01
IMF yields an excellent fit for most Sc-Sd spirals, as well as for
some Sb-Sbc ones. However, it significantly overestimates the UV flux
in early-type disks. The K93 IMF, on the other hand, partly mitigates
the problem in some Sb-Sbc galaxies, but the perfect agreement between
model and observations found in Sc-Sd disks is partially lost. Very
late-type disks, however, are better fitted with the K93 IMF than with
the K01 one.

\begin{figure*}
\begin{center}
\resizebox{0.48\hsize}{!}{\includegraphics{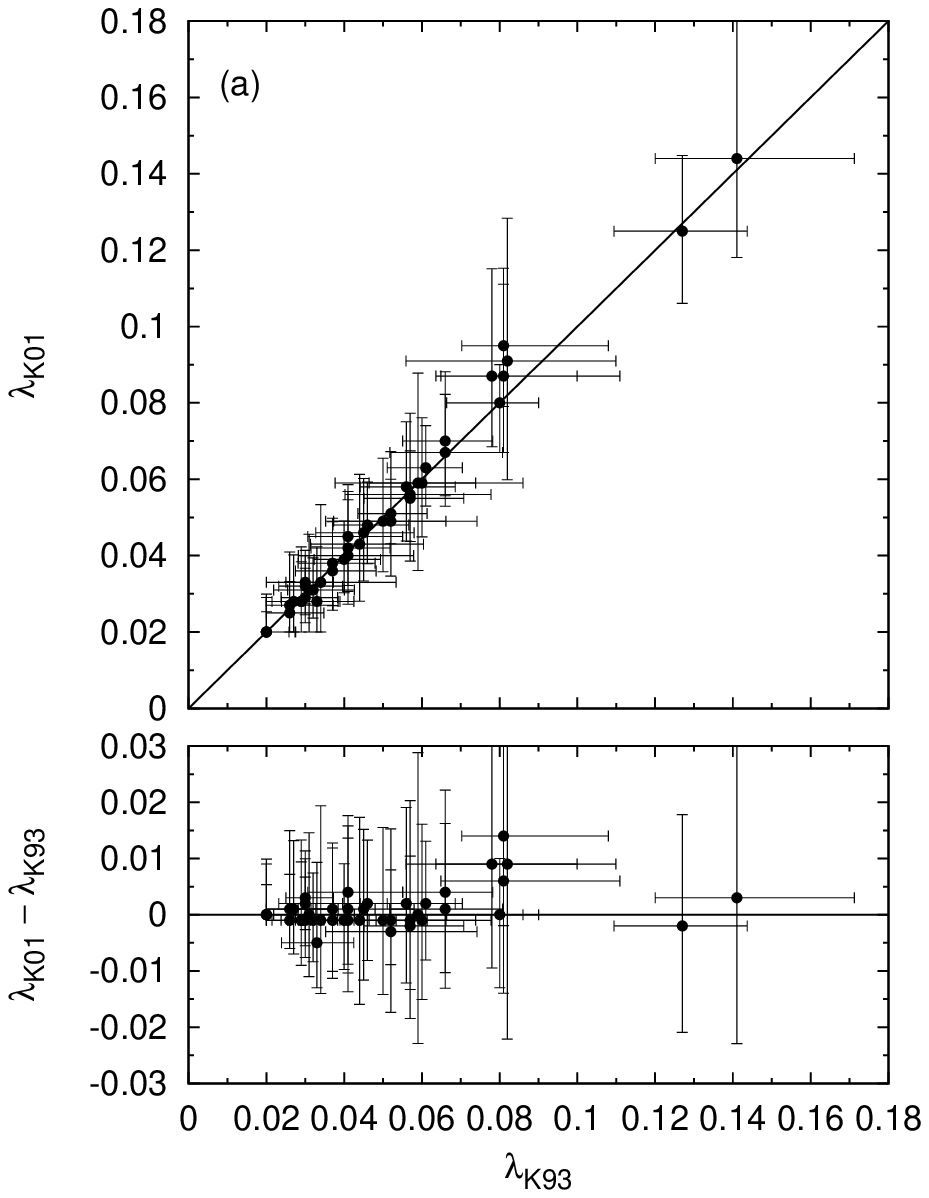}}
\resizebox{0.48\hsize}{!}{\includegraphics{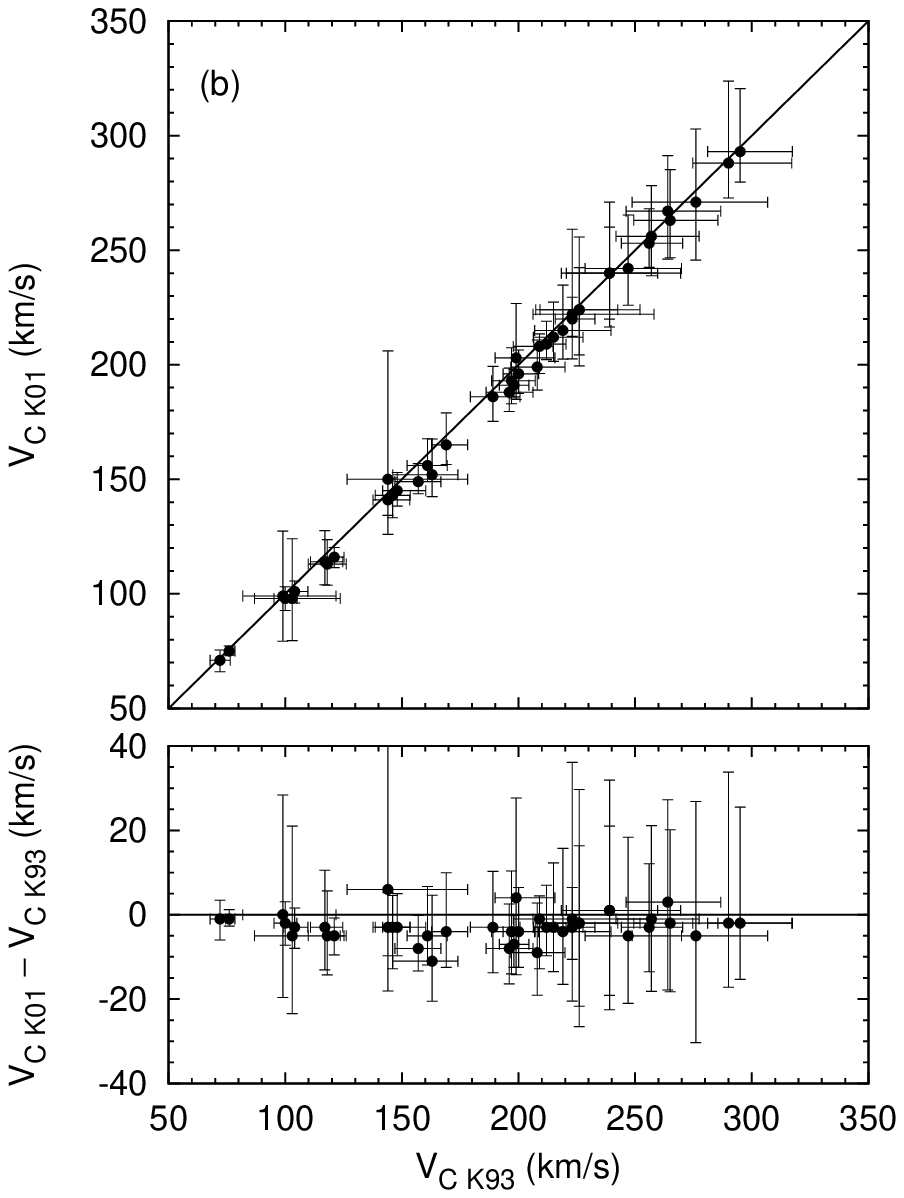}}
\caption{Comparison of the values of the spin parameter and circular
  velocities obtained when using the IMFs of Kroupa et al$.$ (1993)
  and Kroupa (2001). The solid line shows the corresponding 1:1
  relations.\label{params_IMF_check}}
\end{center}
\end{figure*}

Elucidating whether a varying IMF is the actual reason behind these
discrepancies is not straightforward, as other mechanisms could
introduce similar systematic biases in the UV profiles. For instance,
the UV bands can be affected by variations in the SFR taking place on
timescales of $\sim$100\,Myr, whereas the $u$ band will respond to
variations over $\sim$500\,Myr. This increase in the characteristic
timescale between the UV and optical bands seems to be indeed rather
sharp (Boissier et al$.$ 2011, in prep). Therefore, short-scale SFR
variations could partly explain why the discrepancies between the
observed and predicted profiles drops abruptly between the UV and $u$
bands. Also, the calibration of the stellar atmospheres used in the
spectral synthesis models might not be as accurate in the UV as they
are in the optical, since observations of stars in the UV are not so
easily accesible.

It has been argued that the so-called Integrated Galactic Initial Mass
Function (IGIMF), which takes into account the clustered nature of
star formation, varies among galaxies, leading to a scarcity of
massive stars in galaxies with low global SFRs (Weidner \& Kroupa
2005). Krumholz \& McKee (2008) brought forward a physical explanation
to the apparent dearth of massive stars in low-density regions,
arguing that gas column densities of at least 1\,g\,cm$^{-2}$ are
required to halt cloud fragmentation and produce massive
stars. Stochastic formation of high mass stars might also play a role
in galaxies with very low levels of star formation (Lee et al$.$
2009).

These mechanisms have been proposed to explain the observed
discrepancies between the UV and H$\alpha$-derived SFRs in dwarf
galaxies (Pflamm-Altenburg et al$.$ 2007, 2009; Lee et al$.$ 2009) and
in low surface brightness galaxies (Meurer et al$.$ 2009). In this
regard, even though both the K93 and K01 IMFs are canonical, it would
make sense that Sdm-Sm disks are better fitted with a top-light IMF
like the K93 one rather than with the K01 one. However, the effects
described here seem to arise mainly in irregular galaxies with much
lower levels of SF than the disks considered in this work (see also
Boselli et al$.$ 2009). In any case, we cannot appeal to these
mechanisms to explain the discrepancies in the UV bands found in
early-type disks.

\begin{figure*}
\begin{center}
\resizebox{0.48\hsize}{!}{\includegraphics{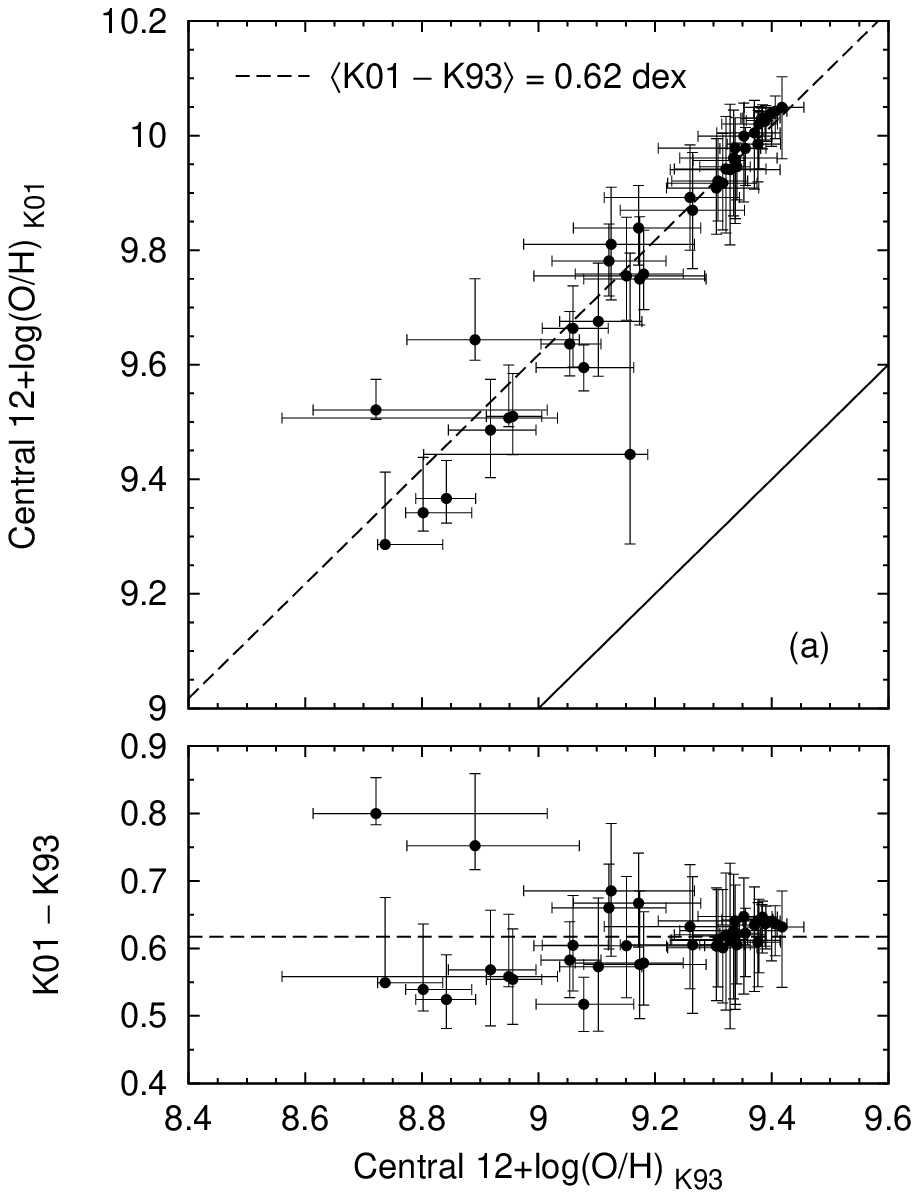}}
\resizebox{0.48\hsize}{!}{\includegraphics{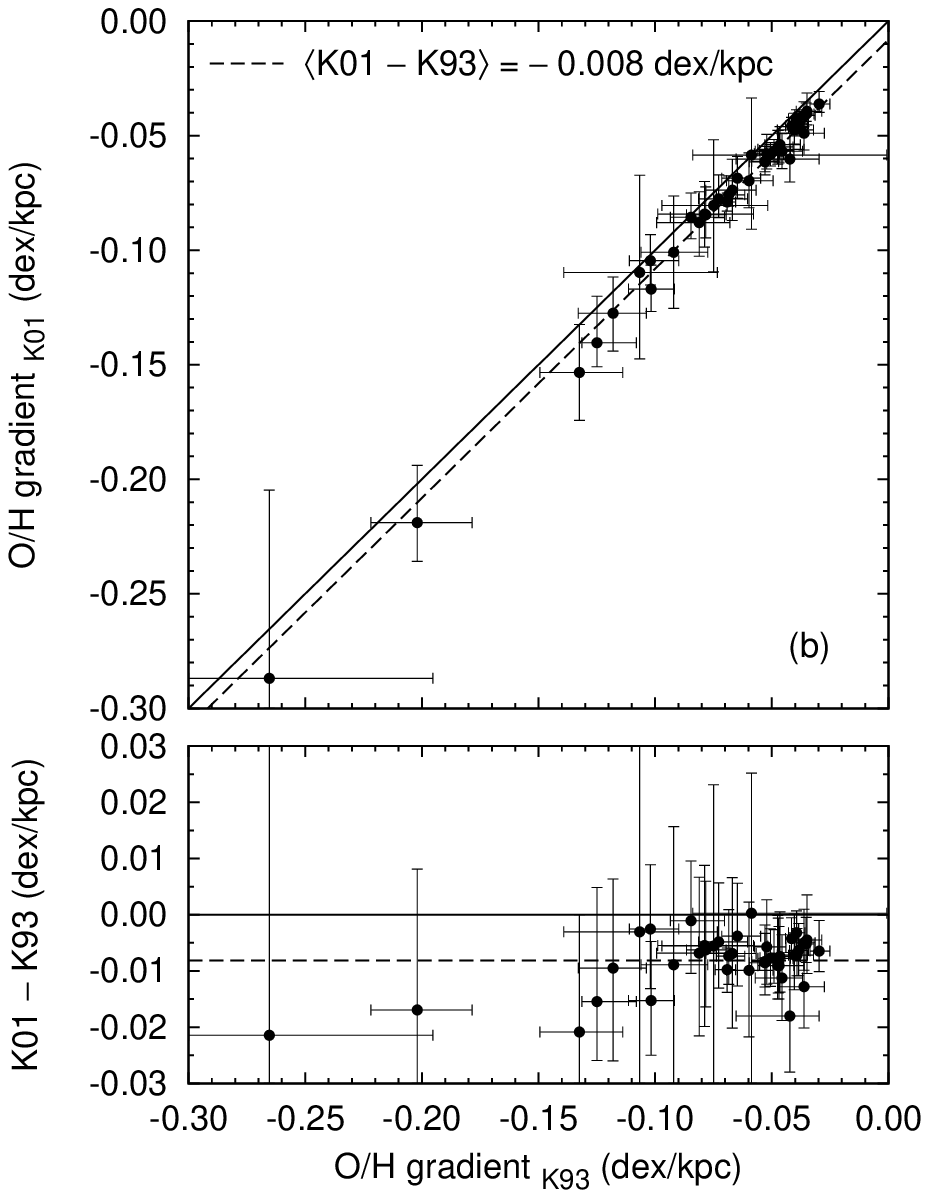}}
\caption{Comparison of the central abundances (a) and gradients (b)
  obtained when using the IMFs of Kroupa et al$.$ (1993) and Kroupa
  (2001). The solid lines in each panel correspond to the 1:1
  relation, while the dashed lines show linear fits where only the
  zero-point has been left as a free parameter. The corresponding
  offset is quoted in each panel.\label{metal_IMF_check}}
\end{center}
\end{figure*}

Nevertheless, it is worth asking whether our particular choice of IMF
affects the results of the fitting. Thanks to our two-step fitting
procedure, which estimates the intrinsic error of the model for each
galaxy and band after a first run, the UV bands are automatically
assigned a relatively large error whenever large discrepancies are
found. This prevents the UV bands from biasing the fit at longer
wavelengths. In Fig.~\ref{params_IMF_check} we compare the values of
$\lambda$ and $V_\mathrm{C}$ obtained with both IMFs. The differences
are obviously negligible and much smaller than the estimated
uncertainties, since the fit is still excellent in the optical and
near-IR bands even when it fails in the UV. Therefore, even if the
very recent level of SF (either massive or total) predicted by the
models for the last few hundreds of Myrs is not entirely reliable, the
overall star formation history across longer timescales can still be
trusted.

Besides modifying the emitted UV flux, changing the IMF also has a
significant impact in the resulting metallicity profiles predicted by
the model. In Fig.~\ref{metal_IMF_check} we compare the central
abundances and radial gradients obtained with both IMFs. Given that
the K01 IMF is richer in massive stars than the K93 one, it produces
more Type II supernovae and metals, leading to oxygen abundances which
are $\sim 0.62$\,dex larger than those resulting from the K93
IMF. Note that the uncertainties associated with different metallicity
calibrations are typically $\sim 0.3$\,dex (see Moustakas et al$.$ 2010). The radial
gradients, on the other hand, remain nearly unchanged, the K01 ones
being just mildly steeper.

\begin{figure*}
\begin{center}
\resizebox{0.7\hsize}{!}{\includegraphics{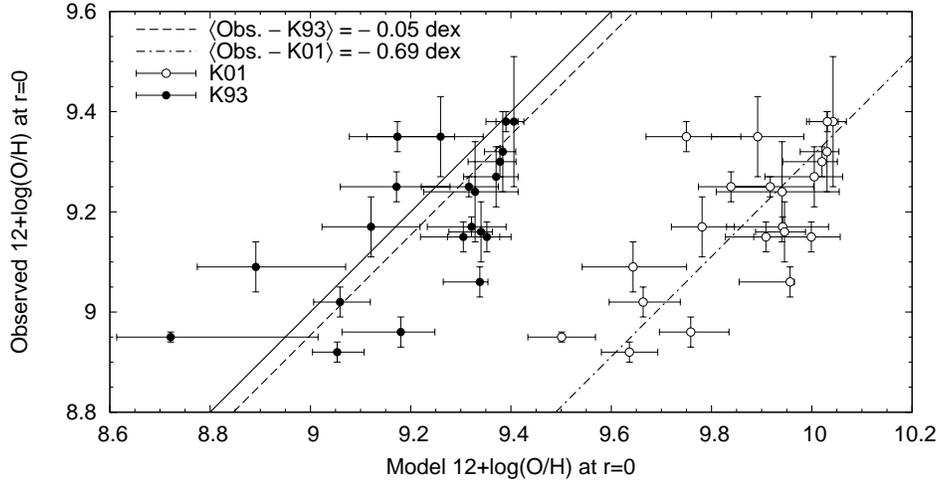}}
\caption{Comparison between the observed and predicted central
  abundances for those SINGS galaxies with metallicity profiles
  computed by Moustakas et al$.$ (2010). The average offsets
  for the K93 and K01 IMFs are shown with a dashed and dot-dashed
  line, respectively, while the solid one corresponds to the line of
  equality.\label{z0_check}}
\end{center}
\end{figure*}

We must also check whether the oxygen abundance profiles predicted by
the model are in agreement with the observed ones. As in Paper~II,
here we rely on the metallicity zero-points and gradients measured by
Moustakas et al$.$ (2010), using the calibration of Kobulnicky \&
Kewley (2004). In Fig.~\ref{z0_check} we plot the central oxygen
abundances as a function of the predicted values using both IMFs. The
values computed with the K93 IMF are just 0.1\,dex larger than the
observed central abundances, whereas those yielded by the K01 IMF are
roughly 0.7\,dex larger. Such high metallicites have never been
observed; moreover, the flat slope for massive stars of the K01 IMF
would lead to a significant depletion of deuterium through astration
in the solar neighborhood.

As pointed out by several authors (see, e.g$.$, Kewley \& Ellison
2008; Moustakas et al$.$ 2010), different methods used to measure the
gas-phase metallicity from observed emission-line spectra can yield
largely discrepant results. Besides the calibration of Kobulnicky \&
Kewley (2004), Moustakas et al$.$ (2010) also derived metallicity
profiles using the calibration of Pilyugin \& Thuan (2005). Both
methods are based on the strong-line $R_{23}$ parameter (Pagel et
al$.$ 1979). The empirical method of Pilyugin \& Thuan (2005) is
calibrated on HII regions having direct abundance measurements based
on the electron temperature. The theoretical calibration of Kobulnicky
\& Kewley (2004), on the contrary, relies on photoionization
models. The method of Pilyugin \& Thuan (2005) leads to oxygen
abundances $\sim$0.6\,dex lower than those obtained with the
Kobulnicky \& Kewley (2004) calibration. Therefore, had we used the
Pilyugin \& Thuan (2005) values in Fig.~\ref{z0_check}, even the
metallicites predicted with the K93 IMF would be too high compared
with the observed ones, and the discrepancy with the K01 IMF would be
even larger than it already is. As thoroughly discussed by Moustakas
et al$.$ (2010), neither empirical nor theoretical strong-line methods
are devoid of problems. On one hand, empirical methods like the
Pilyugin \& Thuan (2005) one might fail outside the metallicity range
spanned by the HII regions used in the calibrations, and can therefore
underestimate the true abundances in the high-metallicity regime. On
the other hand, theoretical methods such as the Kobulnicky \& Kewley
(2004) one adopt several simplifying assumptions on the properties of
the gas clouds and the ionizing stars that can make the derived
metallicities higher than the actual ones. The two calibrations
discussed here bracket the range of abundances usually obtained with
other methods, so the actual metallicities probably lie somewhere in
between.

\begin{figure*}
\begin{center}
\resizebox{0.7\hsize}{!}{\includegraphics{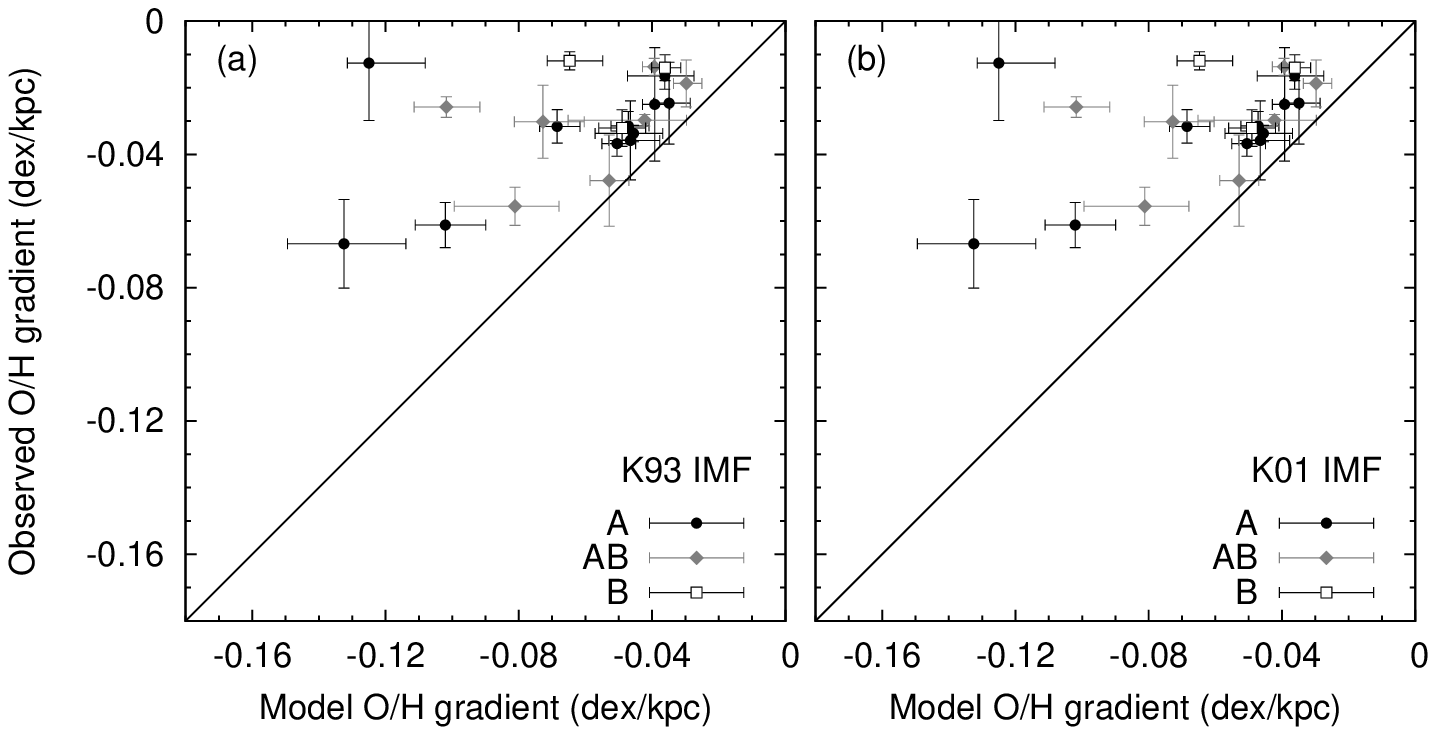}}\\
\resizebox{0.7\hsize}{!}{\includegraphics{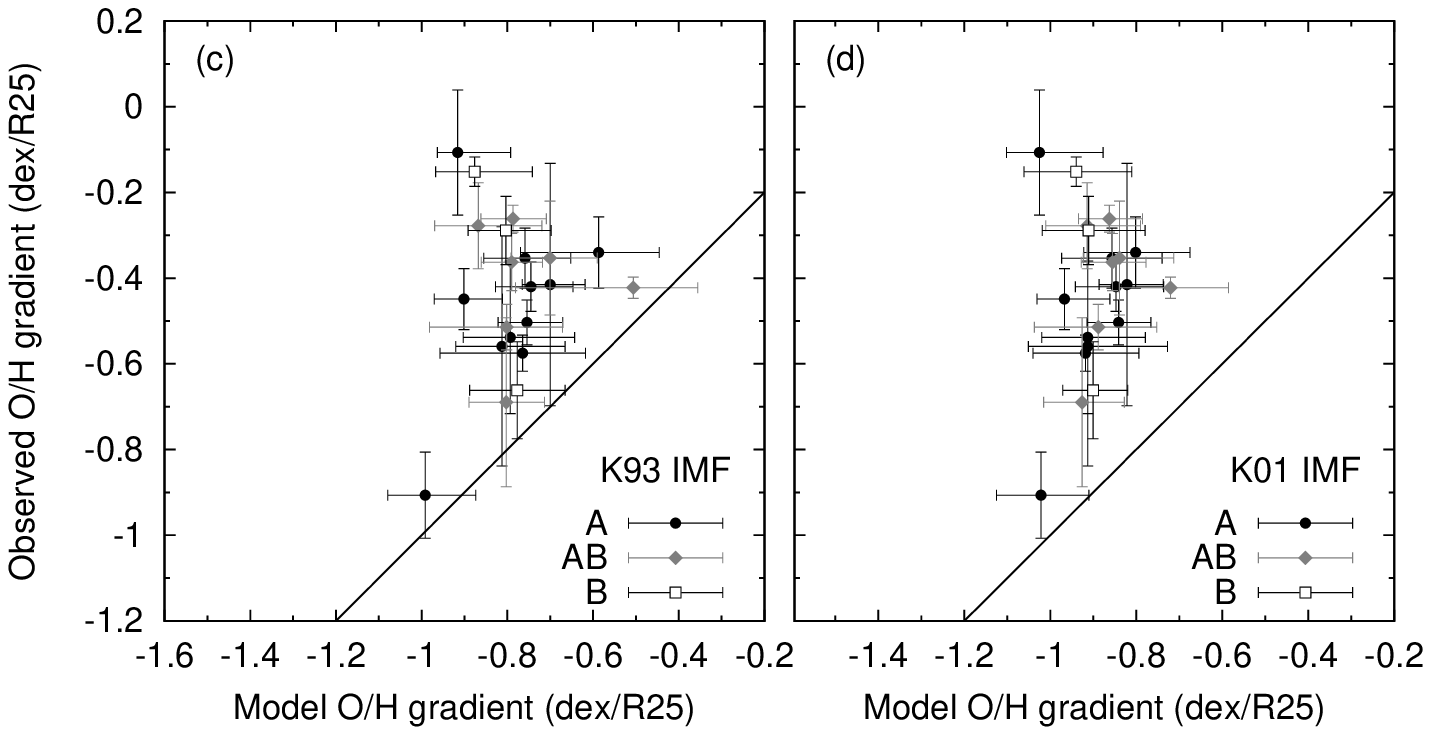}}\\
\caption{Comparison between the observed and predicted abundance
  gradients for those galaxies in the sample of Moustakas et al$.$
  (2010). Panels (a) and (b) show the gradients in units of
  dex\,kpc$^{-1}$ for both IMFs, whereas in panels (c) and (d) the
  gradients are expressed in units of the optical radius R25. Galaxies
  are sorted out into non-barred (A), barred (B) and mixed (AB). The
  solid line marks the 1:1 relation in all cases.\label{zgrad_check}}
\end{center}
\end{figure*}

As for the radial gradients, they are plotted in
Fig.~\ref{zgrad_check}, both in units of dex\,kpc$^{-1}$ (upper
panels) and normalized by the optical radius R25 (lower panels). The
gradients predicted by the model are always steeper than the observed
ones, although for most galaxies the difference is just about $\sim
0.015$\,dex\,kpc$^{-1}$.  When expressed in terms of the optical
radius, the model predicts a roughly constant gradient of
$-0.8$\,dex/R25 for the K93 IMF, and $-0.9$\,dex/R25 for the K01
one. This ``universal'' oxygen gradient, which does not depend on
galaxy mass, is usually found in observations (see e.g$.$ Henry \&
Worthey 1999). However, the spread is larger than that predicted by
the models, with observed gradients ranging from $-1$\,dex/R25 to
almost flat gradients. The fact that the comparison of gradients in
units of dex\,kpc$^{-1}$ is somewhat better than in dex/R25 is not
surprising, since the former are mainly determined by the radius of the
disks, which varies substantially within the sample. Therefore, the
comparison between observed and model gradients in dex\,kpc$^{-1}$ is
more a test of the prediction of disk sizes than of abundance
gradients.

Bars may induce radial gas flows that can yield shallower metallicity
gradients (see, e.g$.$, Martin \& Roy 1994). Two out of the three
barred galaxies in Fig.~\ref{zgrad_check} are among the ones with the
flattest observed gradients, although the subsample considered here is
obviously too small to extract any statistically significant
conclusion. Recent numerical N-body simulations have shown that radial
stellar migration can also flatten the final metallicity profiles
(Ro\u{s}kar et al$.$ 2008; Mart\'inez-Serrano et al$.$ 2009;
S\'anchez-Bl\'azquez et al$.$ 2009), although these results refer to
the stellar metallicity and not the gas-phase one.

To summarize this dicussion on the IMF, most Sc-Sd spirals require the
K01 IMF in order not to underestimate the UV luminosity, but at the
expense of ending up with oxygen abundances much larger than the
observed ones. Conversely, the K93 IMF provides a somewhat better fit
for the remaining Hubble types, and is able to reproduce the correct
present-day abundances. Whether the better agreement with the observed
metallicities should be given a special importance is unclear, since
neither the measured values nor the predicted ones are devoid of
possible sources of large systematic errors. On one hand, the
different existing calibrations used to compute the oxygen abundance
from observed line ratios may lead to systematically different values,
as explained before (Kewley \& Ellison 2008; Moustakas et al$.$
2010). On the other hand, uncertainties in the stellar yields used in
the disk evolution models will affect the predicted
metallicities. Besides, we cannot neglect the possibility that the
library of synthetic spectra used to compute the multiwavelength
profiles could be quite off in the UV range.

With these issues in mind, we opt for a compromise solution and adopt
the K93 IMF as our default choice in this paper, otherwise
stated. This choice is done without any prejudice to the K01 which, we
reiterate, provides an excellent fit for many disks in our sample,
especially Sc-Sd ones.

\subsubsection{Gas content and SF efficiency}\label{gas_content}
Apart from the broadband photometric profiles, the model also outputs
the total gas density profiles, which can be therefore used as another
observational constrain of the accuracy of the model. In particular,
and continuing the discussion presented in the previous section, the
gas fraction might shed some light into the disagreement in abundance
gradients. The baryonic gas fraction is defined as
$f_g=M_g/(M_*+M_g)$, where $M_*$ and $M_g$ are the stellar and gas
masses, respectively. As star formation proceeds, gas is progressively
consumed and transformed into stars, which later enrich the remaining
gas with heay elements. Therefore, $f_g$ is expected to decrease with
O/H, although the particular trend might depend on the presence of
inflows or outflows (Boissier et al$.$ 2001; Garnett 2002).

\begin{figure}
\begin{center}
\resizebox{1\hsize}{!}{\includegraphics{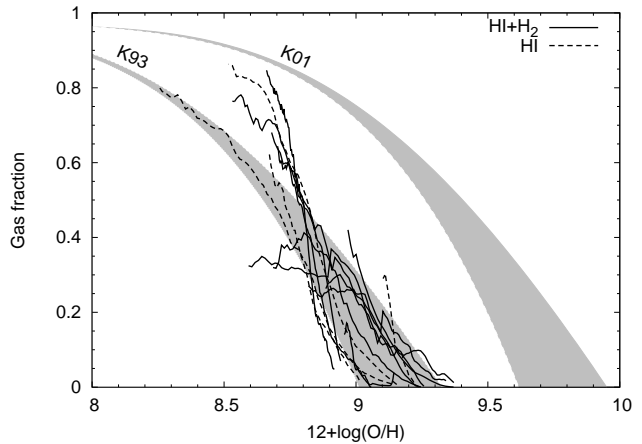}}\\
\caption{Trend between the gas fraction and the gas-phase oxygen
  abundance. The shaded bands show the model predictions for the K93
  and K01 IMFs. The black lines show the observed gas fractions of our
  galaxies, resulting from the combination of atomic and molecular gas
  profiles with stellar mass ones. The latter were derived from the
  observed 3.6\,\micron\ profiles, adopting the M/L ratio yielded by
  the model at this band. Solid lines indicate those galaxies with
  H$_2$ profiles, whereas dashed ones are used when only HI profiles
  were available. The observed gas profiles were multiplied by 1.36
  to include helium and heavy elements. The model profiles already
  show the total gas surface density.\label{gas_fraction}}
\end{center}
\end{figure}

In Fig.~\ref{gas_fraction} we compare the gas fraction profiles
predicted by the model with the actual ones found in our
galaxies. According to the model, $f_g$ decreases with $V_\mathrm{C}$
and increases with $\lambda$ (Boissier et al$.$ 2001). The shaded
areas shown here cover the range of values of $V_\mathrm{C}$ and
$\lambda$ found in our sample. As for the observed quantities, the
atomic gas profiles were measured on HI maps from THINGS, using the
same ellipticity, position angle and radial step as those used when
computing the UV, optical and near-IR profiles. Whenever possible, we
included CO profiles compiled from the literature to account for the
molecular gas content. The CO profiles were converted into H$_2$ ones
by means of a metallicity-dependent X$_{\mathrm{CO}}$ conversion
factor (Boselli et al$.$ 2002), using the local oxygen abundance at
each galactocentric distance. Solid lines are used for those galaxies
with available molecular gas profiles (although most if not all of the
gas content in the outer parts is entirely atomic). Dashed lines
correspond to galaxies for which only HI data were available. In all
cases, we multiplied the hydrogen profiles by 1.36 to account for
helium and heavy elements. Stellar mass profiles for each galaxy were
derived from the 3.6\,$\micron$\ ones, using the $M/L$ ratio yielded
by the model, and combined with the gas profiles to obtain $f_g(r)$.

As already discussed before, the K01 IMF yields excessively large
metallicities, whereas the K93 provides a much better agreement with
the observed values. While the model is successful at reproducing the
observed trend, for some galaxies $f_g$ seems to decrease faster with
O/H compared to the model predictions, consistently with the observed
O/H gradients being flatter than the model ones.

\begin{figure}
\begin{center}
\resizebox{1\hsize}{!}{\includegraphics{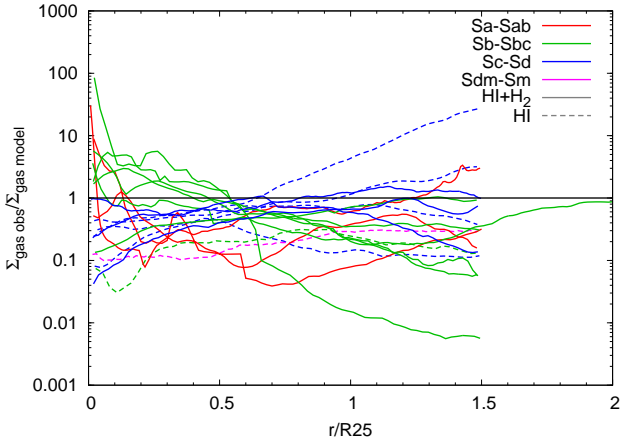}}
\caption{Ratio between the observed gas surface density and the value
  predicted by the model at different radii for each galaxy (see text for
  details). Solid lines correspond to galaxies with both atomic and
  molecular gas profiles available, whereas dashed ones are used for
  those objects laking CO profiles. The observed gas profiles have
  been multiplied by 1.36 to account for helium and heavy
  elements. The model profiles already include the total gas surface
  density.\label{gas_ratio}}
\end{center}
\end{figure}

The gas profiles can also shed light on the discrepancy between the
observed and predicted UV profiles, which may arise from differences
in the amount of gas that settles onto the disk or the way it is
transformed into stars. In Fig.~\ref{gas_ratio} we show the ratio
between the gas surface density predicted by the model and the actual
one for each galaxy as a function of radius. Here we plot the model
gas density obtained with the K93 IMF; the values yielded by the K01
IMF are within $\pm 20$\% of the ones shown here.

\begin{figure}
\begin{center}
\resizebox{1\hsize}{!}{\includegraphics{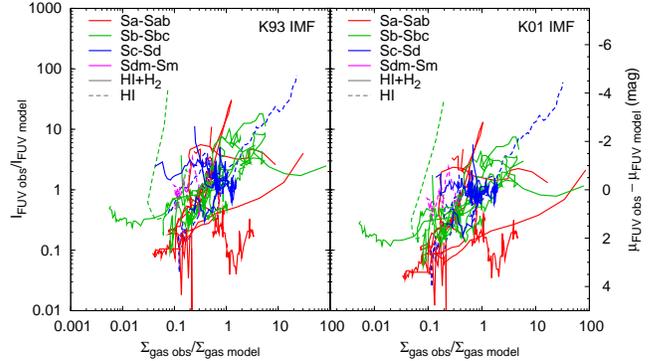}}\\
\caption{The observed-to-predicted gas ratio is plotted against the
  discrepancy between the actual UV surface brightness $-$already
  corrected for internal extinction$-$ and the value predicted by the
  model. This UV discrepancy is expressed in terms of a magnitude
  offset (right $y$ axis) and the corresponding flux ratio (left $y$
  axis). Unlike Fig.~\ref{magdiff_lambda}, where this offset was
  averaged all over the disk, here we plot the full range of values at
  different galactocentric distances, so each galaxy follows a track
  in this plot.\label{gas_ratio_vs_magdiff}}
\end{center}
\end{figure}

In the disks of Sc-Sd spirals the gas profiles predicted by the models
agree with the observed ones. However, in general the model
overestimates the actual gas content by a factor of $\lesssim 10$,
which is considerable. Does this lead to systematic differences in
the emerging FUV luminosity? In Fig.~\ref{gas_ratio_vs_magdiff} we
compare the offset between the observed and predicted FUV brightness
with the corresponding observed-to-predicted gas ratio at each
radius. While there is a non-negligible scatter, a clear trend is
seen, in the sense that the model overestimates the FUV brightness
wherever it also predicts too much gas content.

It is worth investigating whether the star formation efficiency plays
a role in driving the predicted UV profiles away from the actual
ones. After all, even if the gas profiles are correctly reproduced by
the model, the emitted UV light will still depend on the particular
way in which the model handles the process of star formation. As
explicitly stated in Eq.~\ref{eq_SFR_sam_models}, the model relies on
a hybrid star formation law that combines the classical
Schmidt-Kennicutt law with an orbital term (see Leroy et al$.$ (2008)
for an extensive analysis of various other laws). The value of the
efficiency $\alpha$ in Eq.~\ref{eq_SFR_sam_models} was chosen to
reproduce the present-day observed gas fraction in nearby galaxies
(Boissier et al. 2003).

\begin{figure}
\begin{center}
\resizebox{1\hsize}{!}{\includegraphics{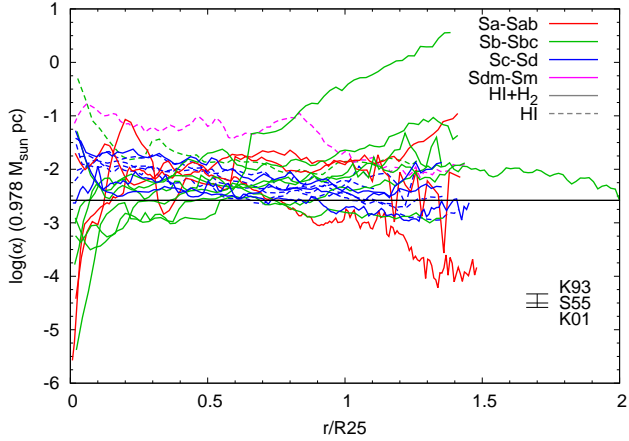}}
\caption{Empirical value of the efficiency parameter $\alpha$ in
  Eq.~\ref{eq_SFR_sam_models} (see text for details on how $\alpha$
  was derived from observations). The units of $\alpha$ are such that
  $\Sigma_{\mathrm{SFR}}(r)$ is measured in
  M$_{\odot}$\,pc$^{-2}$\,Gyr$^{-1}$, $\Sigma_{\mathrm{gas}}(r)$ in
  M$_{\odot}$\,pc$^{-2}$, $r$ in kpc and $V(r)$ in km\,s$^{-1}$. The
  horizontal solid line marks the constant value assumed in the models
  (see Boissier et al. 2003). $\Sigma_{\mathrm{SFR}}(r)$ was derived from the
  extinction-corrected UV profiles using the recipe of Kennicutt
  (1998), which assumes a Salpeter (1955) IMF. The small tics at the
  bottom right show how the plotted curves would shift with the K93
  and K01 IMFs.\label{sf_efficiency}}
\end{center}
\end{figure}

Even though the efficiency $\alpha$ is kept constant in the
models\footnote{Note that although $\alpha$ can be interpreted as the
  efficiency of the SFR, by star formation efficiency (SFE) one
  usually means the ratio of the SFR and the gas mass. In our case,
  $\mathrm{SFE}(t,r) = \alpha
  \Sigma_{\mathrm{g}}(t,r)^{0.5}V(r)r^{-1}$, so even under the
  assumption of a constant value of $\alpha$, SFE is still a varying
  function of radius and time.}, it may actually vary among galaxies
or even within them. This assumption can be tested, given that $\alpha$
depends on the SFR density, the gas density and the rotational
velocity at each radius, all of which can be measured. In
Fig.~\ref{sf_efficiency} we plot the radial variation of $\alpha$ for
each galaxy. The SFR profiles were computed from the
extinction-corrected FUV ones, using the conversion factor of
Kennicutt (1998). This recipe assumes solar metallicity, a constant
SFR over the last few Myrs and a Salpeter (1955) IMF. The small tics
at the bottom right show how the profiles whould shift with the K93
and K01 IMFs. The gas surface density was again obtained from THINGS
and CO profiles. Finally, the circular velocity at each radius was
derived from the fits of Leroy et al$.$ (2008) to the THINGS rotation curves
(Eq.~\ref{eq_vrot}). For the few THINGS galaxies not considered by
Leroy et al$.$ (2008), the rotation curve from the best fitting model was used instead
as a proxy for the actual one.

Fig.~\ref{sf_efficiency} shows that, except for a few objects,
$\alpha$ remains roughly constant with radius within most disks,
although it does vary from galaxy to galaxy. The empirical values of
$\alpha$ are consistent with or somewhat larger than the constant
value assumed in the model, shown here as a horizontal line. The
average scatter is nonetheless considerable, around 1\,dex. There are
several factors contributing to this dispersion in the measured values
of $\alpha$. On one hand, variations in the IMF will affect the
adopted calibration for the SFR as a function of the FUV
luminosity. Moreover, the assumption of a constant SFR over the last
few Myrs might not hold in some cases. Besides, departures from solar
metallicity $-$both among and within galaxies$-$ will also affect the
SFR calibration. Uncertainties in the CO-to-H$_2$ conversion factor
will modify the total gas surface density, although this is only a
concern in the innermost regions. Nevertheless, part of the observed
scatter likely reflects intrinsic variations of $\alpha$ among
galaxies.

Interestingly, when plotting the observed-to-predicted UV offset as a
function of the empirical values of $\alpha$ we do not observe any
significant trend between both parameters, in contrast with
Fig.~\ref{gas_ratio_vs_magdiff}, where the data cloud was clearly
tilted. In other words, the UV discrepancy appears to be independent
of the precise value of $\alpha$, so a morphology-dependent value of
$\alpha$ cannot account for the mismatch between the observed and
theoretical UV profiles. Therefore, the bottom line of this analysis
is that the discrepancies in the UV profiles of early-type disks are
mainly due to an excessive amount of gas retained by the model in
these galaxies, rather than to its subsequent conversion into stars.

Some of the galaxies in our sample with the largest UV discrepancies
belong to the Virgo cluster or the Coma Cloud. Hydrodynamical
interactions with the hot intergalactic medium or gravitational ones
with other cluster members might have removed part of the gas in their
disks, thus quenching the recent star formation activity (see Boselli
\& Gavazzi 2006 for a review). This effect would lead to significantly
lower UV fluxes than those predicted by the model. In order to test
whether this is actually the case, we have computed the so-called HI
deficiency for the galaxies in our sample. This parameter, defined by
Haynes \& Giovanelli (1984), compares the observed HI mass of a galaxy
with the typical HI mass of isolated field galaxies with a similar
morphological type $T^{\mathrm{obs}}$ and linear optical diameter
$D^{\mathrm{obs}}_{\mathrm{opt}}$:
\begin{equation}
\mathrm{HI\ def} = \langle\log M_{\mathrm{HI}}(T^{\mathrm{obs}},D^{\mathrm{obs}}_{\mathrm{opt}})\rangle - \log M^{\mathrm{obs}}_{\mathrm{HI}}\label{eq_HIdef}
\end{equation}
Therefore, positive HI deficiencies correspond to galaxies with lower
gas contents than those of similar but isolated objects. Here we use
the calibration of
$M_{\mathrm{HI}}(T^{\mathrm{obs}},D^{\mathrm{obs}}_{\mathrm{opt}})$
derived by Solanes et al$.$ (1996), since it relies on a larger sample
of galaxies than that used in the seminal paper by Haynes \&
Giovanelli (1984). Following Solanes et al$.$ (2001), the reference HI
masses for galaxies later than Sc have been computed following the
prescriptions for Sc ones. The intrinsic scatter in measurements of
the HI deficiency is typically $\pm 0.2$-$0.3$\,dex (Haynes \&
Giovanelli 1984; Solanes et al$.$ 1996), and it is customary to assume
that galaxies with deficiencies lower than 0.03-0.05 posses normal HI
contents.

\begin{figure}
\begin{center}
\resizebox{1\hsize}{!}{\includegraphics{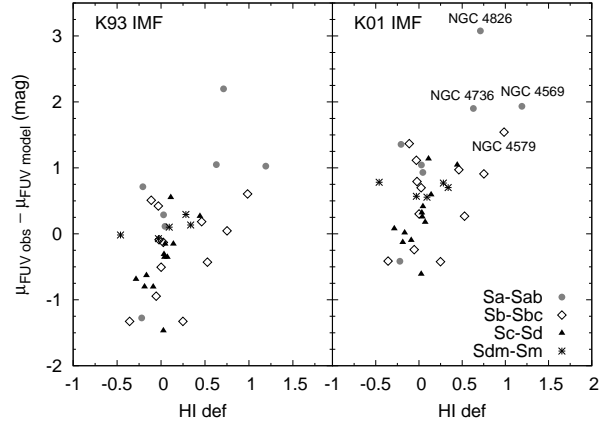}}
\caption{Offset between the observed and model UV surface brightness
  as a function of the HI deficiency. The latter is obtained by
  comparing, in logarithmic scale, the total HI mass of a given galaxy
  and the typical HI mass of isolated galaxies with similar
  morphological types and linear diameters (see
  Eq.~\ref{eq_HIdef}).\label{magdiff_vs_defHI}}
\end{center}
\end{figure}

In Fig.~\ref{magdiff_vs_defHI} we plot the average UV offset between
the observed and predicted UV profiles as a function of the HI
deficiency. The integrated HI masses of the SINGS galaxies have been
taken from Draine et al$.$ (2007). There are indeed some galaxies with
simultaneously large HI deficiencies and UV offsets. In NGC~4569,
NGC~4579 and NGC~4826, the model perfectly fits the optical and
near-IR profiles, but it substantially overestimates the UV ones,
especially in the outer regions.

NGC~4826 is an anemic spiral in the Coma 1 Cloud (van den Bergh 1976;
Boselli \& Gavazzi 2009). It is known to host two counter-rotating
gaseous disks that possibly point toward a past merger event (Braun et
al$.$ 1992). It also exhibits a central dark lane that has earned this
object the nickname ``The black-eye galaxy''. Its stellar component
extends beyond $r \gtrsim 13$\,kpc, as shown by the optical and
near-IR profiles, but most of the star formation activity is currently
restricted to $r \lesssim 3$\,kpc, where most of the gas is located.

Something similar happens with NGC~4569 and NGC~4579, two other anemic
spirals in the Virgo Cluster (van den Bergh 1976). In order to
properly fit the multi-wavelength profiles of NGC~4569, Boselli et
al$.$ (2006) employed a modified version of the BP00 models in which
the gas infall could be tuned to simulate starvation (by simply
stopping gas infall) or ram pressure stripping (by removing gas
already settled onto the disk). These authors concluded that ram
pressure stripping is required to explain the truncated HI and
star-forming disks of this galaxy.

NGC~4736 constitutes another interesting example. At radii larger than
225\arcsec\ ($\sim 6$\,kpc), a pronounced anti-truncation can be
clearly seen in its light profiles at all wavelengths. The bulge
dominates the emission at $r \lesssim 75$\arcsec\ ($\sim 2$\,kpc), and
a prominent star-forming ring is visible at $r \simeq 40$\arcsec
($\sim 1$\,kpc). Both the outer disk and the bulge (together with the
ring) were excluded when performing the fit, and the best-fitting
model is very succesful at reproducing the multi-band profiles of the
inner disk, but again overestimates the true FUV and NUV
profiles. Trujillo et al$.$ (2009) carried out a detailed analysis of
this galaxy by fitting its sepectral energy distribution (SED) at
different radii, and also by performing smoothed particle
hydrodynamics simulations. These authors favor a scenario where an
oval distortion gives rise to both the outer disk and the enhanced
central star formation activity, involving radial flows which of
course the BP00 models do not consider.

Even though the particular objects described above are rather extreme,
Fig.~\ref{magdiff_vs_defHI} shows that the model is still quite off in
the UV for many objects with seemingly normal HI contents, including
early-type spirals like NGC~3031 or NGC~2841 (see
Fig.~\ref{ngc2841_fit}). In these cases, any external interaction
probably plays a minor role compared to the larger failings of the
models for predicting gas contents. As mentioned at the end of
Section~\ref{model_extension}, the BP00 models parameterize the gas
infall rate as a function of both the local mass surface density and
the total mass of the galaxy, so that infall will proceed faster in
the densest parts of disks and, in general, in the most massive
galaxies. Boissier (2000) showed that tuning the mass dependence of
the infall rate can modify the present-day colors of galaxies. Redder
stellar populations can be obtained if gas infall in massive
early-type disks takes place even faster and earlier than in the
finally adopted version of the model. Further investigation in this
direction is left for future papers.

\subsection{Implications for the inside-out growth of disks}

After having analyzed the strenghts and weaknesses of the BP00 models,
here we discuss the implications of the model fitting results
regarding the inside-out growth of spiral disks. First, we will
compare the predicted and observed color profiles in our sample, to
ascertain which aspects of disk evolution can be reproduced by the
models and which ones would require a more complex approach (such as
N-body simulations). With these limitations in mind, we will then
study the size evolution of our disks since $z=1$ assuming that, at
least to first order, they have evolved as dictated by the model that
best fits their current multi-wavelength profiles.

\subsubsection{Color profiles}
\begin{figure*}
\begin{center}
\resizebox{1\hsize}{!}{\includegraphics{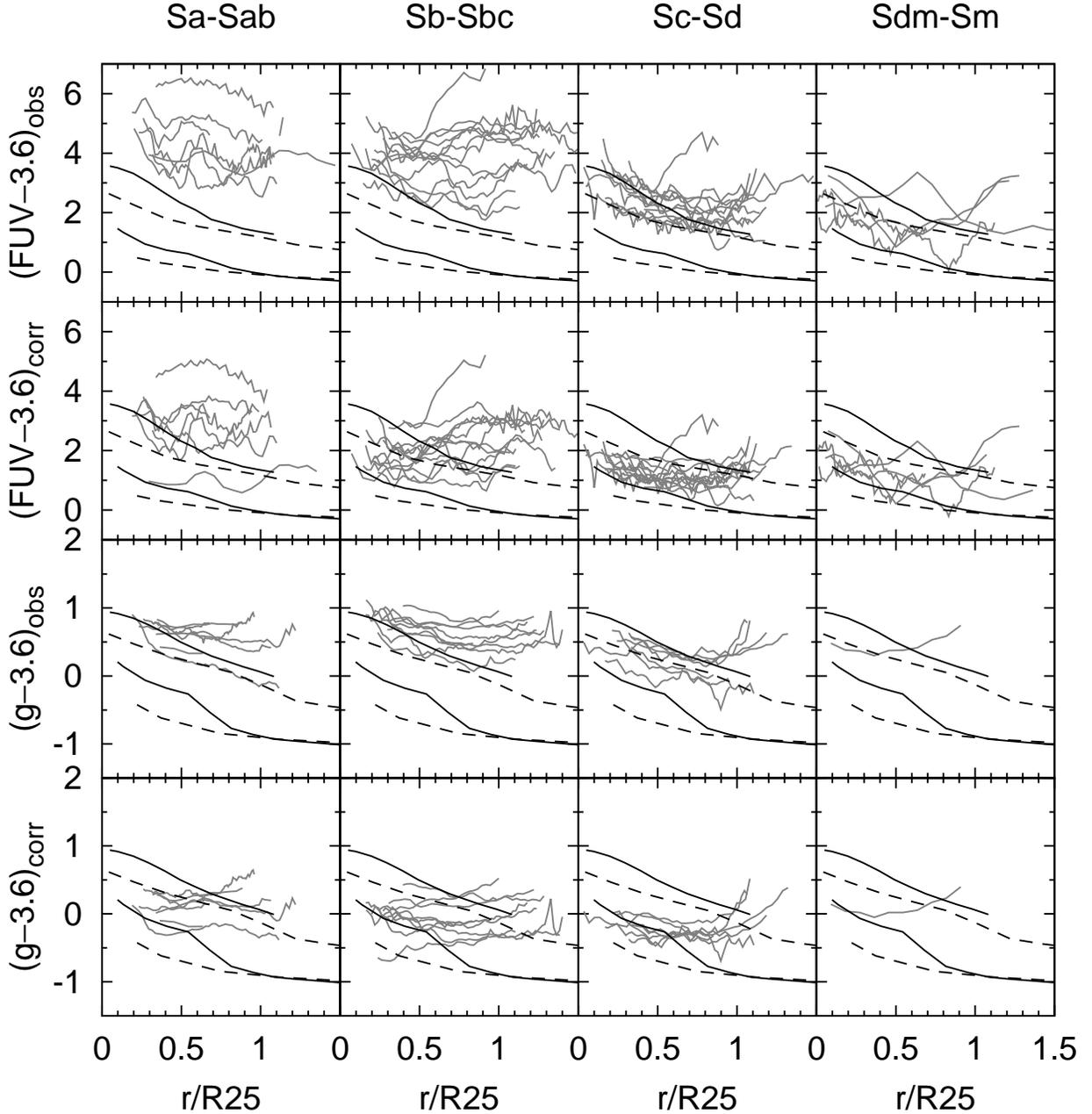}}
\caption{Comparison between the model and observed
  $(\mathrm{FUV}-3.6\,\micron)$ and $(g-3.6\,\micron)$ color profiles
  (only those galaxies with SDSS data appear in the bottom
  panels). Gray lines show the color profiles of our galaxies, both
  before and after correcting for internal extinction (``obs'' and
  ``corr'', respectively). The black lines correspond to model color
  profiles (with the K01 IMF) for values of $\lambda$ and
  $V_\mathrm{C}$ that roughly encompass most values found in our
  sample. Solid lines have $\lambda=0.03$, and dashed ones have
  $\lambda=0.09$. Within each pair of profiles, the upper and redder
  one corresponds to $V_{\mathrm{C}}=360$\,km\,s$^{-1}$, while the
  lower and bluer one has
  $V_{\mathrm{C}}=80$\,km\,s$^{-1}$.\label{color_profs_k01}}
\end{center}
\end{figure*}

In Fig.~\ref{color_profs_k01} we compare the color profiles of our
galaxies with those predicted by the model. The gray lines show the
color profiles within the radial range used in the fitting (that is,
excluding the bulges). The black lines show the model predictions
(using the K01 IMF) for selected values of $\lambda$ and
$V_{\mathrm{C}}$ that roughly bracket the values found for most of our
galaxies. The solid lines correspond to $\lambda=0.03$, and the dashed
ones to $\lambda=0.09$. Within each pair of lines, the upper (that is,
redder) one has $V_{\mathrm{C}}=360$\,km\,s$^{-1}$, whereas the lower
one has $V_{\mathrm{C}}=80$\,km\,s$^{-1}$. Two sets of color profiles
are shown: $(\mathrm{FUV}-3.6\,\micron)$ profiles and
$(g-3.6\,\micron)$ ones, both of them with and without correcting for
internal extinction.

Even after accounting for the effect of dust attenuation, the
$(\mathrm{FUV}-3.6\,\micron)$ profiles of Sa-Sbc galaxies fall outside
the region of the diagram delineated by the model predictions. This is
a direct consequence of the failure of the models at reproducing the
UV profiles in early-type disks, as extensively discussed in the
previous sections. In later types, however, the agreement is
excellent once extinction is taken into account (the difference is
minimal in Sdm-Sm galaxies, since they are not particularly
dusty).

The discrepancies become much less severe redward of the Balmer
break. After correcting for internal extinction, the model predictions
for the $(g-3.6\,\micron)$ profiles nicely encompass the actual color
profiles of our galaxies from early to late Hubble types. As we
already showed in Fig.~\ref{ngc2841_fit} with NGC~2841, the models are
capable of reproducing the optical and near-IR profiles of early-type
disks even when they fail in the UV.

However, the color profiles of our galaxies become redder at large
galactocentric distances, thus exhibiting a U-like shape. Note that
the fact that we can acceptably fit the light profiles but not the
color ones is not necessarily contradictory. Our models can reproduce
the globally exponential nature of disks, but are not sensible to
small variations such as those that can lead to the observed color
profiles.

These U-shaped profiles appear to be common both in nearby (Bakos et
al. 2008) and distant galaxies (Azzollini et al. 2008a). N-body
simulations show that they may result from a combination of a drop in
the SFR (seeded by warps in the gaseous disk, radial distribution of
angular momentum, misalignment between the rotation of the infalling
gas and the disk, etc.) and radial stellar migration, which would
populate the outskirts of disk with old stars formed inwards
(Ro\u{s}kar et al$.$ 2008; Mart\'inez-Serrano et al$.$ 2009;
S\'anchez-Bl\'azquez et al$.$ 2009). Unfortunately, none of these
processes can be easily translated into an analytic 1D scheme such as
ours without introducing too many unconstrained free variables.

However, the simplicity of our models $-$compared to N-body
simulations$-$ comes at the advantage of being able to easily generate
large grid of models over a wide range of halo masses and spins. Until
similar grids of N-body disk simulations become available, the BP00
constitute a reasonable first-order approach to infer a galaxy's past
evolution from its present-day photometric profiles.

\subsubsection{A first look on the past evolution of SINGS disks}
We have already checked that the model is able to reproduce the
observed circular velocities of our galaxies, as well as the expected
values of the spin parameter $-$both of which seem to be unaffected by
our particular choice of IMF. In addition, we have also carefully
explored some physical reasons that may be responsible for the failure
of the model at reproducing the UV profiles of some early-type disks,
which may hamper the study of the recent SFH in these objects. We can
now proceed to study the evolution of the objects in our sample, by
assuming that, at least on timescales of a few Gyr, our disks have
evolved in a similar way as the corresponding model that best fits its
present-day multiwavelength profiles (see also Boissier \& Prantzos
2001 for a detailed analysis of the evolution with redshift of several
physical properties of galaxies).

For each model characterized with a particular pair of values of
$\lambda$ and $V_\mathrm{C}$, we determine the disk scale-length
$R_\mathrm{d}$ at each epoch $t$ by fitting an exponential law to the
total stellar mass density profile:
\begin{equation}
\Sigma_\mathrm{stars}(r,t) = \Sigma_\mathrm{stars}(0,t)e^{-r/R_\mathrm{d}(t)}
\end{equation}

\begin{figure*}
\begin{center}
\resizebox{0.9\hsize}{!}{\includegraphics{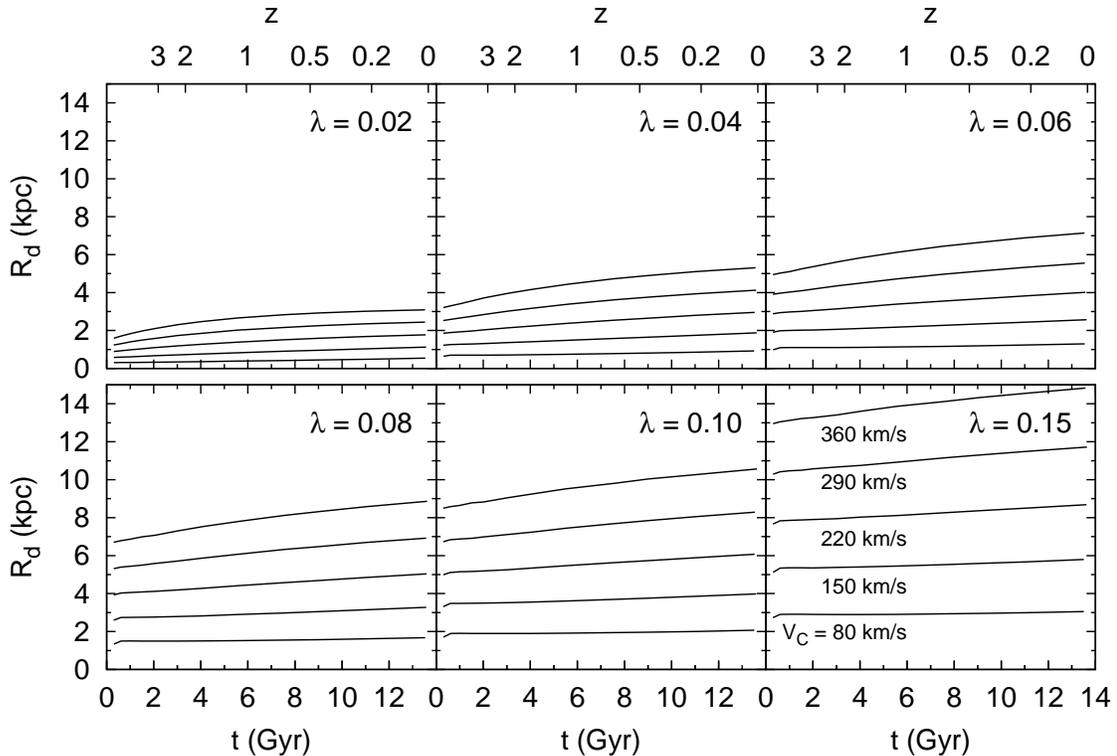}}
\caption{Temporal evolution of the exponential scale-length
  $R_\mathrm{d}$ of the total stellar mass density profiles. For each
  value of the spin parameter $\lambda$ we have plotted the curves
  corresponding to five selected circular velocities, as labeled in
  the bottom right panel.\label{stardens_growth}}
\end{center}
\end{figure*}

In Fig.~\ref{stardens_growth} we show the temporal evolution of the
disk scale-length for selected values of $\lambda$ and
$V_\mathrm{C}$. As expected, $R_\mathrm{d}$ increases with time in all
cases. At any given epoch, the most extended disks are those with
larger values of either parameter, as we already pointed out in
Fig.~\ref{sample_model_profiles}.

The curves describing the growth rate of $R_\mathrm{d}$ seem to get
steeper with increasing $V_\mathrm{C}$ at fixed $\lambda$. In order to
quantify the slope of these curves, we have performed a linear fit to
$R_\mathrm{d}$ as a function of time between $z=1$ and $z=0$. Disk
growth seems to be approximately linear since $z=1$ and, as mentioned
in the previous sections, it is not clear whether the models can
describe disk evolution beyond that redshift, when mergers were more
frequent and the thin disk was not fully assembled.

\begin{figure*}
\begin{center}
\resizebox{0.49\hsize}{!}{\includegraphics{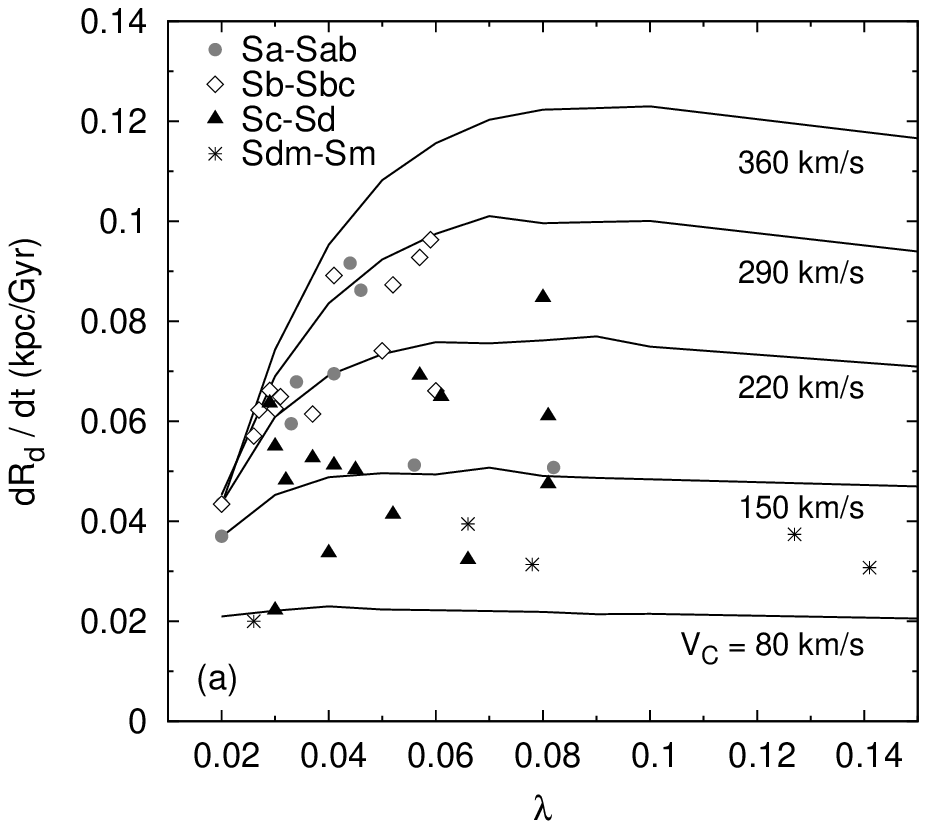}}
\resizebox{0.49\hsize}{!}{\includegraphics{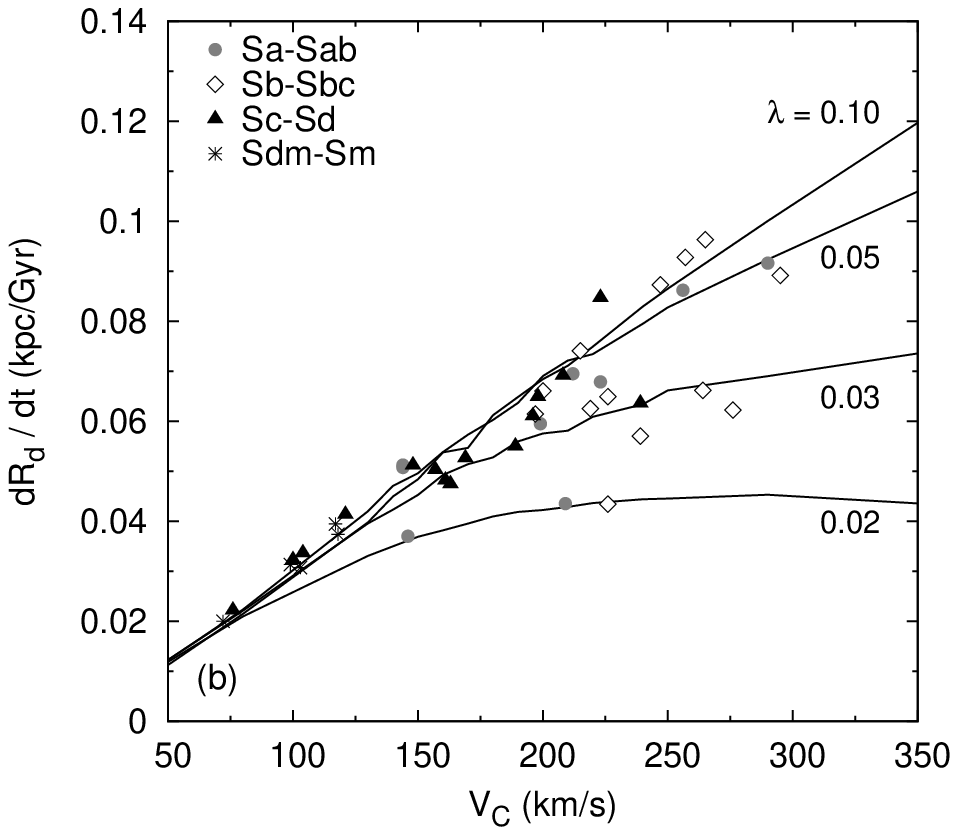}}
\caption{Growth rate of the scale-length of the stellar mass density
  profiles, as a function of $\lambda$ (a) and $V_\mathrm{C}$ (b). The
  growth rate has been computed by fitting $R_\mathrm{d}$ as a
  function of $t$ between $z=1$ and $z=0$. Observed morphological
  types are coded with different symbols.\label{stardens_growth_rate}}
\end{center}
\end{figure*}

In Fig.~\ref{stardens_growth_rate} we plot the disk growth rate
$dR_\mathrm{d}/dt$ as a function of both the spin parameter and the
rotational velocity. The growth rate increases up to $\lambda \sim
0.06$, but for larger values it seems to be largely insensitive to the
particular spin of the galaxy. It is clear that $dR_\mathrm{d}/dt$
mainly depends on the circular velocity, while the spin parameter only
seems to be relevant at large velocities. In general, very late-type
disks appear to grow at a rate of 0.02-0.04\,kpc\,Gyr$^{-1}$;
early-type spirals, on the other hand, can increase their
scale-lengths at a rate up to $\sim 0.1$\,kpc\,Gyr$^{-1}$, depending
on their spin.

\begin{figure*}
\begin{center}
\resizebox{0.49\hsize}{!}{\includegraphics{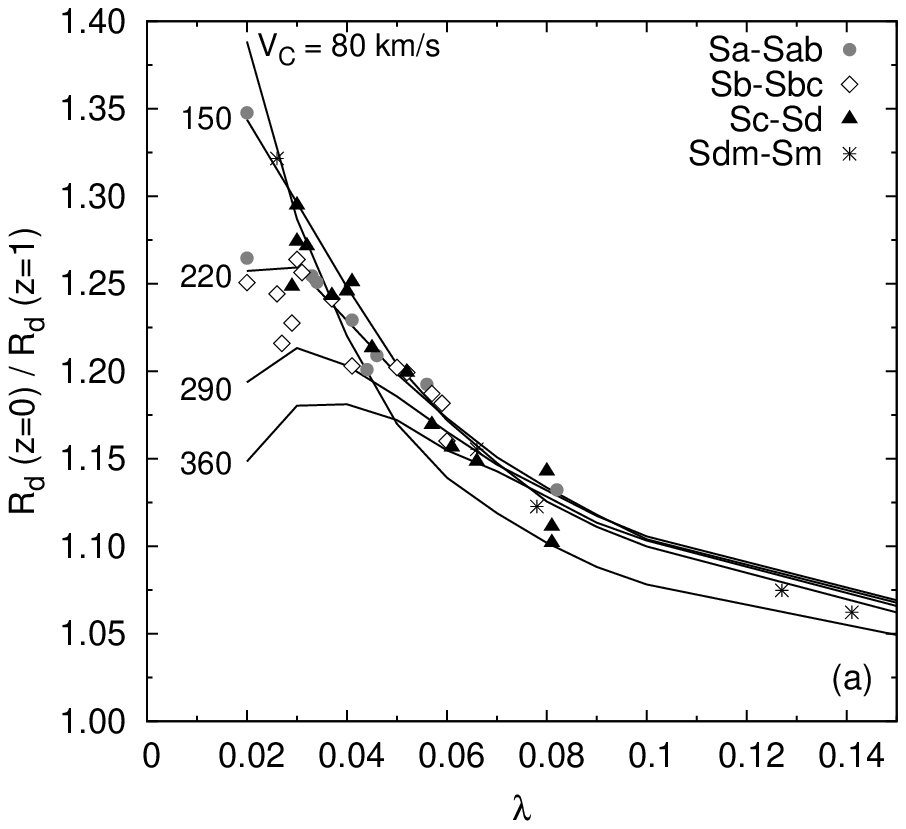}}
\resizebox{0.49\hsize}{!}{\includegraphics{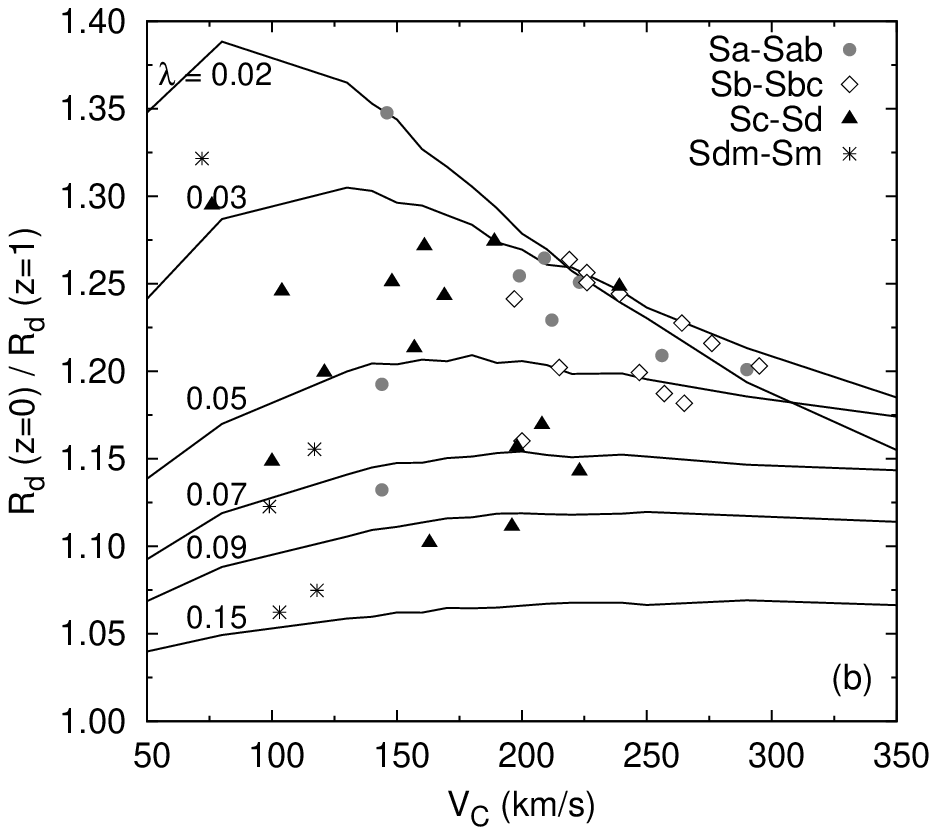}}
\caption{Ratio of the stellar disk scale-lengths at $z=0$ and $z=1$ as
  a function of the spin (a) and the rotational velocity (b). Observed
  morphological types are coded with different
  symbols.\label{size_ratio}}
\end{center}
\end{figure*}

Rather than describing the evolution of disks in terms of their {\it
  absolute} growth rate in kpc\,Gyr$^{-1}$, it is perhaps more
illustrative to focus on their {\it relative} size increase. We have
plotted in Fig.~\ref{size_ratio} the ratio of the scale-length of the
stellar mass profiles at $z=0$ relative to $z=1$. Interestingly, this size
ratio is essentially a unique function of the spin, with almost no
dependence on $V_\mathrm{C}$ $-$and hence on mass. It might seem
striking that high-spin galaxies experience almost no change in size
since $z=1$. However, as Fig.~\ref{stardens_growth} demonstrates,
these galaxies will already exhibit extended stellar mass profiles at
$z=1$.  Therefore, even if the absolute growth rate is high, it will
not have a significant impact on the relative increment in size.

\begin{figure*}
\begin{center}
\resizebox{0.45\hsize}{!}{\includegraphics{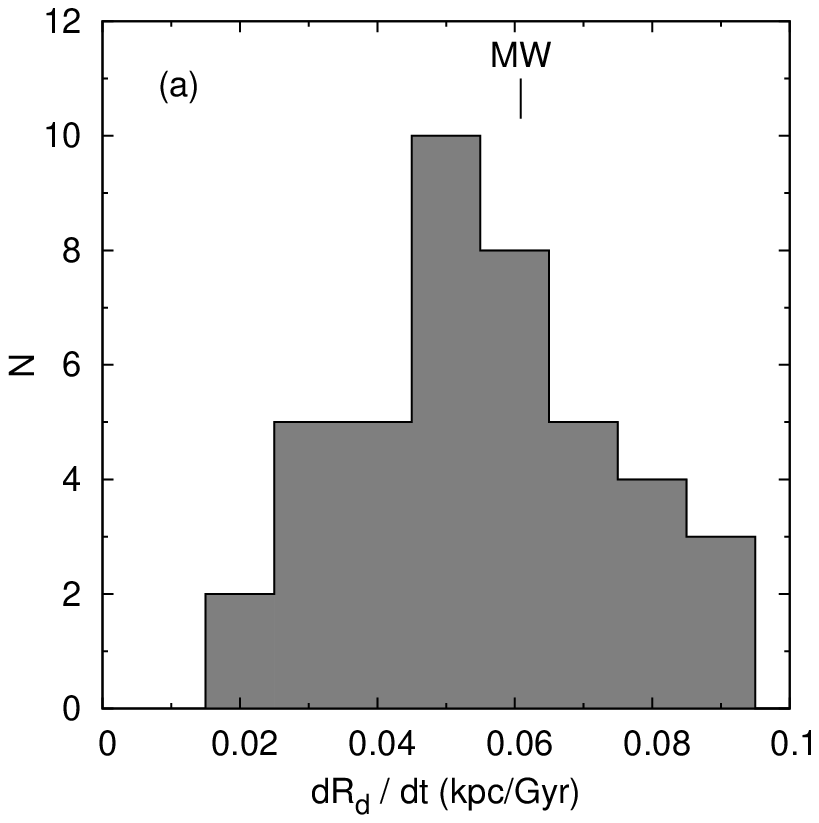}}
\resizebox{0.45\hsize}{!}{\includegraphics{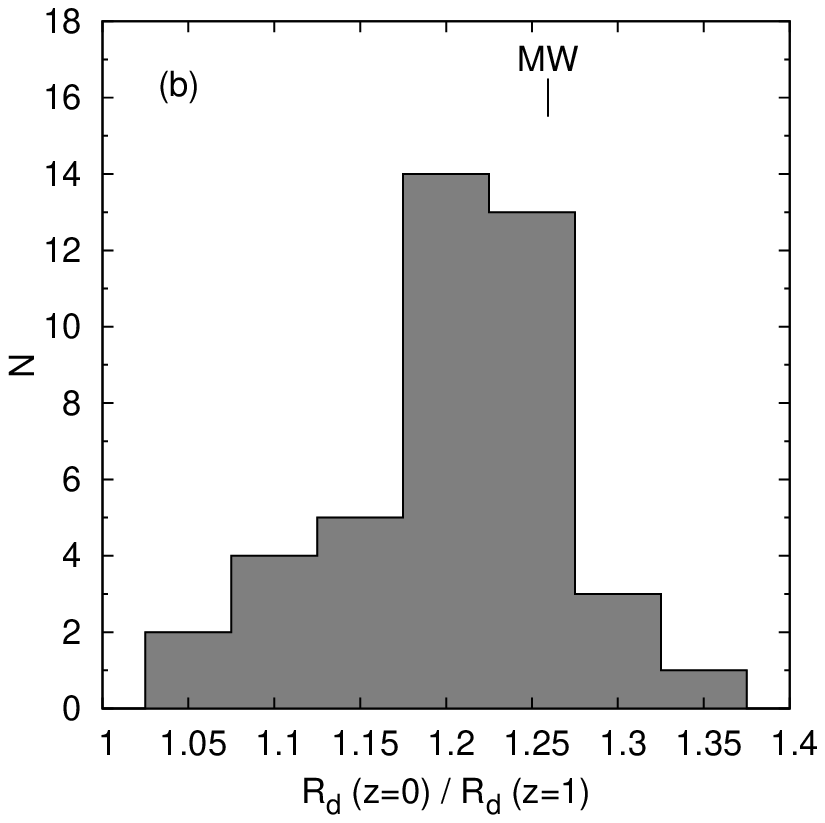}}
\caption{Distribution of the absolute growth rate (a) and relative
  size increase (b) in our sample. The values for a MW-like
  galaxy, with $\lambda = 0.03$ and $V_\mathrm{C}=220$\,km\,s$^{-1}$,
  have been marked as a reference.\label{growth_hist}}
\end{center}
\end{figure*}

Can we extrapolate the conclusions obtained for our sample to the
general population of disk-like galaxies? The histograms in
Fig.~\ref{growth_hist} show the distribution of both the absolute and
relative growth rates in our sample. Both distributions peak around
the values typical for galaxies similar to the Milky Way, with an
absolute growth rate of about 0.05-0.06\,kpc\,Gyr$^{-1}$ and a
relative size increase roughly equal to 20-25\% since $z=1$.

We should not blindly extend these results to the whole population of
spiral galaxies in the Local Universe. The absolute growth rate
depends primarily on $V_\mathrm{C}$ and, as we discussed in
section~\ref{stat_properties}, low-mass disks are considerably
underrepresented in our sample. Therefore, most disks in a
volume-limited sample would likely grow at slower rates than the peak
value in Fig.~\ref{growth_hist}.

However, the situation is different regarding the relative increment
in size. As we stated in section~\ref{stat_properties}, the
distribution of spin values in our sample matches reasonably well the
one found in N-body simulations of disk formation. Therefore, we can
treat our sample as being representative of a complete one regarding
any $\lambda$-dependent quantity. This is precisely the case of the
relative size ratio, which depends almost entirely on $\lambda$ alone
according to Fig.~\ref{size_ratio}. Therefore, we can safely conclude
that most disks have probably undergone an increase of 20-25\% in
their scale-lengths since $z=1$ until now, regardless of their total
mass. This result is in perfect agreement with the growth rate we
estimated in Mu\~noz-Mateos et al$.$ (2007) on a larger sample of
galaxies, but using only extinction-corrected FUV and $K_S$-band
profiles plus a very simple toy model of disk growth.

It is interesting to compare our theoretical expectations for
inside-out disk growth with actual measurements of disk sizes at
different redshifts. From a observational perspective, this issue is
typically addressed by studying the evolution (or lack of thereof) in
the magnitude-size and mass-size relations, the average surface
brightness and the size number density (Schade et al$.$ 1996; Lilly et
al$.$ 1998; Simard et al$.$ 1999; Ravindranath et al$.$ 2004; Trujillo
et al$.$ 2004, 2006; Barden et al$.$ 2005; McIntosh et al$.$ 2005;
Trujillo \& Pohlen 2005; Azzollini et al$.$ 2008b). The results of such
studies are sometimes contradictory, due to selection effects and the
inherent difficulty in disentangling the evolution of individual
galaxies from the evolution of a population of galaxies as a whole.

Boissier \& Prantzos (2001) confronted their predicted size-luminosity
trend with several observed data-sets from the literature. Here we
revisit this issue, and compare our model with the mass-size relation
derived by Barden et al$.$ (2005). These authors determined disk
effective radii at various $z$ by fitting a S\'ersic model (S\'ersic
1968) to HST images from GEMS (Galaxy Evolution from Morphologies and
SEDs; Rix et al$.$ 2004). Stellar masses were derived from SED fitting
to COMBO-17 data for the same objects. As a local anchor for their
study they relied on a sample of nearby galaxies from SDSS. These
authors found little or no evolution with $z$ in the mass-size
relation. Assuming that galaxies can only become more massive with
time, they argued that they should also increase their sizes
accordingly.

\begin{figure}
\begin{center}
\resizebox{1\hsize}{!}{\includegraphics{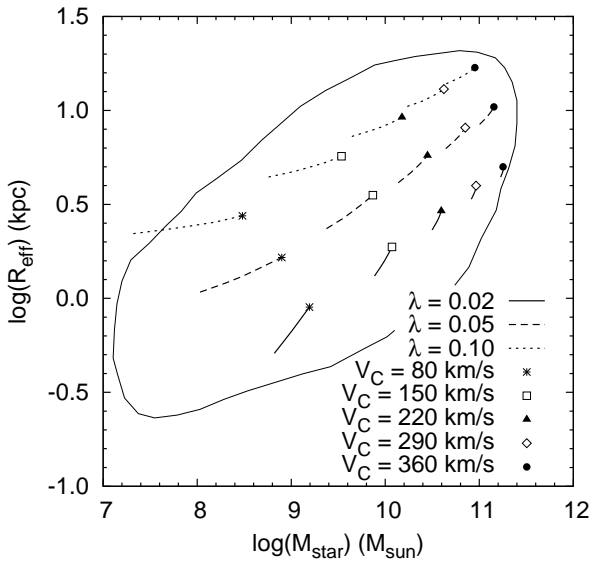}}
\caption{Evolution of the effective radius and total stellar mass of
  our model galaxies. The effective radius was obtained by fitting a
  S\'ersic model to the stellar mass profiles at different $z$. Each
  track follows the evolution of a model disk from $z=1$ to $z=0$ (the
  $z=0$ step is marked with a symbol). The corresponding spins and
  circular velocities of each model are indicated with different line
  styles and symbols, respectively. The closed curve shows the
  observed mass-size relation from Barden et al$.$
  (2005).\label{mass_size}}
\end{center}
\end{figure}

To replicate these measurements, for each one of our model disks we
fitted a S\'ersic profile to the radial distribution of stellar mass at
different epochs. The results are shown in Fig.~\ref{mass_size}. Each track follows
the evolution of the effective radius and total stellar mass of a
model galaxy from $z=1$ to $z=0$ (the $z=0$ step is marked with a
symbol). Models with different velocities and spins are coded with
different symbols and line styles, respectively. The irregular closed
line encompasses the empirical data-points of Barden et al$.$ (2005),
including different redshift bins between $z=0$ and $z=1$ (see their
Fig.~10).

Our model is in perfect agreement with the observed mass-size
trend. Note that the effective radii in Barden et al$.$ (2005) were
corrected to the restframe V band, without any further correction to
get the actual stellar size. S\'ersic fits to our model profiles
indicate that the V-band effective radii are typically 0.05-0.15\,dex
larger than the stellar ones, so the closed curve in Fig.~\ref{mass_size} should be
shifted downwards by that amount. The agreement would be still
excellent; it would actually improve if we note that most galaxies are
expected to lie between the $\lambda=0.02$ and $\lambda=0.05$
tracks, according to the $\lambda$ probability distribution shown in Fig.~4.

\section{Conclusions}\label{conclusions}
In this paper we have fitted the UV, optical and near-IR profiles of
42 disk-like galaxies of the SINGS sample with the models of Boissier
\& Prantzos (1999, 2000). In order to recover the unattenuated
starlight profiles at each wavelength, we have used the
radially-varying TIR/UV ratio as tracer of the internal
extinction. The disk evolution models are calibrated on the Milky Way
(BP99) and further extended to other disk-like galaxies through
scaling laws derived from the $\Lambda$CDM scenario (BP00). By
considering in a consistent framework the gas infall, the star
formation activity and the subsequent chemical evolution, the models
are capable of predicting the current multi-band profiles of spirals
as a function of only two parameters: the maximum circular velocity of
the rotation curve, $V_{\mathrm{C}}$, and the dimensionless spin
parameter, $\lambda$. By fitting the profiles of the SINGS galaxies
with those predicted by the models, we have not only checked the
accuracy of the models themselves, but have also used them to infer
the size evolution of the SINGS galaxies. The main conclusions of this
work can be summarized as follows:

\begin{enumerate}
\item{The rotational velocities are in good agreement with those
  measured from observed HI rotation curves, as well as with those
  estimated from the Tully-Fisher relation. In the latter case, the
  values of $V_{\mathrm{C}}$ derived from the model fitting tend to be
  10-20\% larger than those predicted by the TF relation, but mainly
  for the most massive disks.}

\item{Most galaxies in our sample exhibit spin values of $\lambda \sim
  0.03$. In fact, even though the sample is not volume-limited, its
  statistical distribution of spin values closely resembles the narrow
  distributions usually found in N-body simulations, which typically
  have an almost universal peak at $\lambda \sim 0.03$-$0.04$.}

\item{There is a clear, well-known trend between $V_{\mathrm{C}}$ and
  Hubble type, in the sense that early-type disks have larger circular
  velocities $-$and are hence more massive$-$ than late-type
  ones. There is not, however, any evident trend between the
  morphological type and $\lambda$, which supports the findings of
  numerical simulations that most haloes posses the same spin,
  regardless of their total mass or mass assembly history.}

\item{While there is an excellent agreement between the model
    predictions and the observed profiles in the optical and near-IR
    bands, significant departures may arise in the UV bands, depending
    on the morphological type and the particular choice of IMF. The
    Kroupa (2001) IMF yields excellent results in Sc-Sd spirals, but
    overestimates the UV luminosity in early-type disks, and to a much
    lesser extent in Sdm-Sm ones. The Kroupa et al$.$ (1993) IMF
    brings the UV model profiles into better agreement with the
    observed ones in Sb-Sbc spirals, as well as in Sdm-Sm ones, but at
    the expense of loosing the excellent fits for Sc-Sd disks. While
    differences in the high-mass end of the IMF might indeed play a
    role in very late-type galaxies, it is doubtful that the IMF is
    behind the discrepancies in the UV predictions for early-type
    disks. Anyway, the values of $\lambda$ and $V_{\mathrm{C}}$ are
    largely unaffected by the specific IMF chosen. For the
    HI-deficient galaxies in our sample, gas removal due to
    interactions with the intracluster medium is the most likely
    culprit. For those galaxies with normal HI masses it may be
    necessary to revisit the mass dependence of the gas infall rate,
    since the model seems to retain too much gas in these objects.}

\item{The metallicity gradients predicted by the models are
    $\sim$\,0.015\,dex\,kpc$^{-1}$ steeper than the observed ones. The
    central oxygen abundances depend on the IMF: the values yielded by
    the K93 IMF are in perfect agreement with the observed central
    metallicities, but those obtained with the K01 one overestimate
    the real values by $\sim$\,0.7\,dex.}

\item{According to the models, the absolute growth rate (in
  kpc\,Gyr$^{-1}$) of the exponential scale-length of disks depends
  mainly on $V_{\mathrm{C}}$, with rapidly rotating disks expanding
  faster. In our sample, most galaxies have their scale-lengths
  increased by about 0.05-0.06\,kpc each Gyr. Still, this is not
  representative of the overall population of disks, since low-mass
  ones are underrepresented in our sample.}

\item{The ratio between the current disk scale-length and that at
  $z=1$ is a decreasing function of $\lambda$, with little dependence
  on $V_{\mathrm{C}}$. Even though high-spin disks grow faster in
  absolute terms, such a rapid radial expansion does not significantly
  increase their scale-lengths, which are already considerably large
  at $z=1$. On average, most disks in our sample are now 20-25\%
  larger than at $z=1$. This value can be treated as being
  representative of a volume-limited sample, given that our galaxies
  have the $\lambda$ distribution expected for such a kind of sample.}

\item{The model predicts that disk galaxies should simultaneously increase
their sizes and stellar masses as time goes by. The results of the
model for a grid of values of $\lambda$ and $V_{\mathrm{C}}$ provide a
perfect match to the observed constancy of the mass-size relation between $z=0$ and $z=1$.}

\end{enumerate}

\acknowledgments

JCMM acknowledges the receipt of a Formaci\'on del Profesorado
Universitario fellowship from the Spanish Ministerio de Educaci\'on y
Ciencia, as well as finantial support from NASA JPL/Spitzer grant RSA
1374189 provided for the S$^4$G project. A.G.dP is also financed by
the Spanish Ram\'on y Cajal program. JCMM, AGdP, JZ and JG are
partially financed by the Spanish Programa Nacional de Astronom\'{\i}a
y Astrof\'{\i}sica under grants AYA2006-02358 and AyA2009-10368. They
are also partly supported by the Consolider-GTC program under grant
CSD2006-00070 and the AstroMadrid project (CAM S2009/ESP-1496). Part
of this work was performed during a three-month stay at the
Laboratoire d'Astrophysique de Marseille (LAM). JCMM thanks the
Spanish Ministerio de Educaci\'on y Ciencia for providing the
necessary funds, as well as the LAM staff for their support and warm
hospitality. He also acknowledges support from the National Radio
Astronomy Observatory, which is a facility of the National Science
Foundation operated under cooperative agreement by Associated
Universities, Inc. We also wish to thank THINGS members A$.$ Leroy and
F$.$ Walter for kindly providing the HI radial profiles in advance of
publication. We also thank the anonymous referee for providing
valuable comments that have improved the paper.

GALEX (Galaxy Evolution Explorer) is a NASA Small Explorer, launched
in 2003 April. We gratefully acknowledge NASA's support for
construction, operation, and science analysis for the GALEX mission,
developed in cooperation with the Centre National d'\'Etudes Spatiales
of France and the Korean Ministry of Science and Technology. This work
is part of SINGS, the {\it Spitzer} Infrared Nearby Galaxies
Survey. The {\it Spitzer} Space Telescope is operated by the Jet
Propulsion Laboratory, Caltech, under NASA contract 1403.

Funding for the SDSS and SDSS-II has been provided by the Alfred
P. Sloan Foundation, the Participating Institutions, the National
Science Foundation, the U.S. Department of Energy, the National
Aeronautics and Space Administration, the Japanese Monbukagakusho, the
Max Planck Society, and the Higher Education Funding Council for
England. The SDSS Web Site is http://www.sdss.org/.

The SDSS is managed by the Astrophysical Research Consortium for the
Participating Institutions. The Participating Institutions are the
American Museum of Natural History, Astrophysical Institute Potsdam,
University of Basel, University of Cambridge, Case Western Reserve
University, University of Chicago, Drexel University, Fermilab, the
Institute for Advanced Study, the Japan Participation Group, Johns
Hopkins University, the Joint Institute for Nuclear Astrophysics, the
Kavli Institute for Particle Astrophysics and Cosmology, the Korean
Scientist Group, the Chinese Academy of Sciences (LAMOST), Los Alamos
National Laboratory, the Max-Planck-Institute for Astronomy (MPIA),
the Max-Planck-Institute for Astrophysics (MPA), New Mexico State
University, Ohio State University, University of Pittsburgh,
University of Portsmouth, Princeton University, the United States
Naval Observatory, and the University of Washington.

This publication makes use of data products from the Two Micron All
Sky Survey, which is a joint project of the University of
Massachusetts and the Infrared Processing and Analysis
Center/California Institute of Technology, funded by the National
Aeronautics and Space Administration and the National Science
Foundation.

Finally, we have made use of the NASA/IPAC Extragalactic Database
(NED), which is operated by the Jet Propulsion Laboratory, California
Institute of Technology (Caltech) under contract with NASA.

{\it Facilities:} \facility{GALEX}, \facility{Sloan},
\facility{CTIO:1.5m}, \facility{KPNO:2.1m}, \facility{FLWO:2MASS},
\facility{CTIO:2MASS}, \facility{{\it Spitzer}}, \facility{VLA}

\appendix
\section{Two-dimensional distribution of $\chi^2$ values}\label{app_chi2}
When fitting multi-wavelength profiles with the disk evolution models
of BP00, one must bear in mind that the circular velocity and the spin
may not act as completely independent parameters. Depending on the
particular shape of a galaxy's profile at different wavelengths,
variations in one parameter might be compensated by variations in the
other one while still providing an acceptably good fit.

In order to depict the possible internal degeneracies between
$\lambda$ and $V_{\mathrm{C}}$, in this appendix we present the
two-dimensional $\chi^2$ distributions obtained for each galaxy in our
sample (Fig.~\ref{chi2plots}). Even though we keep track of the
individual $\chi^2$ distributions corresponding to each particular
band for each galaxy, the plots presented here show the distribution
of total $\chi^2$ values taking into account all bands. The best
fitting model is marked with a white dot, and the area encompassing
all models with $\chi \leq 2 \chi_{\mathrm{min}}$ has been delimited
with a white closed line. The same range in $\lambda$ and
$V_{\mathrm{C}}$ is displayed in all panels, except for the small
subset of galaxies requiring larger spins and/or lower circular
velocities than those in our initial grid. For the sake of clarity,
those galaxies have been grouped together at the end.

In general, some galaxies exhibit some degree of anticorrelation
between both parameters for low and intermediate values of $\lambda$,
since the increment in the radial scale-length caused by augmenting
$\lambda$ can be partly compensated by decreasing $V_{\mathrm{C}}$. In
some other objects, the $\chi^2$ distribution around the best fitting
model does not show any significant degeneracy. Finally, for large
values of the spin the correlation is possitive: further incrementing
$\lambda$ significantly decreases the central surface brightness,
which can be compensated to some extent by increasing $V_{\mathrm{C}}$
$-$even though this tends to augment the radial scale-length as well.

The fact that the spin parameter is not as strongly constrained as the
circular velocity is mainly due to the different way in which both
quantities affect the radial profiles (Fig. 1). Modifying Vc will
shift the model profiles above or below the observed ones, thus
rapidly increasing the $\chi^2$ value. Varying $\lambda$, on the other
hand, will mainly change the scale-length alone. Given that the
observed profiles exhibit inhomogeneities that deviate from our smooth
predictions, this leaves some room for varying $\lambda$ while still
obtaining a good fit.

\clearpage
\begin{figure}
\begin{center}
\resizebox{0.75\hsize}{!}{\includegraphics{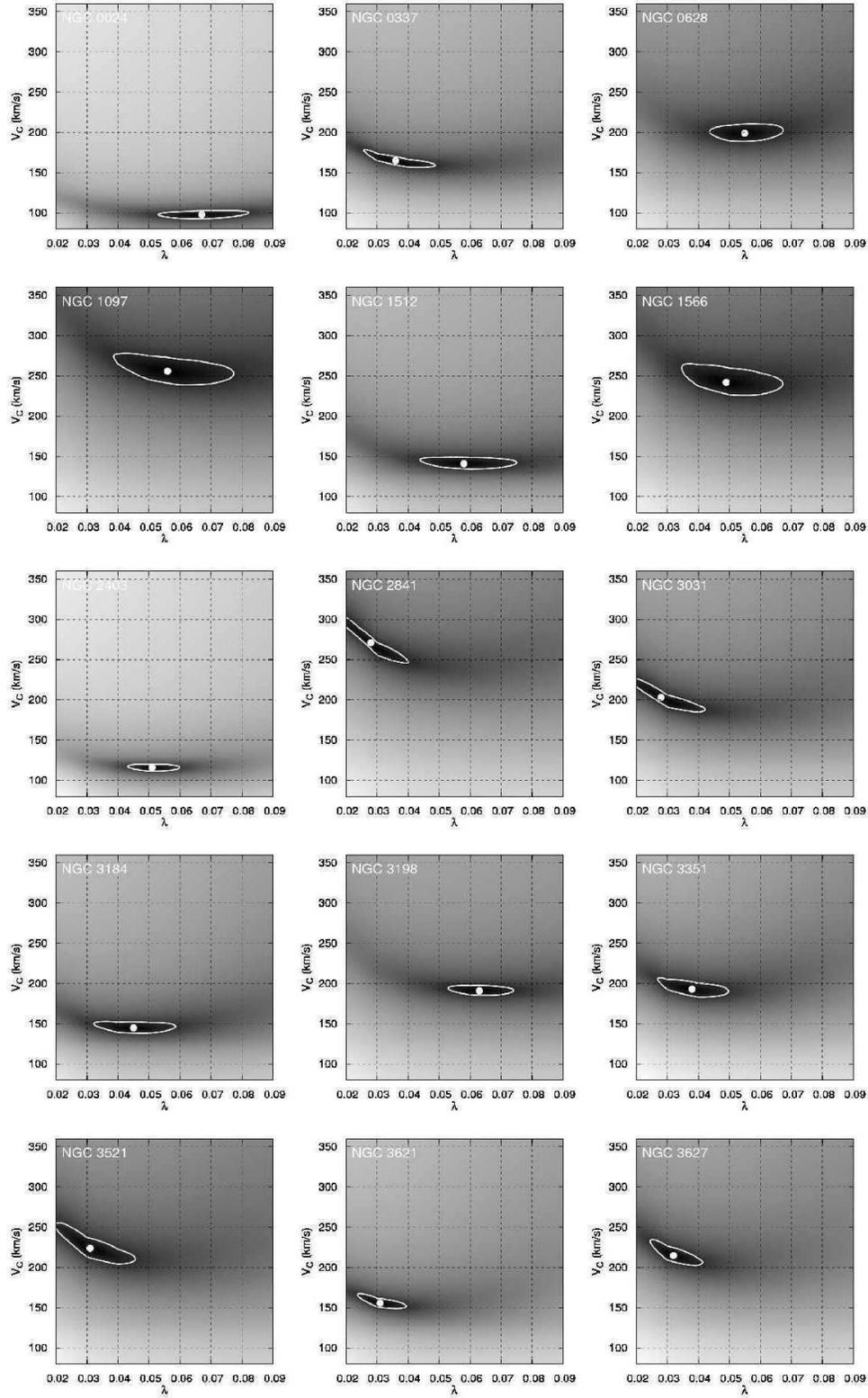}}
\caption{Two-dimensional $\chi^2$ distributions for the galaxies in
  our sample. Darker shades of gray correspond to lower values of
  $\chi^2$. The values of $\lambda$ and $V_{\mathrm{C}}$ corresponding
  to the best-fitting model have been marked with a white dot. The
  white curved line encloses all models that satisfy $\chi^2 \leq
  2\chi^2_{\mathrm{min}}$.\label{chi2plots}}
\end{center}
\end{figure}

\clearpage
\begin{figure}
\begin{center}\figurenum{\ref{chi2plots}}
\resizebox{0.75\hsize}{!}{\includegraphics{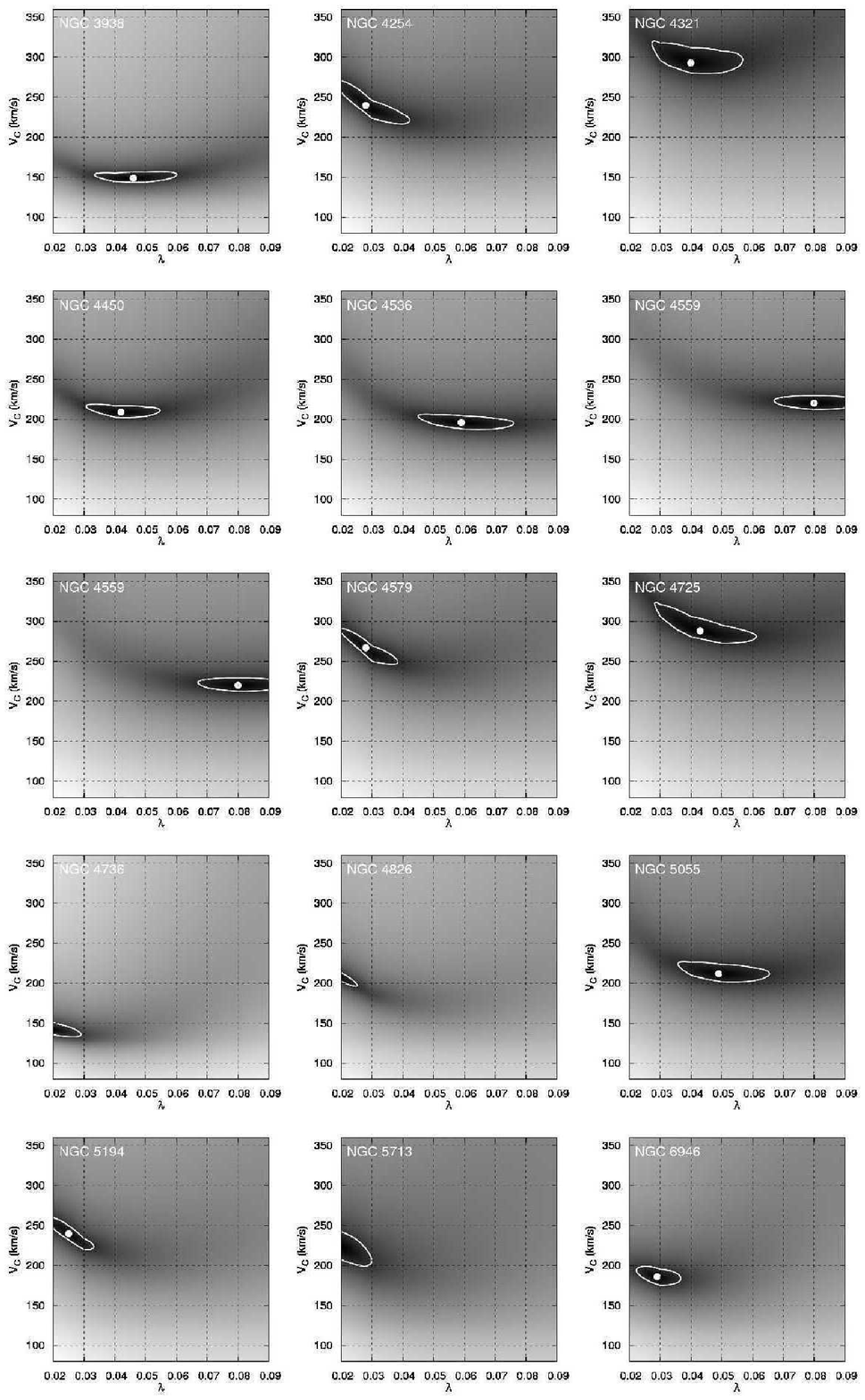}}
\caption{\it Continued}
\end{center}
\end{figure}

\clearpage
\begin{figure}
\begin{center}\figurenum{\ref{chi2plots}}
\resizebox{0.75\hsize}{!}{\includegraphics{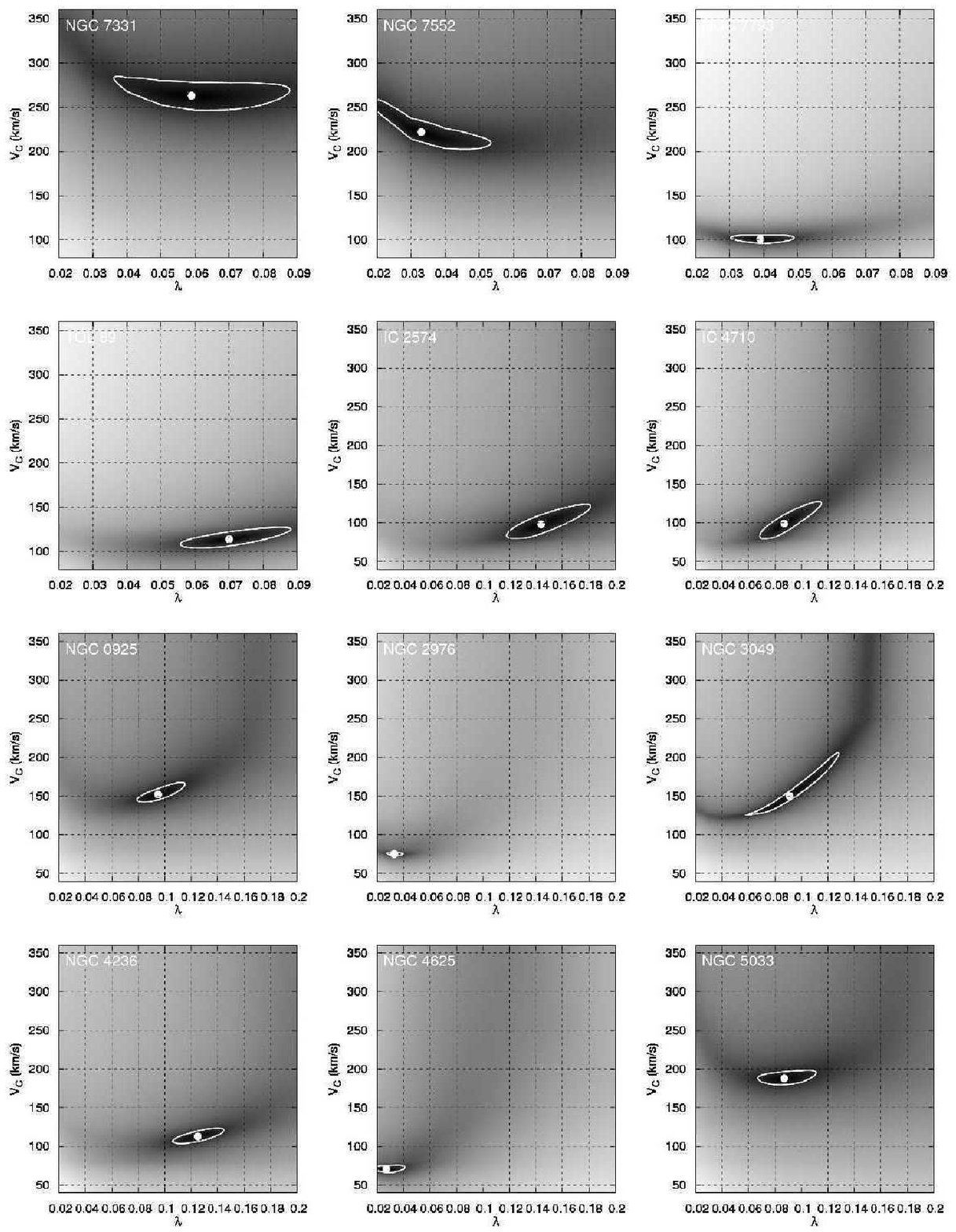}}
\caption{\it Continued}
\end{center}
\end{figure}

\clearpage
\begin{deluxetable*}{lrrrrrcrrccc}
\setlength{\tabcolsep}{3pt}
\tabletypesize{\scriptsize}
\tablecolumns{12}
\tablecaption{Sample.\label{table1}}
\tablewidth{0pt}
\tablehead{
\colhead{Object name} & \colhead{RA$_{2000}$} & \colhead{DEC$_{2000}$} & \colhead{$2a$} & \colhead{$2b$} & \colhead{P.A.} & \colhead{E(B$-$V)} & \colhead{dist} & \colhead{$T$} & \colhead{Morphological} & \colhead{$r_{\mathrm{in}}$} & \colhead{$r_{\mathrm{out}}$}\\
\colhead{} & \colhead{(h:m:s)} & \colhead{(d:m:s)}
 & \colhead{(arcmin)} & \colhead{(arcmin)}
 & \colhead{(deg)} & \colhead{(mag)}
 & \colhead{(Mpc)} & \colhead{type}
 & \colhead{type} & \colhead{(arcsec)}
 & \colhead{(arcsec)}\\
\colhead{(1)} & \colhead{(2)} & \colhead{(3)} & \colhead{(4)} & \colhead{(5)} & \colhead{(6)} & \colhead{(7)} & \colhead{(8)} & \colhead{(9)} & \colhead{(10)} & \colhead{(11)} & \colhead{(12)}}

\startdata
    NGC~0024 & 00 09 56.5 & $-$24 57 47.3 &   5.8 &   1.3 & 46 & 0.020 & 8.2 & 5 & SA(s)                        & 48 & \nodata \\                      
    NGC~0337 & 00 59 50.1 & $-$07 34 40.7 &   2.9 &   1.8 & 310 & 0.112 &  25 & 7 & SB(s)d                      & 30 & \nodata \\                      
%   NGC~0584 & 01 31 20.7 & $-$06 52 05.0 &   4.2 &   2.3 & 55 & 0.042 &  28 & $-$5 & E4                        & \nodata & \nodata \\                 
    NGC~0628 & 01 36 41.8 & 15 47 00.5 &     10.5 &   9.5 & 25 & 0.070 &  11 & 5 & SA(s)c                       & 54 & \nodata \\                      
%   NGC~0855 & 02 14 03.6 & 27 52 37.8 &      2.6 &   1.0 & 60 & 0.072 & 9.7 & $-$5 & E                         & \nodata & \nodata \\                 
    NGC~0925 & 02 27 16.9 & 33 34 45.0 &     10.5 &   5.9 & 282 & 0.076 & 9.3 & 7 & SAB(s)d                     & 66 & \nodata \\                      
    NGC~1097 & 02 46 19.1 & $-$30 16 29.7 &   9.3 &   6.3 & 310 & 0.027 &  15 & 3 & SB(s)b                      & 54 & \nodata \\                      
%   NGC~1266 & 03 16 00.7 & $-$02 25 38.5 &   1.5 &   1.0 & 290 & 0.098 &  31 & $-$2 & (R')SB0(rs) pec          & \nodata & \nodata \\                 
%   NGC~1291 & 03 17 18.6 & $-$41 06 29.1 &   9.8 &   8.1 & 345 & 0.013 & 9.7 & 0 & (R)SB(s)0/a                 & 144 & \nodata \\                     
%   NGC~1316 & 03 22 41.7 & $-$37 12 29.6 &  12.0 &   8.5 & 50 & 0.021 &  19 & $-$2 & SAB(s)0 pec               & \nodata & \nodata \\                 
%   NGC~1377 & 03 36 39.1 & $-$20 54 08.0 &   1.8 &   0.9 & 92 & 0.028 &  24 & $-$2 & S0                        & \nodata & \nodata \\                 
%   NGC~1404 & 03 38 51.9 & $-$35 35 39.8 &   3.3 &   3.0 &   0 & 0.011 &  19 & $-$5 & E1                       & \nodata & \nodata \\                 
%   NGC~1482 & 03 54 38.9 & $-$20 30 08.8 &   2.5 &   1.4 & 283 & 0.040 &  25 & $-$0.8 & SA0$^+$ pec            & \nodata & \nodata \\                 
    NGC~1512 & 04 03 54.3 & $-$43 20 55.9 &   8.9 &   5.6 & 90 & 0.011 &  10 & 1 & SB(r)a                       & 48 & 300 \\                          
    NGC~1566 & 04 20 00.4 & $-$54 56 16.1 &   8.3 &   6.6 & 60 & 0.009 &  17 & 4 & SAB(s)bc                     & 30 & \nodata \\                      
%   NGC~1705 & 04 54 13.5 & $-$53 21 39.8 &   1.9 &   1.4 & 50 & 0.008 & 5.1 & 11 & SA0- pec                    & \nodata & \nodata \\                 
    NGC~2403 & 07 36 51.4 & 65 36 09.2 &     21.9 &  12.3 & 307 & 0.040 & 3.2 & 6 & SAB(s)cd                    & 24 & \nodata \\                      
%Holmberg~II & 08 19 05.0 & 70 43 12.1 &      7.9 &   6.3 & 15 & 0.032 & 3.4 & 10 & Im                          & \nodata & \nodata \\                 
%    M81~Dwa & 08 23 56.0 & 71 01 45.0 &   1.3 &   1.3 &   0 & 0.021 & 3.5 & 10 & I?                            & \nodata & \nodata \\                 
%    DDO~053 & 08 34 07.2 & 66 10 54.0 &   1.5 &   1.3 & 300 & 0.037 & 3.6 & 10 & Im                            & \nodata & \nodata \\                 
%   NGC~2798 & 09 17 23.0 & 41 59 59.0 &   2.6 &   1.0 & 340 & 0.020 &  27 & 1 & SB(s)a pec                     & \nodata & \nodata \\                 
    NGC~2841 & 09 22 02.6 & 50 58 35.5 &   8.1 &   3.5 & 327 & 0.016 &  14 & 3 & SA(r)b                         & 90 & \nodata \\                      
%   NGC~2915 & 09 26 11.5 & $-$76 37 34.8 &   1.9 &   1.0 & 309 & 0.275 & 3.8 & 90 & I0                         & \nodata & \nodata \\                 
% Holmberg~I & 09 40 32.3 & 71 10 56.0 &   3.6 &   3.0 &   0 & 0.048 & 3.8 & 10 & IAB(s)m                       & \nodata & \nodata \\                 
    NGC~2976 & 09 47 15.5 & 67 54 59.0 &   5.9 &   2.7 & 323 & 0.069 & 3.6 & 5 & SAc pec                        & \nodata & \nodata \\                 
    NGC~3049 & 09 54 49.7 & 09 16 17.9 &   2.2 &   1.4 & 25 & 0.038 &  22 & 2 & SB(rs)ab                        & 18 & \nodata \\                      
    NGC~3031 & 09 55 33.2 & 69 03 55.1 &  26.9 &  14.1 & 337 & 0.080 & 3.6 & 2 & SA(s)ab                        & 204 & 900 \\                         
%   NGC~3034 & 09 55 52.2 & 69 40 46.9 &  11.2 &   4.3 & 65 & 0.159 & 3.9 & 90 & I0                             & \nodata & \nodata \\                 
%Holmberg~IX & 09 57 32.0 & 69 02 45.0 &   2.5 &   2.0 & 40 & 0.079 & 3.6 & 10 & Im                             & \nodata & \nodata \\                 
%    M81~Dwb & 10 05 30.6 & 70 21 52.0 &   0.9 &   0.6 & 320 & 0.080 & 5.3 & 10 & Im                            & \nodata & \nodata \\                 
%   NGC~3190 & 10 18 05.6 & 21 49 55.0 &   4.4 &   1.5 & 305 & 0.025 &  17 & 1 & SA(s)a pec                     & 48 & \nodata \\                      
    NGC~3184 & 10 18 17.0 & 41 25 28.0 &   7.4 &   6.9 & 135 & 0.017 & 8.6 & 6 & SAB(rs)cd                      & 48 & \nodata \\                      
    NGC~3198 & 10 19 54.9 & 45 32 59.0 &   8.5 &   3.3 & 35 & 0.012 &  17 & 5 & SB(rs)c                         & 48 & \nodata \\                      
     IC~2574 & 10 28 23.5 & 68 24 43.7 &  13.2 &   5.4 & 50 & 0.036 & 4.0 & 9 & SAB(s)m                         & \nodata & \nodata \\                 
%   NGC~3265 & 10 31 06.8 & 28 47 47.0 &   1.3 &   1.0 & 73 & 0.024 &  20 & $-$5 & E                            & \nodata & \nodata \\                 
%    MRK~ 33 & 10 32 31.9 & 54 24 03.7 &   1.0 &   0.9 &   0 & 0.012 &  24 & 10 & Im pec                        & \nodata & \nodata \\                 
    NGC~3351 & 10 43 57.7 & 11 42 13.0 &   7.4 &   5.0 & 13 & 0.028 &  12 & 3 & SB(r)b                          & 48 & \nodata \\                      
    NGC~3521 & 11 05 48.6 & $-$00 02 09.1 &  11.0 &   5.1 & 343 & 0.058 & 9.0 & 4 & SAB(rs)bc                   & 48 & \nodata \\                      
    NGC~3621 & 11 18 16.5 & $-$32 48 50.6 &  12.3 &   7.1 & 339 & 0.080 & 8.3 & 7 & SA(s)d                      & 48 & 230 \\                         
    NGC~3627 & 11 20 15.0 & 12 59 29.6 &   9.1 &   4.2 & 353 & 0.032 & 9.1 & 3 & SAB(s)b                        & 48 & \nodata \\                      
%   NGC~3773 & 11 38 13.0 & 12 06 42.9 &   1.2 &   1.0 & 345 & 0.027 &  13 & $-$2 & SA0                         & \nodata & \nodata \\                 
    NGC~3938 & 11 52 49.4 & 44 07 14.6 &   5.4 &   4.9 & 15 & 0.021 &  12 & 5 & SA(s)c                          & 48 & \nodata \\                      
%   NGC~4125 & 12 08 06.0 & 65 10 26.9 &   5.8 &   3.2 & 275 & 0.019 &  21 & $-$5 & E6 pec                      & \nodata & \nodata \\                 
    NGC~4236 & 12 16 42.1 & 69 27 45.3 &  21.9 &   7.2 & 342 & 0.015 & 4.5 & 8 & SB(s)dm                        & \nodata & \nodata \\                 
    NGC~4254 & 12 18 49.6 & 14 24 59.4 &   5.4 &   4.7 & 35 & 0.039 &  17 & 5 & SA(s)c                          & 48 & \nodata \\                      
    NGC~4321 & 12 22 54.9 & 15 49 20.6 &   7.4 &   6.3 & 30 & 0.026 &  18 & 4 & SAB(s)bc                        & 60 & \nodata \\                      
    NGC~4450 & 12 28 29.6 & 17 05 05.8 &   5.2 &   3.9 & 355 & 0.028 &  17 & 2 & SA(s)ab                        & 48 & \nodata \\                      
    NGC~4536 & 12 34 27.1 & 02 11 16.4 &   7.6 &   3.2 & 310 & 0.018 &  15 & 4 & SAB(rs)bc                      & 48 & \nodata \\                      
%   NGC~4552 & 12 35 39.8 & 12 33 22.8 &   5.1 &   4.7 &   0 & 0.041 &  15 & $-$5 & E0                          & \nodata & \nodata \\                 
    NGC~4559 & 12 35 57.7 & 27 57 35.1 &  10.7 &   4.4 & 330 & 0.018 &  17 & 6 & SAB(rs)cd                      & 48 & \nodata \\                      
    NGC~4569 & 12 36 49.8 & 13 09 46.3 &   9.5 &   4.4 & 23 & 0.046 &  17 & 2 & SAB(rs)ab                       & 48 & \nodata \\                      
    NGC~4579 & 12 37 43.6 & 11 49 05.1 &   5.9 &   4.7 & 275 & 0.041 &  17 & 3 & SAB(rs)b                       & 48 & \nodata \\                      
%   NGC~4594 & 12 39 59.4 & $-$11 37 23.0 &   8.7 &   3.5 & 90 & 0.051 & 9.1 & 1 & SA(s)a                       & \nodata & \nodata \\                    
    NGC~4625 & 12 41 52.7 & 41 16 25.4 &   2.2 &   1.9 & 330 & 0.018 & 9.5 & 9 & SAB(rs)m pec                   & \nodata & 50 \\                 
%   NGC~4631 & 12 42 08.0 & 32 32 29.4 &  15.5 &   2.7 & 86 & 0.017 & 9.0 & 7 & SB(s)d                          & 144 & \nodata \\                     
    NGC~4725 & 12 50 26.6 & 25 30 02.7 &  10.7 &   7.6 & 35 & 0.012 &  17 & 2 & SAB(r)ab pec                    & 96 & \nodata \\                      
    NGC~4736 & 12 50 53.1 & 41 07 13.6 &  11.2 &   9.1 & 285 & 0.018 & 5.2 & 2 & (R)SA(r)ab                     & 75 & 230 \\                          
%    DDO~154 & 12 54 05.3 & 27 08 58.7 &   3.0 &   2.2 & 35 & 0.009 & 4.3 & 10 & IB(s)m                         & \nodata & \nodata \\                 
    NGC~4826\ \ddag & 12 56 43.8 & 21 40 51.9 &  10.0 &   5.4 & 295 & 0.041 &  7.5 & 2 & (R)SA(rs)ab                    & 96 & \nodata \\                      
%    DDO~165 & 13 06 24.9 & 67 42 25.0 &   3.5 &   1.9 & 90 & 0.024 & 4.6 & 10 & Im                             & \nodata & \nodata \\                 
    NGC~5033 & 13 13 27.5 & 36 35 38.0 &  10.7 &   5.0 & 170 & 0.011 &  13 & 5 & SA(s)c                         & 96 & \nodata \\                      
    NGC~5055 & 13 15 49.3 & 42 01 45.4 &  12.6 &   7.2 & 285 & 0.018 & 8.2 & 4 & SA(rs)bc                       & 96 & \nodata \\                      
    NGC~5194\ \dag & 13 29 52.7 & 47 11 42.6 &  11.2 &   9.0 &   0 & 0.035 & 8.4 & 4 & SA(s)bc pec              & 48 & 400 \\                          
%   NGC~5195 & 13 29 59.6 & 47 15 58.1 &   5.8 &   4.6 & 79 & 0.036 & 8.4 & 90 & SB0 pec                        & \nodata & \nodata \\                 
      TOL~89 & 14 01 21.6 & $-$33 03 49.6 &   2.8 &   1.7 & 352 & 0.066 &  16 & 8.1 & (R')SB(s)dm pec           & \nodata & \nodata \\                 
%   NGC~5408 & 14 03 20.9 & $-$41 22 40.0 &   1.6 &   0.8 & 12 & 0.069 & 4.5 & 9.7 & IB(s)m                     & \nodata & \nodata \\                 
%   NGC~5474 & 14 05 01.6 & 53 39 44.0 &   4.8 &   4.3 &   0 & 0.011 & 6.8 & 6 & SA(s)cd pec                    & 48 & \nodata \\                      
    NGC~5713 & 14 40 11.5 & $-$00 17 21.2 &   2.8 &   2.5 & 10 & 0.039 &  27 & 4 & SAB(rs)bc pec                & \nodata & \nodata \\                 
%   NGC~5866 & 15 06 29.6 & 55 45 47.9 &   4.7 &   1.9 & 308 & 0.013 &  15 & $-$1 & SA0                         & \nodata & \nodata \\                 
     IC~4710 & 18 28 38.0 & $-$66 58 56.0 &   3.6 &   2.8 &  5 & 0.089 & 8.5 & 9 & SB(s)m                       & \nodata & \nodata \\                 
%   NGC~6822 & 19 44 56.6 & $-$14 47 21.4 &  15.5 &  13.5 &  7 & 0.236 & 0.60 & 10 & IB(s)m                     & \nodata & \nodata \\                 
    NGC~6946 & 20 34 52.3 & 60 09 14.2 &  11.5 &   9.8 & 75 & 0.342 & 5.5 & 6 & SAB(rs)cd                       & 48 & 400 \\                          
    NGC~7331 & 22 37 04.1 & 34 24 56.3 &  10.5 &   3.7 & 351 & 0.091 &  15 & 3 & SA(s)b                         & 96 & \nodata \\                      
    NGC~7552 & 23 16 10.8 & $-$42 35 05.4 &   3.4 &   2.7 &  1 & 0.014 &  22 & 2 & (R')SB(s)ab                  & 24 & \nodata \\                      
    NGC~7793\ \ddag & 23 57 49.8 & $-$32 35 27.7 &   9.3 &   6.3 & 278 & 0.019 & 3.9 & 7 & SA(s)d                      & 48 & \nodata \\                      
\enddata
\tablecomments{Main properties of the sample. (1): Galaxy name. (2), (3): RA(J2000) and DEC(J2000) of the galaxy center. (4), (5): Apparent major and minor isophotal diameters at $\mu_{B}$=25\,mag\,arcsec$^{-2}$ from the RC3 catalog. (6): Position angle from RC3. \dag The PA and axis ratio of NGC~5194 adopted here differ from those in the RC3, which are affected by the presence of NGC~5195. (7): Galactic color excess from Schlegel et al$.$ (1998). (8): Distance to the galaxy, rounded to the nearest Mpc when larger than 10 Mpc, taken from Gil de Paz et al$.$ (2007) and Kennicutt et al$.$ (2003). \ddag The distances to NGC~4826 and NGC~7793 have been updated with respect to those used in Papers I and II. (9): Morphological type $T$ as given in the RC3 catalog.  (10): Full description of the morphological type from the RC3. (11), (12): Inner and outer limits along the semimajor axis used to restrict the fitting procedure.}
\end{deluxetable*}

\clearpage
\setlength{\tabcolsep}{1.8pt}
\begin{deluxetable*}{lrrrrrrrrrcc}

\tabletypesize{\scriptsize}
\tablecolumns{12}
\tablecaption{Model results.\label{table2}}
\tablewidth{0pt}
\tablehead{
\colhead{} & \multicolumn{4}{c}{Kroupa et al$.$ (1993) IMF} & \colhead{} & \multicolumn{4}{c}{Kroupa (2001) IMF} & \colhead{} & \colhead{}\\[0.5ex]
\cline{2-5} \cline{7-10}\\[-1.5ex]
\colhead{Object name} & \colhead{$\lambda$} & \colhead{$V_{\mathrm{C}}$} & \colhead{Metallicity} & \colhead{Gradient} & \colhead{} & \colhead{$\lambda$} & \colhead{$V_{\mathrm{C}}$} & \colhead{Metallicity} & \colhead{Gradient} & \colhead{$dR_{\mathrm{d}}/dt$} & \multirow{2}{*}{$\frac{R_{\mathrm{d}}(z=0)}{R_{\mathrm{d}}(z=1)}$}\\
\colhead{} & \colhead{} & \colhead{(km\,s$^{-1}$)} & \colhead{at $r=0$} & \colhead{(dex\,kpc$^{-1}$)} & \colhead{} & \colhead{} & \colhead{(km\,s$^{-1}$)} & \colhead{at $r=0$} & \colhead{(dex\,kpc$^{-1}$)} & \colhead{(kpc\,Gyr$^{-1}$)} & \colhead{}\\
\colhead{(1)} & \colhead{(2)} & \colhead{(3)} & \colhead{(4)} & \colhead{(5)} & \colhead{} & \colhead{(6)} & \colhead{(7)} & \colhead{(8)} & \colhead{(9)} & \colhead{(10)} & \colhead{(11)}}

\startdata
NGC~0024 & 0.066$^{+0.015}_{-0.014}$ & 100$^{+  5}_{-  5}$ & 8.92$^{+0.08}_{-0.07}$ & $-$0.118$^{+ 0.014}_{- 0.015}$ & 	 & 0.067$^{+0.015}_{-0.014}$ &  98$^{+  5}_{-  5}$ & 9.49$^{+0.09}_{-0.08}$ & $-$0.128$^{+ 0.016}_{- 0.017}$ & 	0.029 & 1.14\\[0.8ex]
NGC~0337 & 0.037$^{+0.011}_{-0.010}$ & 169$^{+  9}_{-  6}$ & 9.26$^{+0.09}_{-0.12}$ & $-$0.085$^{+ 0.014}_{- 0.009}$ & 	 & 0.036$^{+0.013}_{-0.010}$ & 165$^{+ 14}_{-  9}$ & 9.87$^{+0.10}_{-0.10}$ & $-$0.086$^{+ 0.011}_{- 0.009}$ & 	0.052 & 1.26\\[0.8ex]
NGC~0628 & 0.057$^{+0.014}_{-0.012}$ & 208$^{+ 12}_{- 11}$ & 9.17$^{+0.11}_{-0.11}$ & $-$0.046$^{+ 0.009}_{- 0.012}$ & 	 & 0.055$^{+0.012}_{-0.011}$ & 199$^{+ 12}_{- 10}$ & 9.84$^{+0.07}_{-0.06}$ & $-$0.057$^{+ 0.008}_{- 0.008}$ & 	0.069 & 1.19\\[0.8ex]
NGC~0925 & 0.081$^{+0.027}_{-0.011}$ & 163$^{+ 11}_{- 17}$ & 8.72$^{+0.29}_{-0.11}$ & $-$0.042$^{+ 0.013}_{- 0.023}$ & 	 & 0.095$^{+0.020}_{-0.016}$ & 152$^{+ 16}_{- 10}$ & 9.50$^{+0.07}_{-0.07}$ & $-$0.060$^{+ 0.011}_{- 0.010}$ & 	0.050 & 1.11\\[0.8ex]
NGC~1097 & 0.057$^{+0.021}_{-0.017}$ & 257$^{+ 20}_{- 15}$ & 9.32$^{+0.07}_{-0.09}$ & $-$0.036$^{+ 0.005}_{- 0.004}$ & 	 & 0.056$^{+0.021}_{-0.017}$ & 256$^{+ 22}_{- 17}$ & 9.94$^{+0.09}_{-0.11}$ & $-$0.041$^{+ 0.006}_{- 0.005}$ & 	0.086 & 1.18\\[0.8ex]
NGC~1512 & 0.056$^{+0.013}_{-0.012}$ & 144$^{+  9}_{-  6}$ & 9.10$^{+0.07}_{-0.07}$ & $-$0.078$^{+ 0.009}_{- 0.008}$ & 	 & 0.058$^{+0.017}_{-0.014}$ & 141$^{+  8}_{-  7}$ & 9.68$^{+0.10}_{-0.10}$ & $-$0.085$^{+ 0.012}_{- 0.010}$ & 	0.047 & 1.18\\[0.8ex]
NGC~1566 & 0.052$^{+0.022}_{-0.017}$ & 247$^{+ 23}_{- 18}$ & 9.33$^{+0.07}_{-0.09}$ & $-$0.040$^{+ 0.008}_{- 0.007}$ & 	 & 0.049$^{+0.018}_{-0.014}$ & 242$^{+ 23}_{- 16}$ & 9.96$^{+0.08}_{-0.10}$ & $-$0.048$^{+ 0.007}_{- 0.006}$ & 	0.080 & 1.20\\[0.8ex]
NGC~2403 & 0.052$^{+0.009}_{-0.008}$ & 121$^{+  4}_{-  4}$ & 9.05$^{+0.05}_{-0.05}$ & $-$0.102$^{+ 0.010}_{- 0.010}$ & 	 & 0.051$^{+0.009}_{-0.008}$ & 116$^{+  4}_{-  5}$ & 9.64$^{+0.06}_{-0.06}$ & $-$0.117$^{+ 0.010}_{- 0.010}$ & 	0.037 & 1.19\\[0.8ex]
NGC~2841 & 0.027$^{+0.011}_{-0.007}$ & 276$^{+ 31}_{- 27}$ & 9.41$^{+0.02}_{-0.04}$ & $-$0.039$^{+ 0.005}_{- 0.004}$ & 	 & 0.028$^{+0.012}_{-0.008}$ & 271$^{+ 32}_{- 25}$ & 10.04$^{+0.03}_{-0.05}$ & $-$0.046$^{+ 0.005}_{- 0.004}$ & 	0.063 & 1.22\\[0.8ex]
NGC~2976 & 0.030$^{+0.007}_{-0.005}$ &  76$^{+  2}_{-  2}$ & 9.04$^{+0.06}_{-0.05}$ & $-$0.202$^{+ 0.024}_{- 0.020}$ & 	 & 0.033$^{+0.007}_{-0.006}$ &  75$^{+  2}_{-  2}$ & 9.62$^{+0.05}_{-0.04}$ & $-$0.219$^{+ 0.025}_{- 0.017}$ & 	0.021 & 1.26\\[0.8ex]
NGC~3049 & 0.082$^{+0.028}_{-0.026}$ & 144$^{+ 34}_{- 17}$ & 8.95$^{+0.08}_{-0.39}$ & $-$0.059$^{+ 0.058}_{- 0.025}$ & 	 & 0.091$^{+0.037}_{-0.031}$ & 150$^{+ 56}_{- 24}$ & 9.50$^{+0.10}_{-0.04}$ & $-$0.058$^{+ 0.025}_{- 0.032}$ & 	0.049 & 1.11\\[0.8ex]
NGC~3031 & 0.033$^{+0.009}_{-0.009}$ & 199$^{+ 17}_{-  9}$ & 9.35$^{+0.05}_{-0.08}$ & $-$0.068$^{+ 0.007}_{- 0.005}$ & 	 & 0.028$^{+0.014}_{-0.008}$ & 203$^{+ 24}_{- 18}$ & 10.00$^{+0.06}_{-0.11}$ & $-$0.076$^{+ 0.008}_{- 0.005}$ & 	0.055 & 1.27\\[0.8ex]
NGC~3184 & 0.041$^{+0.014}_{-0.010}$ & 148$^{+ 12}_{-  6}$ & 9.17$^{+0.11}_{-0.10}$ & $-$0.081$^{+ 0.013}_{- 0.018}$ & 	 & 0.045$^{+0.014}_{-0.013}$ & 145$^{+  8}_{-  7}$ & 9.75$^{+0.11}_{-0.08}$ & $-$0.088$^{+ 0.013}_{- 0.015}$ & 	0.048 & 1.22\\[0.8ex]
NGC~3198 & 0.061$^{+0.009}_{-0.010}$ & 198$^{+  6}_{-  6}$ & 9.12$^{+0.10}_{-0.10}$ & $-$0.049$^{+ 0.007}_{- 0.007}$ & 	 & 0.063$^{+0.011}_{-0.010}$ & 191$^{+  7}_{-  5}$ & 9.78$^{+0.06}_{-0.06}$ & $-$0.057$^{+ 0.005}_{- 0.004}$ & 	0.066 & 1.17\\[0.8ex]
IC~2574 & 0.141$^{+0.030}_{-0.021}$ & 103$^{+ 21}_{- 16}$ & 8.52$^{+0.08}_{-0.08}$ & $-$0.075$^{+ 0.023}_{- 0.022}$ & 	 & 0.144$^{+0.037}_{-0.026}$ &  98$^{+ 26}_{- 18}$ & 9.03$^{+0.10}_{-0.10}$ & $-$0.081$^{+ 0.029}_{- 0.029}$ & 	0.027 & 1.06\\[0.8ex]
NGC~3351 & 0.037$^{+0.011}_{-0.009}$ & 197$^{+ 10}_{-  9}$ & 9.32$^{+0.06}_{-0.09}$ & $-$0.065$^{+ 0.010}_{- 0.007}$ & 	 & 0.038$^{+0.012}_{-0.011}$ & 193$^{+ 14}_{- 10}$ & 9.92$^{+0.09}_{-0.08}$ & $-$0.069$^{+ 0.009}_{- 0.009}$ & 	0.061 & 1.24\\[0.8ex]
NGC~3521 & 0.031$^{+0.012}_{-0.009}$ & 226$^{+ 26}_{- 17}$ & 9.37$^{+0.04}_{-0.06}$ & $-$0.053$^{+ 0.006}_{- 0.006}$ & 	 & 0.031$^{+0.015}_{-0.011}$ & 224$^{+ 32}_{- 20}$ & 10.00$^{+0.06}_{-0.10}$ & $-$0.061$^{+ 0.006}_{- 0.006}$ & 	0.062 & 1.25\\[0.8ex]
NGC~3621 & 0.032$^{+0.008}_{-0.007}$ & 161$^{+  8}_{-  9}$ & 9.30$^{+0.07}_{-0.09}$ & $-$0.102$^{+ 0.012}_{- 0.009}$ & 	 & 0.031$^{+0.008}_{-0.007}$ & 156$^{+ 12}_{-  7}$ & 9.91$^{+0.09}_{-0.08}$ & $-$0.105$^{+ 0.011}_{- 0.011}$ & 	0.048 & 1.29\\[0.8ex]
NGC~3627 & 0.030$^{+0.010}_{-0.007}$ & 219$^{+ 21}_{- 12}$ & 9.35$^{+0.03}_{-0.04}$ & $-$0.053$^{+ 0.006}_{- 0.007}$ & 	 & 0.032$^{+0.009}_{-0.008}$ & 215$^{+ 20}_{- 13}$ & 9.98$^{+0.04}_{-0.06}$ & $-$0.061$^{+ 0.005}_{- 0.004}$ & 	0.061 & 1.25\\[0.8ex]
NGC~3938 & 0.045$^{+0.013}_{-0.012}$ & 157$^{+ 10}_{- 10}$ & 9.15$^{+0.13}_{-0.16}$ & $-$0.079$^{+ 0.021}_{- 0.020}$ & 	 & 0.046$^{+0.014}_{-0.013}$ & 149$^{+  8}_{-  5}$ & 9.75$^{+0.10}_{-0.08}$ & $-$0.084$^{+ 0.014}_{- 0.014}$ & 	0.049 & 1.22\\[0.8ex]
NGC~4236 & 0.127$^{+0.017}_{-0.018}$ & 118$^{+  8}_{-  8}$ & 8.65$^{+0.07}_{-0.06}$ & $-$0.067$^{+ 0.010}_{- 0.011}$ & 	 & 0.125$^{+0.020}_{-0.019}$ & 113$^{+ 11}_{-  9}$ & 9.19$^{+0.08}_{-0.07}$ & $-$0.074$^{+ 0.013}_{- 0.013}$ & 	0.033 & 1.07\\[0.8ex]
NGC~4254 & 0.029$^{+0.014}_{-0.009}$ & 239$^{+ 30}_{- 21}$ & 9.38$^{+0.03}_{-0.06}$ & $-$0.047$^{+ 0.006}_{- 0.005}$ & 	 & 0.028$^{+0.014}_{-0.008}$ & 240$^{+ 31}_{- 24}$ & 10.02$^{+0.03}_{-0.08}$ & $-$0.055$^{+ 0.007}_{- 0.006}$ & 	0.060 & 1.24\\[0.8ex]
NGC~4321 & 0.041$^{+0.017}_{-0.013}$ & 295$^{+ 22}_{- 14}$ & 9.38$^{+0.03}_{-0.04}$ & $-$0.030$^{+ 0.005}_{- 0.004}$ & 	 & 0.040$^{+0.017}_{-0.013}$ & 293$^{+ 27}_{- 13}$ & 10.03$^{+0.02}_{-0.05}$ & $-$0.036$^{+ 0.005}_{- 0.004}$ & 	0.084 & 1.20\\[0.8ex]
NGC~4450 & 0.041$^{+0.011}_{-0.010}$ & 212$^{+  8}_{-  6}$ & 9.31$^{+0.05}_{-0.08}$ & $-$0.052$^{+ 0.008}_{- 0.005}$ & 	 & 0.042$^{+0.013}_{-0.011}$ & 209$^{+ 10}_{-  7}$ & 9.92$^{+0.07}_{-0.07}$ & $-$0.058$^{+ 0.008}_{- 0.007}$ & 	0.068 & 1.23\\[0.8ex]
NGC~4536 & 0.060$^{+0.014}_{-0.014}$ & 200$^{+  9}_{-  7}$ & 9.12$^{+0.14}_{-0.15}$ & $-$0.047$^{+ 0.011}_{- 0.009}$ & 	 & 0.059$^{+0.017}_{-0.014}$ & 196$^{+ 10}_{-  9}$ & 9.81$^{+0.10}_{-0.10}$ & $-$0.056$^{+ 0.007}_{- 0.006}$ & 	0.067 & 1.18\\[0.8ex]
NGC~4559 & 0.080$^{+0.010}_{-0.014}$ & 223$^{+ 10}_{-  7}$ & 9.18$^{+0.07}_{-0.12}$ & $-$0.039$^{+ 0.004}_{- 0.004}$ & 	 & 0.080$^{+0.010}_{-0.013}$ & 220$^{+  9}_{-  8}$ & 9.76$^{+0.08}_{-0.06}$ & $-$0.043$^{+ 0.004}_{- 0.004}$ & 	0.076 & 1.13\\[0.8ex]
NGC~4569 & 0.046$^{+0.011}_{-0.009}$ & 256$^{+ 14}_{- 12}$ & 9.38$^{+0.04}_{-0.05}$ & $-$0.042$^{+ 0.003}_{- 0.003}$ & 	 & 0.048$^{+0.011}_{-0.010}$ & 253$^{+ 15}_{- 11}$ & 9.99$^{+0.06}_{-0.07}$ & $-$0.046$^{+ 0.004}_{- 0.003}$ & 	0.082 & 1.20\\[0.8ex]
NGC~4579 & 0.029$^{+0.010}_{-0.008}$ & 264$^{+ 23}_{- 18}$ & 9.38$^{+0.02}_{-0.03}$ & $-$0.038$^{+ 0.004}_{- 0.004}$ & 	 & 0.028$^{+0.010}_{-0.008}$ & 267$^{+ 24}_{- 21}$ & 10.03$^{+0.02}_{-0.03}$ & $-$0.045$^{+ 0.005}_{- 0.004}$ & 	0.063 & 1.22\\[0.8ex]
NGC~4625 & 0.026$^{+0.011}_{-0.006}$ &  72$^{+  4}_{-  4}$ & 9.08$^{+0.07}_{-0.09}$ & $-$0.265$^{+ 0.070}_{- 0.063}$ & 	 & 0.027$^{+0.014}_{-0.007}$ &  71$^{+  4}_{-  5}$ & 9.67$^{+0.09}_{-0.11}$ & $-$0.287$^{+ 0.082}_{- 0.073}$ & 	0.019 & 1.30\\[0.8ex]
NGC~4725 & 0.044$^{+0.016}_{-0.013}$ & 290$^{+ 27}_{- 15}$ & 9.42$^{+0.04}_{-0.06}$ & $-$0.035$^{+ 0.003}_{- 0.003}$ & 	 & 0.043$^{+0.018}_{-0.015}$ & 288$^{+ 36}_{- 15}$ & 10.05$^{+0.05}_{-0.09}$ & $-$0.041$^{+ 0.005}_{- 0.003}$ & 	0.086 & 1.20\\[0.8ex]
NGC~4736 & 0.020$^{+0.008}_{-0.001}$ & 146$^{+  8}_{-  7}$ & 9.34$^{+0.02}_{-0.07}$ & $-$0.125$^{+ 0.017}_{- 0.006}$ & 	 & 0.020$^{+0.009}_{-0.001}$ & 143$^{+  8}_{- 10}$ & 9.96$^{+0.01}_{-0.10}$ & $-$0.140$^{+ 0.020}_{- 0.010}$ & 	0.036 & 1.35\\[0.8ex]
NGC~4826 & 0.020$^{+0.006}_{-0.001}$ & 209$^{+  6}_{- 11}$ & 9.39$^{+0.00}_{-0.01}$ & $-$0.069$^{+ 0.004}_{- 0.005}$ & 	 & 0.020$^{+0.005}_{-0.001}$ & 208$^{+  5}_{- 12}$ & 10.03$^{+0.00}_{-0.03}$ & $-$0.079$^{+ 0.004}_{- 0.004}$ & 	0.043 & 1.27\\[0.8ex]
NGC~5033 & 0.081$^{+0.030}_{-0.016}$ & 196$^{+ 10}_{- 10}$ & 8.89$^{+0.18}_{-0.12}$ & $-$0.036$^{+ 0.009}_{- 0.011}$ & 	 & 0.087$^{+0.024}_{-0.020}$ & 188$^{+ 11}_{-  8}$ & 9.64$^{+0.11}_{-0.10}$ & $-$0.049$^{+ 0.008}_{- 0.007}$ & 	0.063 & 1.12\\[0.8ex]
NGC~5055 & 0.050$^{+0.016}_{-0.013}$ & 215$^{+ 13}_{-  9}$ & 9.26$^{+0.08}_{-0.15}$ & $-$0.047$^{+ 0.009}_{- 0.006}$ & 	 & 0.049$^{+0.017}_{-0.013}$ & 212$^{+ 15}_{- 10}$ & 9.89$^{+0.09}_{-0.09}$ & $-$0.054$^{+ 0.008}_{- 0.006}$ & 	0.072 & 1.21\\[0.8ex]
NGC~5194 & 0.026$^{+0.009}_{-0.006}$ & 239$^{+ 21}_{- 18}$ & 9.39$^{+0.02}_{-0.04}$ & $-$0.051$^{+ 0.006}_{- 0.004}$ & 	 & 0.025$^{+0.008}_{-0.005}$ & 240$^{+ 20}_{- 20}$ & 10.03$^{+0.02}_{-0.04}$ & $-$0.058$^{+ 0.005}_{- 0.005}$ & 	0.054 & 1.24\\[0.8ex]
TOL~89 & 0.066$^{+0.012}_{-0.011}$ & 117$^{+  8}_{-  6}$ & 8.96$^{+0.05}_{-0.05}$ & $-$0.092$^{+ 0.015}_{- 0.014}$ & 	 & 0.070$^{+0.018}_{-0.014}$ & 114$^{+ 14}_{- 10}$ & 9.51$^{+0.07}_{-0.07}$ & $-$0.101$^{+ 0.025}_{- 0.025}$ & 	0.035 & 1.14\\[0.8ex]
NGC~5713 & 0.020$^{+0.007}_{-0.001}$ & 226$^{+ 17}_{- 19}$ & 9.40$^{+0.00}_{-0.03}$ & $-$0.060$^{+ 0.010}_{- 0.010}$ & 	 & 0.020$^{+0.010}_{-0.001}$ & 224$^{+ 18}_{- 25}$ & 10.04$^{+0.00}_{-0.06}$ & $-$0.070$^{+ 0.012}_{- 0.012}$ & 	0.044 & 1.25\\[0.8ex]
IC~4710 & 0.078$^{+0.022}_{-0.014}$ &  99$^{+ 23}_{- 17}$ & 8.84$^{+0.05}_{-0.06}$ & $-$0.107$^{+ 0.033}_{- 0.033}$ & 	 & 0.087$^{+0.028}_{-0.018}$ &  99$^{+ 28}_{- 20}$ & 9.37$^{+0.07}_{-0.08}$ & $-$0.110$^{+ 0.042}_{- 0.038}$ & 	0.029 & 1.10\\[0.8ex]
NGC~6946 & 0.030$^{+0.008}_{-0.006}$ & 189$^{+ 12}_{- 10}$ & 9.34$^{+0.02}_{-0.06}$ & $-$0.073$^{+ 0.012}_{- 0.009}$ & 	 & 0.029$^{+0.008}_{-0.007}$ & 186$^{+ 13}_{- 11}$ & 9.95$^{+0.04}_{-0.06}$ & $-$0.078$^{+ 0.011}_{- 0.008}$ & 	0.053 & 1.28\\[0.8ex]
NGC~7331 & 0.059$^{+0.027}_{-0.021}$ & 265$^{+ 20}_{- 16}$ & 9.33$^{+0.09}_{-0.10}$ & $-$0.035$^{+ 0.006}_{- 0.005}$ & 	 & 0.059$^{+0.029}_{-0.023}$ & 263$^{+ 22}_{- 16}$ & 9.94$^{+0.11}_{-0.13}$ & $-$0.039$^{+ 0.008}_{- 0.006}$ & 	0.089 & 1.17\\[0.8ex]
NGC~7552 & 0.034$^{+0.019}_{-0.014}$ & 223$^{+ 35}_{- 17}$ & 9.34$^{+0.05}_{-0.13}$ & $-$0.048$^{+ 0.007}_{- 0.008}$ & 	 & 0.033$^{+0.020}_{-0.013}$ & 222$^{+ 37}_{- 19}$ & 9.98$^{+0.05}_{-0.13}$ & $-$0.057$^{+ 0.009}_{- 0.006}$ & 	0.064 & 1.25\\[0.8ex]
NGC~7793 & 0.040$^{+0.009}_{-0.008}$ & 104$^{+  6}_{-  5}$ & 9.06$^{+0.06}_{-0.05}$ & $-$0.133$^{+ 0.019}_{- 0.017}$ & 	 & 0.039$^{+0.010}_{-0.009}$ & 101$^{+  5}_{-  5}$ & 9.66$^{+0.07}_{-0.07}$ & $-$0.153$^{+ 0.021}_{- 0.021}$ & 	0.031 & 1.24\\[0.8ex]
\enddata
\tablecomments{Results from the model fitting. (1): Galaxy name. (2), (6): Dimensionless spin parameter. (3), (7): Maximum circular velocity. (4), (8): Central value of $12+\log(O/H)$. (5), (9): Radial metallicity gradient. (10): Temporal growth rate of the stellar disk scale-length, obtained by fitting $R_\mathrm{d}(t)$ between $z=1$ and $z=0$. (11): Ratio of the stellar disk scale-lengths at $z=0$ and $z=1$. Neither (10) nor (11) vary noticeably with the IMF.}
\end{deluxetable*}

\end{document}